\documentclass[usenatbib,useAMS]{mnras}
\usepackage{tablefootnote}
\usepackage{multirow}
\usepackage{xcolor}
\usepackage{graphicx}	
\usepackage{gensymb}
\usepackage{amsmath}	
\usepackage{amssymb}	
\usepackage{multicol}   
\usepackage{bm}		
\usepackage[normalem]{ulem} 
\usepackage{caption} 
\usepackage{url}
\usepackage{longtable}
\usepackage{pdfpages}
\usepackage{subcaption}
\usepackage{hyperref}
\usepackage{pdflscape}
\usepackage{orcidlink}

\usepackage[T2A]{fontenc}
\usepackage[utf8x]{inputenc}
\usepackage[russian,english]{babel}

\usepackage{xspace}
\usepackage[export]{adjustbox}
\newcommand{\pks}{PKS\,0735+178\xspace}
\usepackage{upgreek}

\title[Multiwavelength variability of \pks]{Multiwavelength variability of the high-energy neutrino candidate \pks over three decades}

\author[Mufakharov et al.]{%
T.~V.~Mufakharov,$^{1,2}\orcidlink{0000-0001-9984-127X}$\thanks{E-mail: timur.mufakharov@gmail.com}  
Yu.~V.~Sotnikova,$^{2,3}\orcidlink{0000-0001-9172-7237}$
V.~V.~Vlasyuk,$^{2}\orcidlink{0009-0002-6596-7274}$
D.~O.~Kudryavtsev,$^{2}\orcidlink{0000-0003-2461-1570}$
\newauthor
A.~B.~Pushkarev,$^{4,5}\orcidlink{0000-0002-9702-2307}$
A.~G.~Mikhailov,$^{2}\orcidlink{0000-0002-0279-0777}$
M.~L.~Khabibullina,$^{2}\orcidlink{0000-0001-9515-4552}$
Yu.~A.~Kovalev,$^{5,3}\orcidlink{0000-0002-8017-5665}$
\newauthor
Y.~Y.~Kovalev,$^{6}\orcidlink{0000-0001-9303-3263}$
A.~V.~Popkov,$^{7,3,5}\orcidlink{0000-0002-0739-700X}$
A.~K.~Erkenov,$^{2}\orcidlink{0000-0002-6086-9299}$
O.~I.~Spiridonova,$^{2}\orcidlink{0009-0007-7315-3090}$
\newauthor
T.~A.~Semenova,$^{2}\orcidlink{0000-0002-2902-5426}$
P.~G.~Tsybulev,$^{2,3}\orcidlink{0000-0001-5600-8018}$
D.~S.~Nezamov$^{8}\orcidlink{0009-0001-0597-5793}$
\\
$^{1}$ State Key Laboratory of Radio Astronomy and Technology, Xinjiang Astronomical Observatory, CAS, 150 Science 1-Street, \\ Urumqi 830011, China \\
$^{2}$ Special Astrophysical Observatory of the Russian Academy of Sciences, Nizhny Arkhyz, 369167, Russia\\
$^{3}$ Institute for Nuclear Research, Russian Academy of Sciences, 60th October Anniversary Prospect 7a, Moscow 117312, Russia\\
$^{4}$ Crimean Astrophysical Observatory of the Russian Academy of Sciences, 298409, Nauchny, Russia \\
$^{5}$ Astro Space Center, Lebedev Physical Institute, Russian Academy of Sciences, 117997, Moscow, Russia \\
$^{6}$ Max-Planck-Institut f\"ur Radioastronomie, Auf dem H\"ugel 69, Bonn 53121, Germany\\
$^{7}$ Moscow Institute of Physics and Technology, Institutsky per. 9, Dolgoprudny 141700, Russia\\
$^{8}$ Kazan Federal University, 18 Kremlyovskaya St, Kazan 420008, Russia\\
}
\date{Accepted 2026 June 15. Received 2026 June 15; in original form 2026 May 11}
\pubyear{2026}

\begin{document}
\label{firstpage}
\pagerange{\pageref{firstpage}--\pageref{lastpage}}
\maketitle 

\begin{abstract} 
We present the multiwavelength variability of the BL Lac object \pks, associated with the high-energy neutrino event IC211208A. The light curves cover the radio (1--230\,GHz), optical, and {\it Fermi}-LAT $\gamma$-ray bands 
over a $\sim$30 years time-scale. The light curves are correlated, with delays 
from 0 to 1200\,days 
increasing towards lower frequencies, consistent with 
emission from an opacity-stratified jet. The bright flare 
after the IC211208A event indicates emission from a compact, 
optically thick region
with enhanced activity and more efficient particle acceleration. Short optical and $\gamma$-ray bursts have been detected very close to the neutrino event, within a few days. The $\sim$12-day lag between the $\gamma$-ray and optical emissions 
is detected for the first time, suggesting that the emission regions are not co-spatial.
A characteristic variability time-scale of $\sim$10--11 yr is robustly detected in the radio--mm bands ($\geq3\sigma$), while the optical and $\gamma$-ray data show weaker, 
shorter-period signals. The independent constraints from the VLBI core-shift measurements and radio time delays yield consistent estimates of the jet geometry and disturbance propagation, supporting variability governed by jet propagation effects.
The long-term modulation is consistent with a slow variation in energy release at the jet base, while individual flares arise from shocks propagating downstream. 
Jet precession may contribute to the long-term modulation; however, the required viewing angles are inconsistent with the VLBI constraints, indicating that precession alone cannot explain the observed variability.
\end{abstract}

\maketitle 
\section{Introduction}
Blazars, a highly energetic subclass of active galactic nuclei (AGNs), are defined by a relativistic plasma jet oriented close to the observer's line of sight. This geometric alignment results in dramatic Doppler boosting, making them the most luminous persistent sources in the non-thermal sky. Study of flux variability and jet kinematics provides some of the most direct constraints on particle acceleration and energy dissipation processes within relativistic jets.

\pks is a classical BL Lac object, notable for its extreme multiwavelength (MW) variability and a relativistic jet pointed close to the line of sight. Despite the object is featureless in the optical spectrum, its redshift is generally taken as $z = 0.45$ \citep{2012A&A...547A...1N}, though recent estimates suggest $z \sim 0.65$ \citep{2021ATel15132....1F}. Very-long-baseline interferometry (VLBI) observations reveal a one-sided parsec-scale jet with superluminal components, viewed at small angles ($\theta \sim$ 1\fdg09--3\fdg36, \citealt{2022ApJS..260...12W}), and at least two bends \citep{2006A&A...453..477A}, while the optical and radio fluxes show large-amplitude long-term variability with occasional violent flares \citep{1988AJ.....95..374W, 2007A&A...467..465C, 2022ApJ...933..224F}.

\pks is of particular interest as a possible 
source of a high-energy neutrino (171 TeV) associated with the track-like neutrino event IC211208A \citep{2021ATel15105....1K} that has a $\sim$30 per cent probability of astrophysical origin and an angular separation of $\sim$2.2$\degree$ from \pks \citep{2021GCN.31191....1I}. Enhanced emission was detected in December 2021 across the whole spectrum, from radio to $\gamma$-rays \citep{2023ApJ...954...70A}.
A significant rise in {\it Fermi}-LAT $\gamma$-ray flux (0.1--300\,GeV) was accompanied by a hardening of the spectral slope. Swift/XRT recorded a $>2\times$ X-ray flux increase within $\sim$1.4\,h \citep{2023MNRAS.519.1396S}. 
Coincident with this activity, several high-energy neutrino observatories reported events from the vicinity of \pks. Approximately four hours later, the Baikal-GVD experiment reported a 43 TeV cascade-like event with a $\sim 50$ per cent probability of astrophysical origin \citep{2021ATel15112....1D}. Notably, four days earlier
the Baksan Underground Scintillation Telescope detected a muon neutrino with energy above 1 GeV from a direction consistent with the direction to \pks \citep{2021ATel15143....1P}, and the KM3NeT/ARCA array identified an 18 TeV event in the archival data for December 2021 with a marginal association ($p$-value $\sim 0.14$) \citep{2022ATel15290....1F}. These multiple detections reported by independent facilities make \pks the only known blazar with such spatial and temporal clustering of neutrino events.

The radio light curves of \pks demonstrate a very high level of flaring activity. The first and largest synchrotron flare was detected in 1989--1991 with flux density reaching its highest historical level of more than 5\,Jy \citep{2009MNRAS.399.1622G,1985ApJ...298..296A}. In 2003, 2013, and 2023, \pks underwent three intense flares with flux densities of $\sim 2$\,Jy at radio frequencies. The last MW flare began before IC211208A in 2021.

In some studies the quasi-periodic behavior of \pks was detected in the radio and optical bands, with characteristics time-scales from a few years to 13--14 years \citep{1997A&AS..125..525F,2004IJMPD..13..771D,2004PASP..116..161Q,2007A&A...467..465C}. The presence of 
a
twisted relativistic jet \citep{1994AJ....108...56P,1994ApJ...435..128G,2001MNRAS.328..873G,2001MNRAS.328..719G} is often considered as a possible physical mechanism responsible for the observed periodicity. This assumption requires further verification using more extended, modern multiwavelength datasets, particularly in the radio and optical bands, to separate the contribution of geometric and intrinsic physical drivers.

Multiwavelength modelling of the \pks SED during the 2021 activity suggests that purely leptonic scenarios may be insufficient to reproduce the observed high-energy emission and that additional hadronic or external photon contributions may be required (e.g., \citealt{2023ApJ...954...70A, 2024MNRAS.529.3503B,2024MNRAS.527.8746P}). However, these studies focus primarily on short-term spectral properties and do not address the long-term temporal behaviour of the source.

Despite extensive studies of \pks, its long-term multiwavelength variability and the physical connection between the emission regions at different frequencies remain poorly understood. In particular, it is still unclear how the observed time delays, spectral evolution, and quasi-periodic behavior are linked to the structure and dynamics of the relativistic jet.

In this work we present a comprehensive analysis of \mbox{$\sim 30$}~years of multiwavelength observations of \pks, combining radio, optical, and $\gamma$-ray data. We investigate temporal correlations between the bands, the evolution of radio spectra, and the presence of quasi-periodic variability. By combining an analysis of time delays with the VLBI core-shift measurements, we aim to constrain the geometry and physical conditions of the jet and to test a unified scenario of shock propagation in an opacity-stratified flow.

\begin{figure}
\begin{minipage}[h]{1.0\linewidth}
\center{\includegraphics[width=1\linewidth]{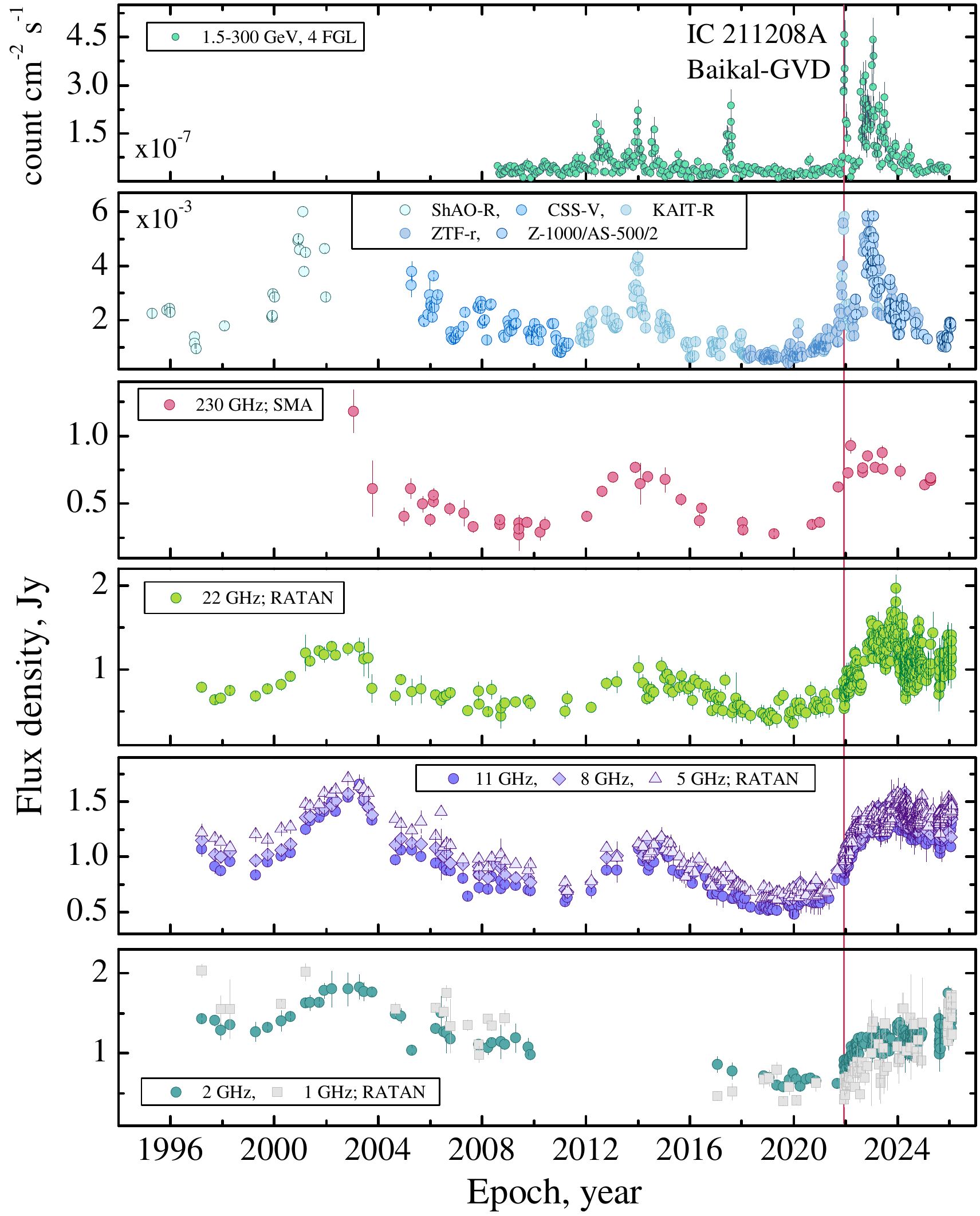}}
\end{minipage}
\hfill
\begin{minipage}[h]{1.0\linewidth}
\center{\includegraphics[width=1\linewidth]{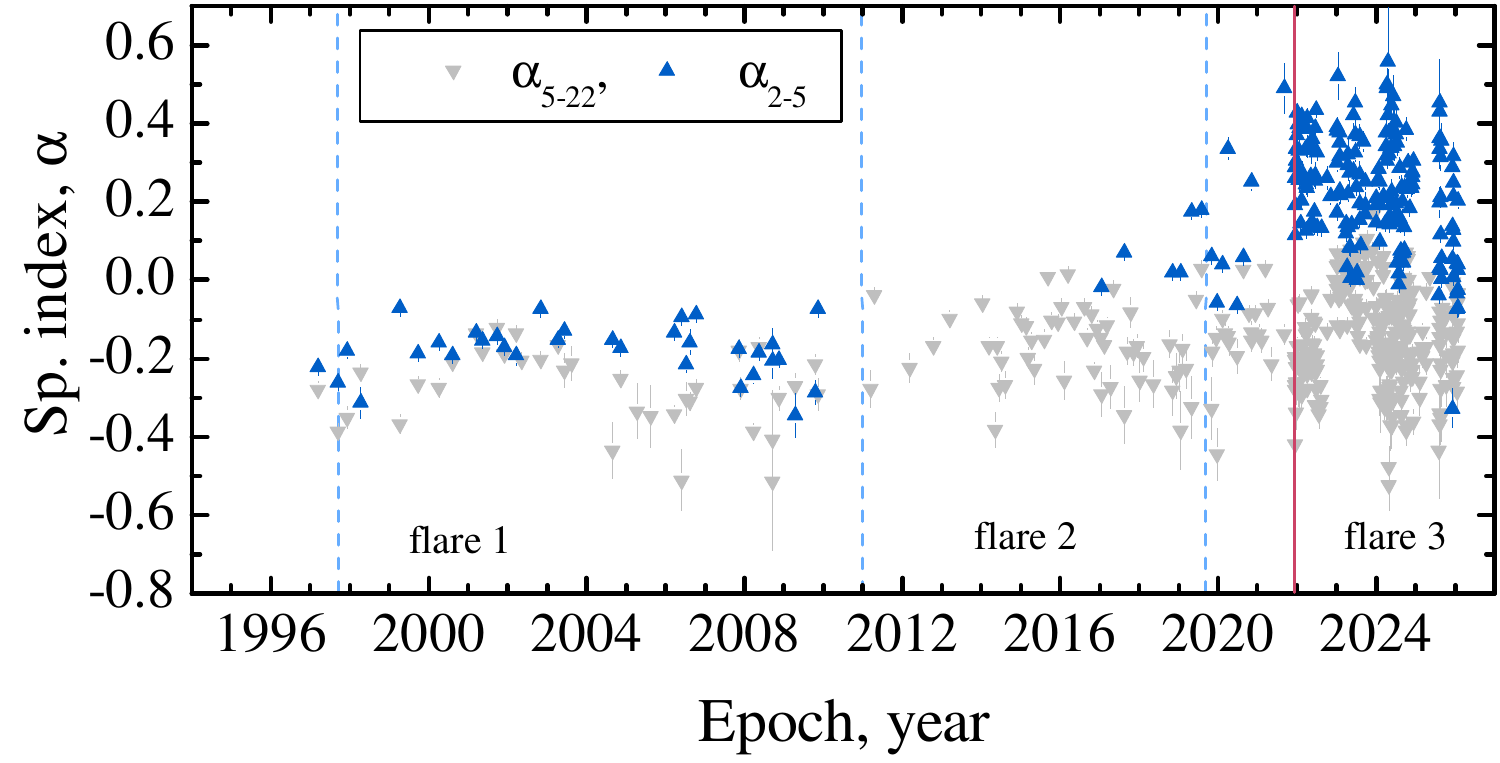}}
\end{minipage}
\caption{Multiband light curves of \pks in 1995--2026 (upper panel), combined with the corresponding low- and high-frequency spectral indices, $\alpha_{2-5}$ and $\alpha_{5-22}$, in the same period (lower panel). Areas of the three flares are marked with vertical dashed lines. The epoch of the neutrino event IC211208A/Baikal-GVD is marked by a vertical line over all panels.}
\label{fig:light}
\end{figure}

\section{Observed data}

For the multiwavelength variability analysis (Table~\ref{tab:faci}), we use both the new observed data from the RATAN-600 (1.2, 2.3, 4.7, 8.2, 11.2, 22.3 GHz), SMA (230 GHz), and Zeiss-1000 and AS-500/2 (optical $R$ band) telescopes as well as previously published measurements. The light curves span 1997--2026 for the RATAN-600 data, 2003--2025 for the SMA observations, 1995--2026 for the optical band fluxes, and 2008--2026 for the Fermi Large Area Telescope (Fermi LAT) data in the 1.5-300 GeV range. The combined radio-to-$\gamma$-ray light curves are presented in Fig.~\ref{fig:light}. The neutrino event IC211208A/Baikal-GVD is marked by a vertical line over all panels.

\begin{table}
\caption{The instruments used in this study.} 
\label{tab:faci}
\centering
\begin{tabular}{llcl}
\hline
Telescope & Institute & Epoch & Band \\
\hline
RATAN-600  &  SAO RAS  & 1997--2026 & 1--22 GHz \\ 
SMA        &  SAO \& ASIAA & 2003--2025 & 230 GHz \\ 
Zeiss-1000 & SAO RAS  & 2022--2026 & $R$ band \\ 
AS-500/2   & SAO RAS  & 2022--2026 & $R$ band \\ 
Catalina 0.7-m & CRTS & 2005--2011 & $V$ band \\ 
KAIT 0.76-m & Berkeley Univ. & 2011--2024 & $R$ band \\ 
48$''$ Schmidt & ZTF Palomar & 2018--2024 & $r$ band \\ 
1.56-m &ShAO CAS & 1995-2001 & $R$ band \\ 
Fermi-LAT & NASA  & 2008--2025 & $\gamma$-ray \\ 
\hline
\end{tabular}
\end{table} 

\subsection{RATAN-600}

\pks is systematically observed with RATAN-600 at six frequencies quasi-simultaneously thanks to
the telescope ring geometry \citep{1979S&T....57..324K,1993IAPM...35....7P,2020gbar.conf...32S}. The procedure of data reduction and calibration is described by \citet{1999A&AS..139..545K,2011AstBu..66..109T,2016AstBu..71..496U,2018AstBu..73..494T,2019AstBu..74..497S}. 
We used the following list of secondary calibrators for flux density calibration to cover a wide declination range from $-35^\circ$ to $+49^\circ$: J1154$-$35, J0240$-$23, 3C\,161, 4C$+$12.50, 3C\,286, 3C\,48, NGC\,7027, and 3C\,147 \citep{1977A&A....61...99B,1980A&AS...39..379T,1985BSAO...19...59A,1994A&A...284..331O,2013ApJS..204...19P,2017ApJS..230....7P}. The characteristic cadence of the RATAN-600 observations (Fig.~\ref{fig:light}) 
ranges from several times per year (1997--2021) to once per month (since 2022). In this paper we denote the RATAN-600 frequencies by their rounded values: 1, 2, 5, 8, 11, and 22 GHz.
The flux densities at 1-22 GHz $S_{\nu}$, their errors $\sigma$, and average observing epochs (yyyy.dd.mm, JD and yyyy.yyyy) are presented in Table~\ref{TableA1_part} (a fragment).

\begin{table*}
\caption{\label{TableA1_part} A fragment of the RATAN-600 measurements of \pks in 1997--2026: epochs in yyyy.mm.dd (Col.~1), Julian Date (JD) (Col.~2), epoch in yyyy.yyyy (Col.~3), flux densities at 22, 11, 8, 5, 2, and 1 GHz and their errors in Jy (Cols.~4--15). The full version is available as online supplementary material.} 
\begin{tabular}{|l|c|c|c|c|c|c|c|c|c|c|c|c|c|c|c|}
\hline
yyyy.mm.dd & JD  &  yyyy.yyyy & $S_{22}$ & $\sigma$ & $S_{11}$ & $\sigma$ & $S_{8}$ & $\sigma$ & $S_{5}$ & $\sigma$ & $S_{2}$ & $\sigma$ & $S_{1}$ & $\sigma$  \\
(1) & (2) & (3) & (4) & (5) & (6) & (7) & (8) & (9) & (10) & (11) & (12) & (13) & (14) & (15) \\
\hline
2025.12.12 & 2461021 &	2025.9452 &	1.24 &	0.20 &	1.34 &	0.09 &	1.35 &	0.03 &	1.40 &	0.05 &	1.27 &	0.07 &	1.15 &	0.09 \\
2025.12.13 & 2461022 &	2025.9479 &	1.11 &	0.18 &	1.23 &	0.07 &	1.36 &	0.03 &	1.38 &	0.20 &	1.75 &	0.09 &	1.62 &	0.09 \\
2025.12.16 & 2461025 &	2025.9562 &	1.41 &	0.16 &	1.31 &	0.06 &	1.38 &	0.03 &	1.42 &	0.05 &	1.41 &	0.18 &	1.51 &	0.07 \\
\hline
\end{tabular}
\end{table*}
   
\subsection{SMA observations}

At 230 GHz (1.3 mm) we exploited the measurements from the Submillimeter Array\footnote{\url{http://sma1.sma.hawaii.edu/callist/callist.html}}
(SMA; \citealt{2007ASPC..375..234G}), obtained during the period from January 2003 to April 2025. The measurements have been made with a relatively low cadence: one measurement every 160 days in average.

\subsection{Optical photometry}

The optical $R$ band data,
covering the observing period from 2005 to 2024, have been collected from several instruments: the SAO RAS 1-metre Zeiss-1000 and \mbox{0.5-metre} \mbox{AS-500/2} optical reflectors (since May 2022), the Zwicky Transient Facility (ZTF) \citep{2019PASP..131a8002B}, which has been operating since 2018, 
the Catalina Real-time Transient Survey (CRTS) \citep{2009ApJ...696..870D}, and the Katzman Automatic Imaging Telescope (KAIT) \citep{2001ASPC..246..121F}. The earliest optical data were taken from \citep{2004PASP..116..161Q}. Details on the instruments of the SAO RAS telescopes are described in \cite{2022Photo...9..950V} and \cite{2024MNRAS.535.2775V}. All the data were averaged over individual nights, converted into the $R$~band if necessary, and the resulting values were transformed into flux densities according to the constant from \cite{1990A&AS...83..183M}. The $R$ band fluxes  $R_{\rm flux}$, their errors $\sigma$, and average observing epochs for SAO RAS data are presented in Table~\ref{TableA3_part} (a fragment).

\begin{table}
\caption{\label{TableA3_part} 
A fragment of the SAO RAS $R$ band measurements
of \pks in 2022--2026: epoch in yyyy.mm.dd (Col.~1), Julian Date (JD) (Col.~2), epoch in yyyy.yyyy (Col.~3), flux density and its error in mJy (Cols.~4--5) and the telescope designation (Col.~6). The full version is available as online supplementary material.} 
\begin{tabular}{|c|c|c|c|c|c|}
\hline
yyyy.mm.dd & JD  &  yyyy.yyyy & $R_{\rm flux}$ & $\sigma$ & Instrum \\
(1) & (2) & (3) & (4) & (5) & (6) \\
\hline
2022.12.23 & 2459937 & 2022.9765 & 4.33 & 0.04 & AS-500/2\\
2022.12.25 & 2459939 & 2022.9820 & 3.95 & 0.03 & AS-500/2\\
2023.01.13 & 2459958 & 2023.0338 & 3.81 & 0.01 & Z-1000\\
2023.01.14 & 2459959 & 2023.0366 & 4.18 & 0.02 & AS-500/2\\
2023.01.23 & 2459968 & 2023.0612 & 4.46 & 0.04 & AS-500/2\\
2023.01.24 & 2459969 & 2023.0640 & 4.19 & 0.07 & Z-1000\\

\hline
\end{tabular}
\end{table}

\subsection{\it {Fermi}-LAT}

The radio object \pks is positionally associated with the $\gamma$-ray source 4FGL J0738.1+1742 detected by the Large Area Telescope (LAT) onboard the Fermi Gamma-ray Space Telescope. The angular distance between the Fermi-LAT and VLBI positions is $0\farcm46$ which is within the 95 per cent error ellipse of the $\gamma$-ray position \citep{4FGL-DR3,RFC}.

We have used the Fermi-LAT raw data since August 4, 2008 to produce a $\gamma$-ray light curve of the source. To construct the light curve with a detailed structure during flaring states, which is important for the purposes of our study, we applied the adaptive binning technique developed by \cite{Lott12} that allows one to extract more information from the data compared to the fixed binning approach. The brighter the source, the more accurate the timing of the light curve fine structure can be derived. For the binning we (i) adopted a target constant flux uncertainty at a level of 25 per cent in each bin, (ii) used a normal time arrow, and (iii) considered an integral flux within 1.5--300~GeV. In total, the obtained light curve comprises 346 bins, the widths of which range from 2 to 63 days with a median of 18 days. The corresponding test statistic ($\mathrm{TS}=2\ln(L_1/L_0)$, where $L_1$ and $L_0$ are the likelihoods for the models with and without a point source at a given position), as a measure of detection significance, ranges from 7 to 352 with a median of 87. The source energy spectrum is assumed to be a fixed power-law with a photon spectral index 1.98. The Region-Of-Interest was set at a radius of $20^\circ$ around the target source. It contains 136 point sources. To derive more accurate fluxes of the source of interest, the emission from two bright confusing sources was taken into account, namely 4FGL J0739.7+1743 ($0\fdg4$ from the target source) and 4FGL J0725.2+1425 ($4\fdg5$), each having a radio counterpart: J0738+1742 and J0725+1425, respectively \citep{RFC}. 

\section{Radio spectra}
\label{sec:spectra}

\begin{figure*}
    \centering
    \includegraphics[width=1.0\linewidth]{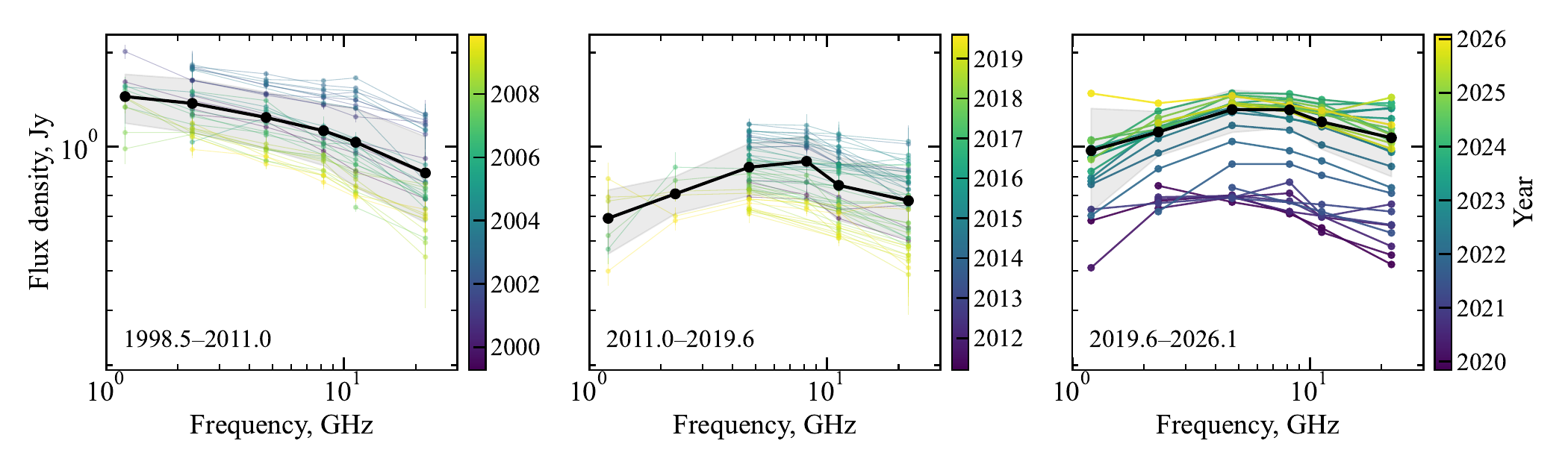}
    \caption{The RATAN-600 radio spectra of \pks. Individual spectra are shown for the first two flaring episodes. For the third episode (2019.6--2026.1), 3-month averaged spectra are displayed to reduce crowding. Thick black curves represent the mean spectrum of each flaring episode.}
    \label{fig:spectra}
\end{figure*}

To compare the spectral states of \pks, we divided the long-term activity into three major flaring episodes using local minima in the radio and optical light curves as the boundaries. The minima are not strictly simultaneous in different bands, mainly due to uneven sampling and opacity-related frequency-dependent evolution; therefore, the adopted boundaries should be regarded as approximate divisions of activity states rather than sharply determined physical start times. For the first episode, the optical and radio minima occur at 1996.9 and 1998.5, respectively. The pre-1998 data are retained in Fig.~\ref{fig:light} for completeness, but are excluded from the quantitative comparison of the radio spectral states because they may still include the declining stage of an earlier radio flare. The subsequent minima are observed in the optical light curve at 2011.0 and 2019.6. The multi-epoch radio spectra, grouped according to these intervals (Fig.~\ref{fig:spectra}), reveal systematic differences in spectral shape between the three flaring episodes.
During the first (1998.5--2011.0) and second (2011.0--2019.6) flares, the spectra are predominantly steep to moderately flat across the 1–-22 GHz range, indicating that the emission is largely optically thin or only weakly affected by synchrotron self-absorption. This is consistent with the corresponding spectral index $\alpha$ evolution (Fig.~\ref{fig:light}, bottom panel), where $\alpha$ remains close to zero or slightly negative ($\alpha \sim 0$ to $-0.2$, here we adopt $S\propto\nu^\alpha$).

In contrast, the third flare (2019.6--2026.1), which coincides with the period of enhanced activity around the IceCube-211208A event, exhibits a qualitatively different behavior. The radio spectra frequently display flattened or inverted shapes, particularly in the 2–-5 GHz range, indicating a significant contribution from optically thick emission. This is reflected in the spectral index evolution, where $\alpha$ reaches positive values and shows larger variability compared to earlier epochs. The improved temporal sampling during this period allows the spectral changes to be traced in greater detail, revealing transitions between optically thick and thin regimes.

The spectral index was derived for two selected frequency ranges: 2–-5 and 5–-22 GHz (Table~\ref{tab:alpha}). According to the Kolmogorov--Smirnov test, the low-frequency spectral index $\alpha_{2-5}$ distributions are statistically different between all three flaring epochs. 
This indicates systematic evolution of the low-frequency spectrum shape, consistent with an increasing contribution of optically thick emission during the most recent flare. The high-frequency index $\alpha_{5-22}$ differs significantly between the first epoch and the two later epochs, but the distributions for the second and third epochs are statistically consistent. This suggests that the strongest spectral evolution during the recent activity is concentrated at lower radio frequencies, while the 5--22 GHz spectral regime remains broadly similar after 2011.
The systematic differences in $\alpha$ between the three flaring epochs suggest that the physical conditions in the emitting regions, such as optical depth, magnetic field, and particle density, vary between events. In particular, the more frequent occurrence of inverted spectra during the third flare indicates a more compact and optically thick emission region, consistent with enhanced activity and more efficient particle acceleration in the inner jet.

\begin{table}
\caption{The median values of $\alpha_{2-5}$ and $\alpha_{5-22}$ in the individual flaring episodes.} 
\label{tab:alpha}
\centering
\begin{tabular}{ccc}
\hline
Epoch & $\alpha_{2-5}$ & $\alpha_{5-22}$ \\
\hline
1998.5--2011.0 & -0.16 & -0.27 \\ 
2011.0--2019.6 & +0.05 & -0.17 \\ 
2019.6--2026.1 & +0.25 & -0.15 \\ 
\hline
\end{tabular}
\end{table} 

\section{Analysis of the multiwavelength light curves}

The MW light curves of \pks demonstrate three bright and prolonged radio and optical flares in 1995--2026 (Fig.~\ref{fig:light}). Two of them coincide with the elevated activity states in $\gamma$-rays in 2012--2015 and 2022-2025. Additionally, two short $\gamma$-ray flares with a characteristic timescale of a few dozen days are seen in 2017 and in late 2021. The late 2021 short $\gamma$-ray flare coincides in time with the short optical flare detected very close to IC211208A. 
 
\subsection{Short $\gamma$-ray and $R$ band flares in 2021--2022}

\begin{figure}
    \centering
    \includegraphics[width=0.9\linewidth]{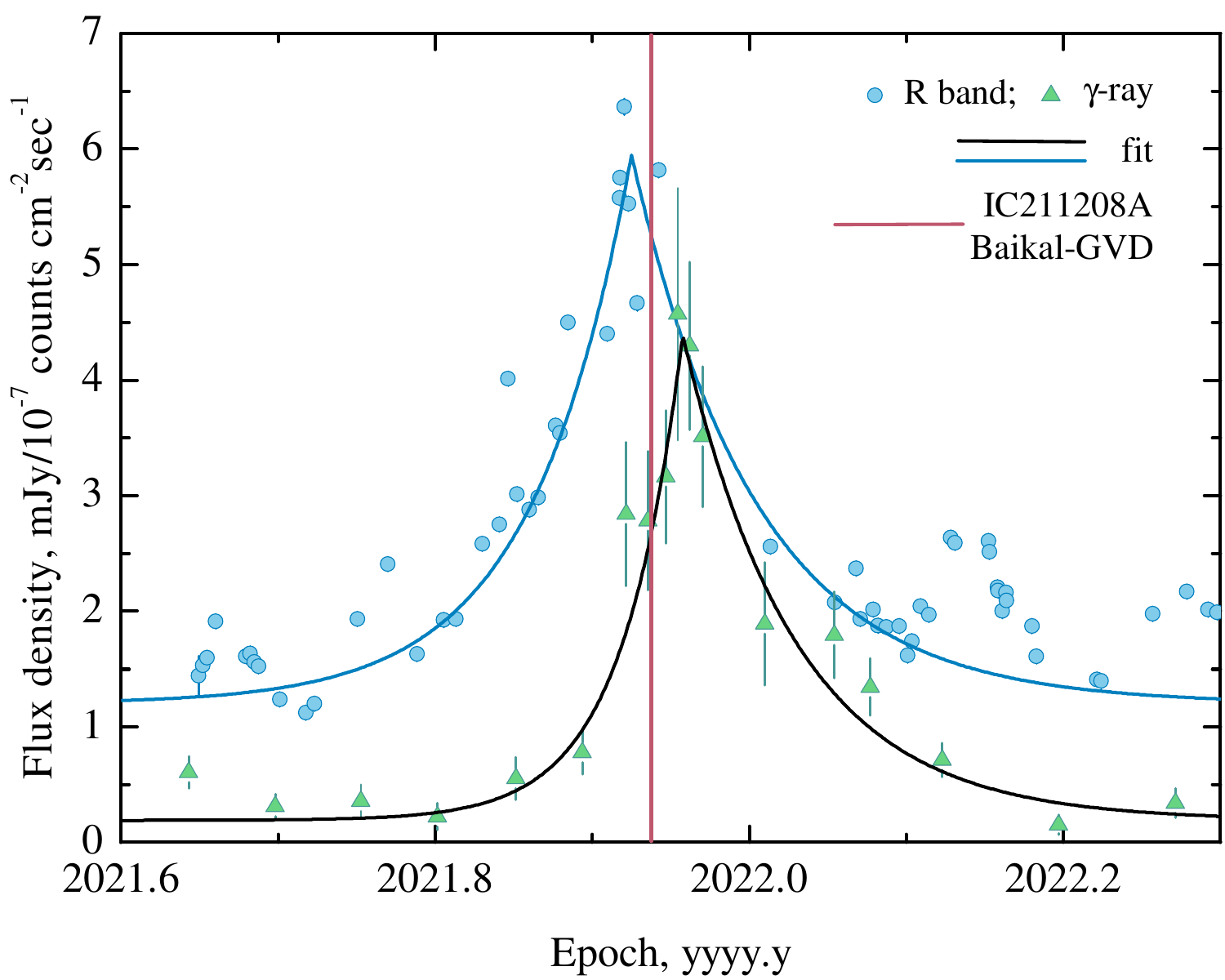}
    \caption{Fine structure of the late-2021 $\gamma$-ray and optical pre-flares, coincided with the IC211208A and Baikal-GVD events (marked by solid vertical line) in 2021--2022.}
    \label{fig:short_flares_vv}
\end{figure}

The short bright $\gamma$-ray and optical $R$ band flares near the IC211208A event (during 2021.5--2022.5) are of particular interest. The emission in these bands strongly correlates, but the time delay between the flares has not been accurately determined in previous studies \citep{2023MNRAS.519.1396S,2024MNRAS.529.3503B,2025A&A...695A.266O}. We fitted the flares (Fig.~\ref{fig:short_flares_vv}) with exponential functions as follows \citep{1999ApJS..120...95V,2009A&A...494..527H}:
\begin{equation}
\label{eq:exp}
\Delta S(t) = 
 \begin{cases}
   \Delta S_{\mathrm{max}}\, e^{{(t-t_{\mathrm{max}})/\tau_{1}}}, & t<t_{\mathrm{max}}, \\
   \Delta S_{\mathrm{max}}\, e^{(t_{\mathrm{max}}-t)/\tau_{2}}, & t>t_{\mathrm{max}}, \\
 \end{cases}
 \end{equation}
where $\Delta S_{\mathrm{max}}$ is the maximum amplitude of a flare (in mJy for the $R$ band and in counts\,cm$^{-2}$\,s$^{-1}$ for $\gamma$-rays), $t_{\mathrm{max}}$ is the epoch of the flare maximum, and $\tau_{1}$ and $\tau_{2}$ are the characteristic times of flare rising and falling. 

The exponential model parameters were calculated by minimization of the $\chi^2$ statistic over the data range considered after removing sub-outbursts that hinder the analysis (e.g., the features in the optical light curve at 2022.15 and 2022.27). 
We estimated $\tau_{\rm R,\,1}=23\pm3$, $\tau_{\rm R,\,2}=29\pm3$, $\tau_{\gamma,\rm\,1}=14\pm2$ and $\tau_{\gamma,\rm\,2}=26\pm3$~days, respectively. The flare maxima are on $t_{\rm max,\,R}= 2021.12.04\,(2021.925,\mathrm{yr})$ and $t_{\rm max,\,\gamma}=2021.12.15\,(2021.958,\mathrm{yr})$. The uncertainty of the position of the flare maximum, equal to 2.5 and 1.5 days for the optical and $\gamma$-ray peaks, was calculated from the $\chi^2$ statistic.

Thus, we conclude that the moment of the maximum in the optical band precede that in $\gamma$-rays by $12\pm3$ days. This fact as well as the proximity of the characteristic times for the brightness rise and fall ($\sim 25$ days, except the shorter time of the flare rise in $\gamma$-rays) need further explanation.

The reliably defined 12-day lag between the optical and $\gamma$-ray emission suggests that the emission in these bands is not strictly co-spatial or instantaneous. In particular, this behavior is difficult to reconcile with a simple one-zone synchrotron self-Compton scenario, which generally predicts nearly simultaneous variability. Instead, the observed delay may indicate a multi-zone emission structure or the contribution of additional processes, such as external Compton scattering or evolving shock conditions within the jet.

Under the assumption of isotropic emission, we estimated the energy released during the short flares in the optical and \mbox{$\gamma$-ray} bands as $E_{R}\approx10^{49}$~erg and $E_{\gamma}\approx3.7\times10^{50}$~erg, respectively. If the $\gamma$-ray photons were generated as a result of inverse Compton scattering of the optical photons, then we can estimate the Lorentz factor of scattering relativistic electrons as $\Gamma_{\rm e} \sim 2\times10^{4}$ from the $\nu_{\gamma}\sim\Gamma^{2}_{\rm e}\nu_{R}$ relation. Next, one scattered photon per $10^{7}$ optical photons is needed to provide the above-estimated energetics of this short $\gamma$-ray flare.

\subsection{Cross-correlations between the light curves}
\label{sec:DCF}

We analysed the similarity between the light curves at different frequencies and measured the time lags between them using discrete correlation functions (DCF, \citealt{1988ApJ...333..646E}). A detailed description of our approach to estimate the lags along with their uncertainties and significances is given in \cite{2026MNRAS.tmp..392S}. The method combines the ideas and software from \cite{1998PASP..110..660P, 2013MNRAS.433..907E, 2015MNRAS.453.3455R, 2015arXiv150306676C, 2021JOSS....6.3001B}.

The measurement results are listed in Table~\ref{tab:ccf-lags}, and the corresponding DCF curves are shown in Appendix~\ref{app:dcf_figs}. The lags were measured both for the entire available time interval and separately for the recent flare. The start of the recent-flare interval, 2019.6, was chosen because it corresponds to the local flux-density minimum preceding the onset of the third major activity episode, consistent with the flare definition adopted in Section~\ref{sec:spectra}. This flare appears to have already passed its maximum in the $\gamma$-ray and optical bands, but is still developing at radio frequencies. This ongoing character makes it impossible to robustly measure the lags relative to 1 and 2\,GHz during the recent-flare interval, because the corresponding low-frequency light curves have not yet formed well-defined peaks. The lags involving these frequencies can still be estimated from the full time interval, although the DCFs with respect to 1 and 2\,GHz are broad (see Appendix~\ref{app:dcf_figs}), leading to larger uncertainties. Our estimate of the time lag for the ``$\gamma$--$R$ band'' pair is consistent with that of \cite{2022ApJ...933..224F}, who reported a 22-day lag between the $R$ and $\gamma$-ray bands.

\begin{table}
\centering
\caption{Time lags between the light curves at different frequencies
for the entire observed time interval and for the recent flare since 2019.6.}
\label{tab:ccf-lags}
\begin{tabular}{ccccc}
\hline
\multirow{2}{*}{Bands} & \multicolumn{2}{c}{Full time range} & \multicolumn{2}{c}{after 2019.6}\\
& Lag (days) & Signif. & Lag (days) & Signif.\\
\hline
$\gamma$ vs $R$     & $0\pm20$ & $3\sigma$ & $-10\pm20$ & $3\sigma$ \\
$\gamma$ vs 230 GHz & $10\pm120$ & $2\sigma$ & $70\pm150$ & $2\sigma$ \\
$\gamma$ vs 22 GHz  & $360\pm100$ & $3\sigma$ & $330\pm120$ & $2\sigma$ \\
$\gamma$ vs 11 GHz  & $450\pm80$ & $3\sigma$ & $420\pm90$ & $3\sigma$ \\
$\gamma$ vs 8 GHz   & $420\pm90$ & $2\sigma$ & $430\pm100$ & $2\sigma$ \\
$\gamma$ vs 5 GHz   & $560\pm110$ & $2\sigma$ & $710\pm230$ & $2\sigma$ \\
$\gamma$ vs 2 GHz   & $1230\pm90$ & $3\sigma$ & $1200\pm40$ & $3\sigma$ \\
$\gamma$ vs 1 GHz   & $1200\pm170$ & $3\sigma$ & -- & -- \\
\hline
$R$ vs 230 GHz & $-60\pm110$ & $1\sigma$ & $60\pm110$ & $1\sigma$ \\
$R$ vs 22 GHz  & $300\pm50$ & $3\sigma$ & $310\pm40$ & $3\sigma$ \\
$R$ vs 11 GHz  & $380\pm20$ & $3\sigma$ & $400\pm30$ & $3\sigma$ \\
$R$ vs 8 GHz   & $390\pm30$ & $2\sigma$ & $400\pm20$ & $2\sigma$ \\
$R$ vs 5 GHz   & $390\pm30$ & $2\sigma$ & $460\pm60$ & $2\sigma$ \\
$R$ vs 2 GHz   & $700\pm260$ & $2\sigma$ & -- & -- \\
$R$ vs 1 GHz   & $1020\pm180$ & $3\sigma$ & -- & -- \\
\hline
230 GHz vs 22 GHz & $270\pm150$ & $2\sigma$ & $220\pm170$ & $2\sigma$ \\
230 GHz vs 11 GHz & $280\pm100$ & $3\sigma$ & $330\pm120$ & $3\sigma$ \\
230 GHz vs 8 GHz  & $350\pm110$ & $3\sigma$ & $420\pm140$ & $2\sigma$ \\
230 GHz vs 5 GHz  & $340\pm140$ & $3\sigma$ & $350\pm170$ & $2\sigma$ \\
230 GHz vs 2 GHz  & $890\pm270$ & $2\sigma$ & -- & -- \\
230 GHz vs 1 GHz  & $1200\pm200$ & $2\sigma$ & -- & -- \\
\hline
22 GHz vs 11 GHz & $140\pm90$ & $3\sigma$ & $130\pm90$ &  $3\sigma$ \\
22 GHz vs 8 GHz  & $110\pm170$ & $2\sigma$ & $90\pm100$ & $2\sigma$ \\
22 GHz vs 5 GHz  & $280\pm80$ & $2\sigma$ & $330\pm110$ & $2\sigma$ \\
22 GHz vs 2 GHz  & $1070\pm100$ & $3\sigma$ & -- & -- \\
22 GHz vs 1 GHz  & $1020\pm150$ & $3\sigma$ & -- & -- \\
\hline
11 GHz vs 8 GHz & $0\pm30$ & $3\sigma$ & $-10\pm20$ & $3\sigma$ \\
11 GHz vs 5 GHz & $0\pm30$ & $3\sigma$ & $0\pm20$ & $3\sigma$ \\
11 GHz vs 2 GHz & $750\pm240$ & $3\sigma$ & -- & -- \\
11 GHz vs 1 GHz & $910\pm130$ & $3\sigma$ & -- & -- \\
\hline
8 GHz vs 5 GHz & $0\pm20$ & $3\sigma$ & $0\pm20$ & $3\sigma$ \\
8 GHz vs 2 GHz & $1160\pm210$ & $3\sigma$ & -- & -- \\
8 GHz vs 1 GHz & $980\pm170$ & $3\sigma$ & -- & -- \\
\hline
5 GHz vs 2 GHz & $570\pm280$ & $3\sigma$ & -- & -- \\
5 GHz vs 1 GHz & $850\pm230$ & $3\sigma$ & -- & -- \\
\hline
2 GHz vs 1 GHz & $720\pm230$ & $3\sigma$ & --& -- \\
\hline
\end{tabular}
\end{table}

\subsection{Frequency-dependent time delays and jet stratification}
\label{sec:delays}

\begin{figure}
\centerline{\includegraphics[width=1.0\linewidth]{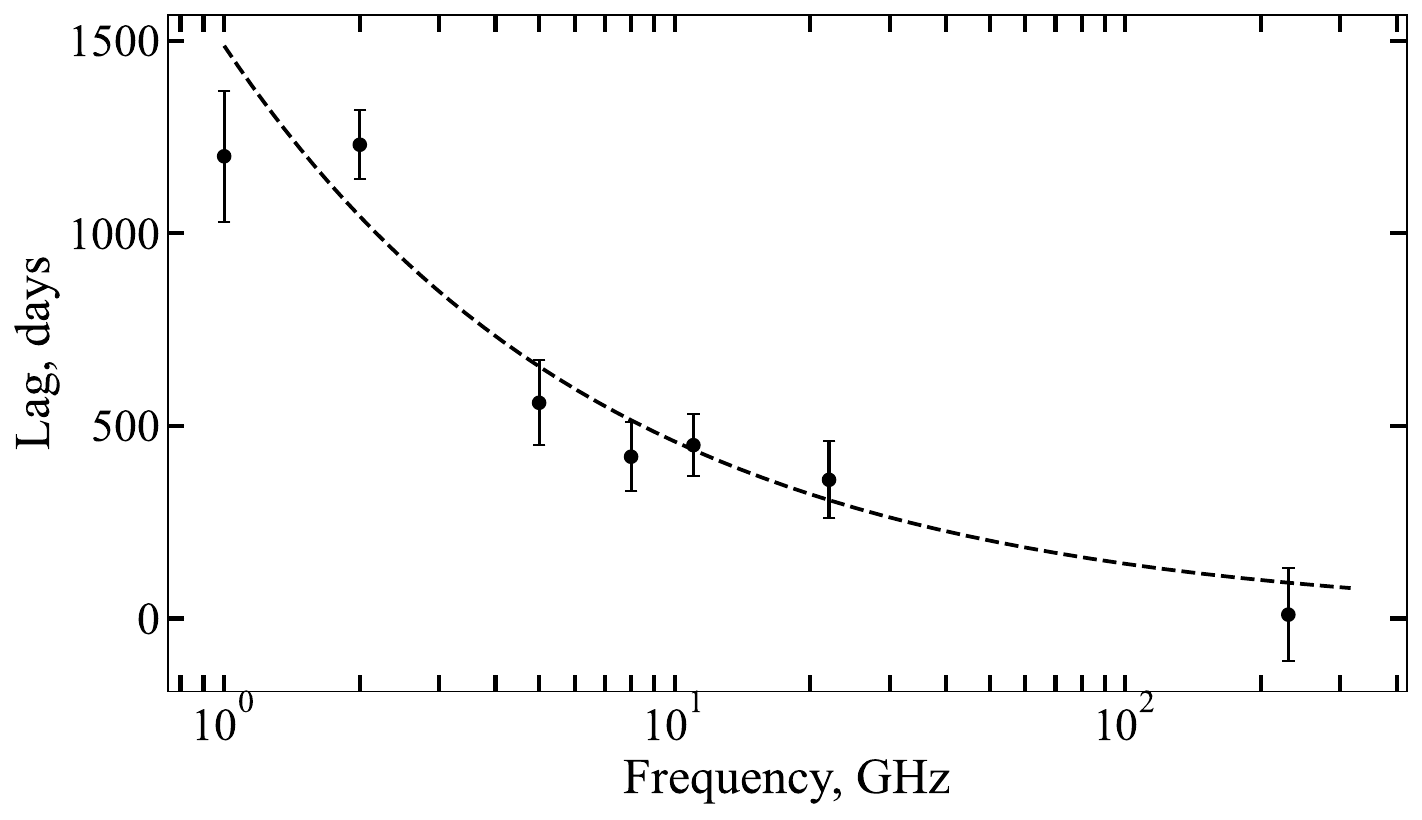}}
\caption{Frequency dependence of the time delays relative to the $\gamma$-ray band.} 
\label{fig:lags}
\end{figure}

The frequency dependence of the time delays relative to the $\gamma$-ray band is shown in Fig.~\ref{fig:lags} (see also Table~\ref{tab:ccf-lags}). The delays increase systematically toward lower frequencies, reaching values up to $\sim1200$ days at 1--2~GHz.

To quantify this behavior, we fit the observed delays with a power-law dependence of the form
\begin{equation}
    \Delta t(\nu) = \tau_0 \, \nu^{-1/k_r},
\label{eq:opacity_fit}
\end{equation}
where $\nu$ is the observing frequency, $k_r$ characterizes the frequency dependence, and $\tau_0$ sets the overall delay scale.
The best-fit parameters are $k_r = 1.96 \pm 0.24$ and $A = 1500 \pm 130$ days.

Using the observed time delays, we estimate the projected separation between emission regions at different frequencies as
\begin{equation}
D = \beta_{\rm app} \, c \, t_{\mathrm{lag,obs}},
\end{equation}
where $c$ is the speed of light, $t_{\mathrm{lag,obs}}$ is the observed time delay, and $\beta_{\rm app}$ is the apparent jet speed. Adopting $\beta \approx 5$ \citep{2025A&A...699A.381K}, we obtain propagation distances inferred from the observed time delays in the range $D \sim 0.03$--$3$ pc.

These results indicate systematic stratification of the emission regions along the jet, with lower-frequency emission originating at larger distances downstream, consistent with a stratified jet structure (see Discussion).

\section{Search for quasi-periodicity}
\label{sec:wwz}

The Weighted Wavelet Z-transform (WWZ) introduced by \citet{1996AJ....112.1709F} provides a powerful framework for analysing unevenly sampled astronomical time series. The method is based on fitting a localised sinusoidal model

\begin{equation}
x(t) \approx A_{\rm WWZ} \cos(\omega t) + B_{\rm WWZ} \sin(\omega t) + C_{\rm WWZ},
\end{equation}

using weighted least squares, where the weight of each data point is defined by a Gaussian window centered at 
a time $\tau$:
\begin{equation}
w(t) = \exp\left[- \kappa\, \omega_{\rm WWZ}^2 (t - \tau)^2\right].
\end{equation}
Here $\omega$ is the trial angular frequency, $\tau$ is the local time shift, and $\kappa$ is a decay parameter controlling the balance between 
the temporal and frequency resolutions.
The resulting statistic $Z(\tau,\omega)$ quantifies the local strength of a sinusoidal oscillation of 
a frequency $\omega$ around a time $\tau$.

For the numerical implementation of the method, we used the open-source Python package developed by Sebastian Kiehlmann,\!\footnote{\url{https://github.com/skiehl/wwz}} which closely follows the formalism of \citet{1996AJ....112.1709F}. For each band, the red-noise background was estimated using surrogate light curves generated following \citet{2013MNRAS.433..907E}, assuming a power-law power spectral density, $P(f)\propto f^{-\beta}$, with $\beta$ estimated from the corresponding observed light curve. The observed sampling and flux distribution were preserved.

Averaging the WWZ power over all time shifts yields the global spectrum:
\begin{equation}
Z_{\rm global}(\omega) = \frac{1}{N_\tau}\sum_{i=1}^{N_\tau} Z(\tau_i,\omega),
\end{equation}
which provides a frequency-domain representation analogous to a periodogram but naturally adapted to irregular sampling.

In this work we adopt the recommended decay constant $\kappa = 1/(8\pi^2)$, which ensures the optimal compromise between temporal localisation and frequency resolution for AGN light curves \citep{1996AJ....112.1709F,2008A&A...488..897H}. Trial periods were scanned on a logarithmic grid spanning $P_{\rm min} = 0.05$~yr to $P_{\rm max} = 15$~yr, with 1000 trial frequencies and 1000 time shifts across the observational baseline.

The statistical significance of peaks in $Z_{\rm global}(\omega)$ was assessed through Monte Carlo simulations of red-noise light curves following the method of \citet{2013MNRAS.433..907E}. For each of $N_{\rm sim} = 1000$ realisations, we computed the WWZ spectrum and determined the 95 per cent and 99 per cent significance thresholds as empirical percentiles:
\begin{equation}
Z_{95\%}(\omega) = \mathrm{percentile}_{95}\left\{Z_{\rm global}^{(k)}(\omega)\right\}_{k=1}^{N_{\rm sim}}.
\end{equation}

The quoted significance values should therefore be interpreted as local significance estimates at the corresponding trial periods, rather than as global false-alarm probabilities over the full searched period range.

In addition to the global spectrum, we examined the two-dimensional $Z(\tau,\omega)$ maps to identify transient, localised oscillatory features. This capability is particularly relevant for AGN light curves, where characteristic timescales may evolve or appear intermittently over the course of long-term monitoring \citep{2008A&A...488..897H}.

Table~\ref{tab:wwz} summarizes the periods detected in \pks using the WWZ technique. 

Across the radio and millimetre bands, quasi-periodic variations are detected with different levels of significance. The 5, 11, 22, and 230 GHz light curves all show a dominant periodicity in the range of 10.9–11.1 yr with high significance ($6.4\sigma$), indicating a coherent and statistically robust modulation across these frequencies. The 8 GHz band similarly reveals a periodicity at 10.7 yr with lower significance ($2.9\sigma$), consistent with the overall trend.

At 2 GHz, two features are observed at 12.6 yr ($2.2\sigma$) and 8.9 yr ($1.3\sigma$), both with lower statistical significance and restricted to the most recent epochs in the light curve. These may indicate either emerging variability components or enhanced noise due to irregular sampling and reduced amplitude at low frequencies.

In the optical $R$ band, two quasi-periodic signals are identified: one at 9.0 yr ($2.1\sigma$) and another at 5.5 yr ($1.7\sigma$). Both features are confined to the last decade of the light curve, suggesting either temporally localised modulation or limited sensitivity to longer-term periodicities due to sampling inhomogeneity and noise contamination.

In the $\gamma$-ray band, two quasi-periodic features are detected: a 5.0 yr signal ($3.1\sigma$) localised in the post-2018 interval, and a broader 8.3 yr modulation ($2.8\sigma$) spanning over a decade. Both exceed the adopted red-noise confidence envelopes; however, given the shorter duration and strongly flaring character of the $\gamma$-ray light curve, these features should be regarded as tentative compared to the radio-band modulation.

In summary, the most robust and recurrent feature across the electromagnetic spectrum is the $\sim$10.7--11.1 yr modulation, firmly detected in multiple radio bands and partially echoed in the optical. Additional shorter periods appear across several bands with moderate significance, including the 5.0 yr signal seen both in the optical band and in $\gamma$-rays, potentially indicating 
multi-timescale modulation mechanisms within the jet or accretion flow. Features with lower local confidence, especially those confined to short temporal intervals, are not considered robust evidence for a coherent long-term modulation within the present analysis.

\begin{table*}
\centering
\caption{Variability timescales detected in the light curves of \pks using the WWZ method. The columns ``Band'' and ``No.'' indicate the observing frequency and the period label. ``Epoch'' shows the time interval where the WWZ signal is 
the
strongest, while ``Period$_{\rm max}$ (yr)'' and ``Sign.'' give the period and the corresponding significance from the WWZ global power spectrum}
\label{tab:wwz}
\begin{tabular}{ccccc}
\hline
\multirow{2}{*}{Band}  & \multirow{2}{*}{No.} & \multicolumn{3}{c}{WWZ} \\
  & & Epoch & Period$_{\rm max}$ (yr) &Sign.\\
\hline
2 GHz  & P1 & 1999.0 – 2026.1 &  12.6 (2009.9) & 2.2\\
  & P2 &  2021.4 – 2026.1 & 8.9 (2026.1) & 1.3 \\
\hline
5 GHz  & P1 &2021.9 – 2026.1 & 11.1 (2026.1) & 6.4 \\
\hline
8 GHz  & P1 & 2002.9 – 2026.1& 10.7 (2015.2)& 2.9 \\
\hline
11 GHz  & P1 & 2000.5 – 2026.1& 10.9 (2020.3)& 6.4 \\
\hline
22 GHz    & P1  &1997.2 – 2026.1 & 10.9 (2011.6)& 6.4 \\
\hline
230 GHz    & P1 & 2003.1 – 2025.3& 10.0 (2019.6)& 6.4 \\
\hline
$R$ band  & P1 &2013.8 – 2026.1 &9.0 (2021.8) &2.1\\
   & P2 &2018.9 – 2026.1 &5.5 (2026.1) &1.7\\
\hline
$\gamma$-ray  & P1 & 2018.6 – 2025.9 & 5.0 (2025.9) & 3.1 \\
  & P2 & 2013.9 – 2025.9 & 8.3 (2021.2) & 2.8 \\
\hline
\end{tabular}
\end{table*}

\begin{figure*}
\centerline{\includegraphics[height=5cm]{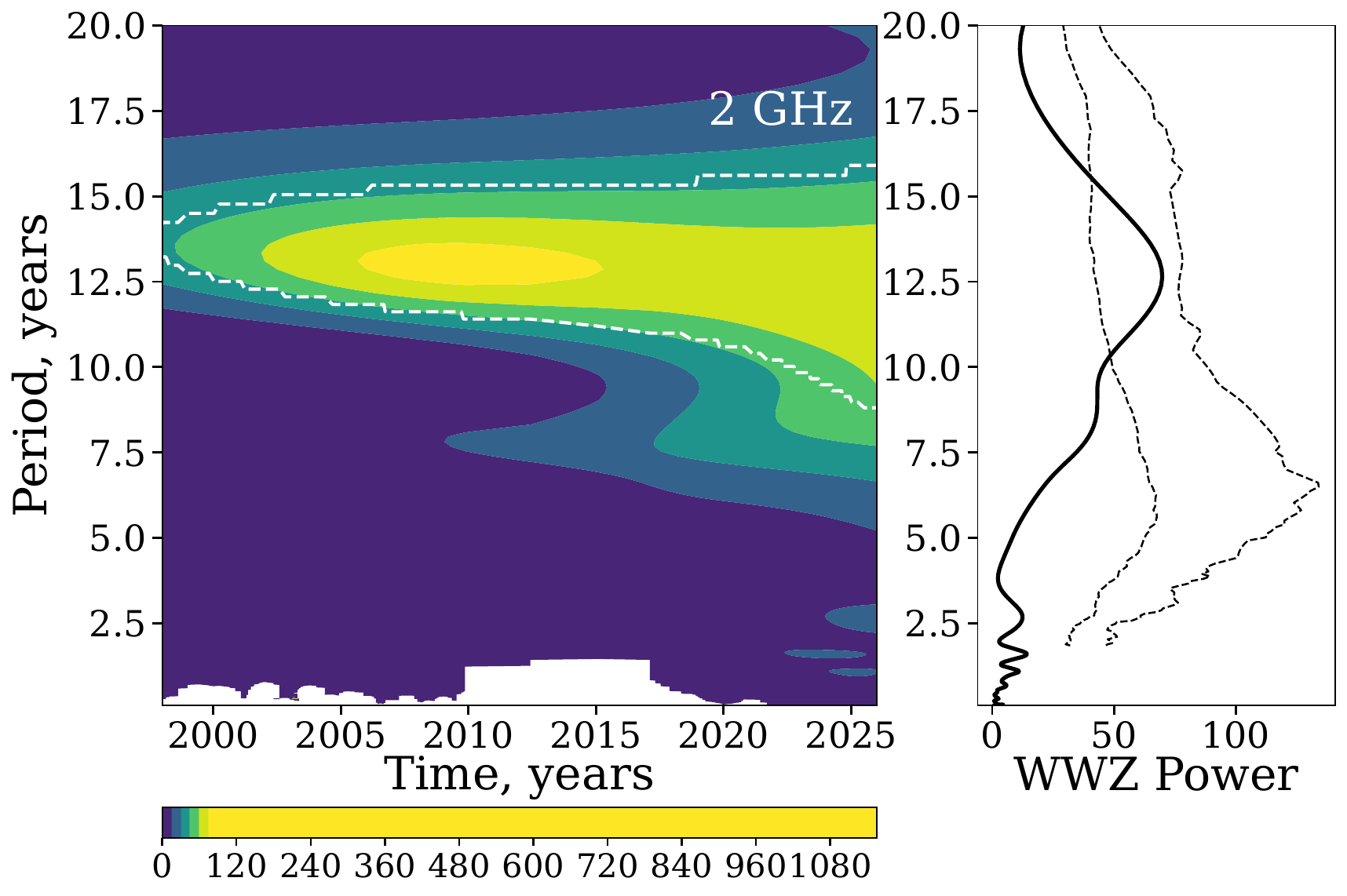}\includegraphics[height=5cm]{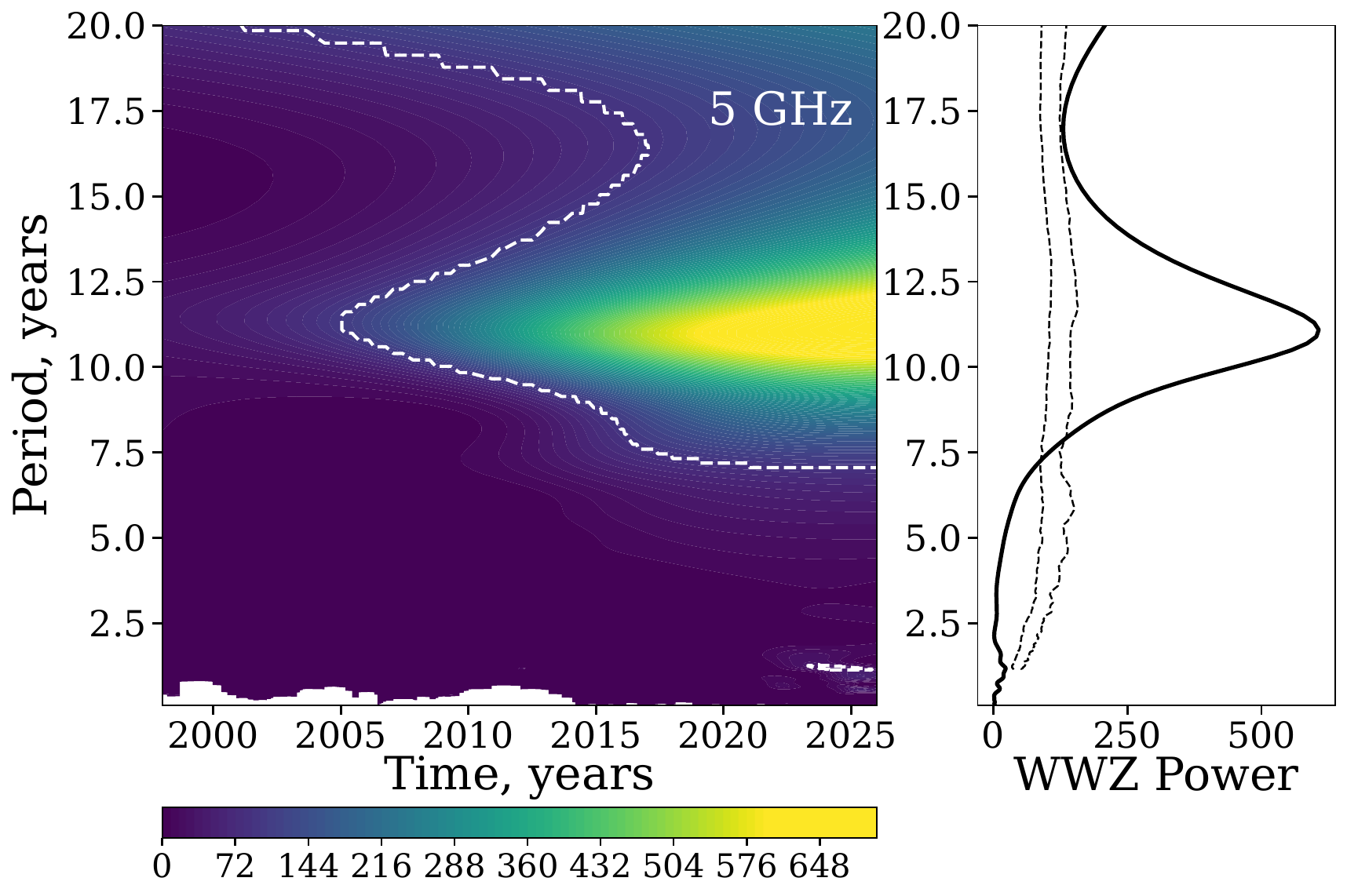}}\centerline{\includegraphics[height=5cm]{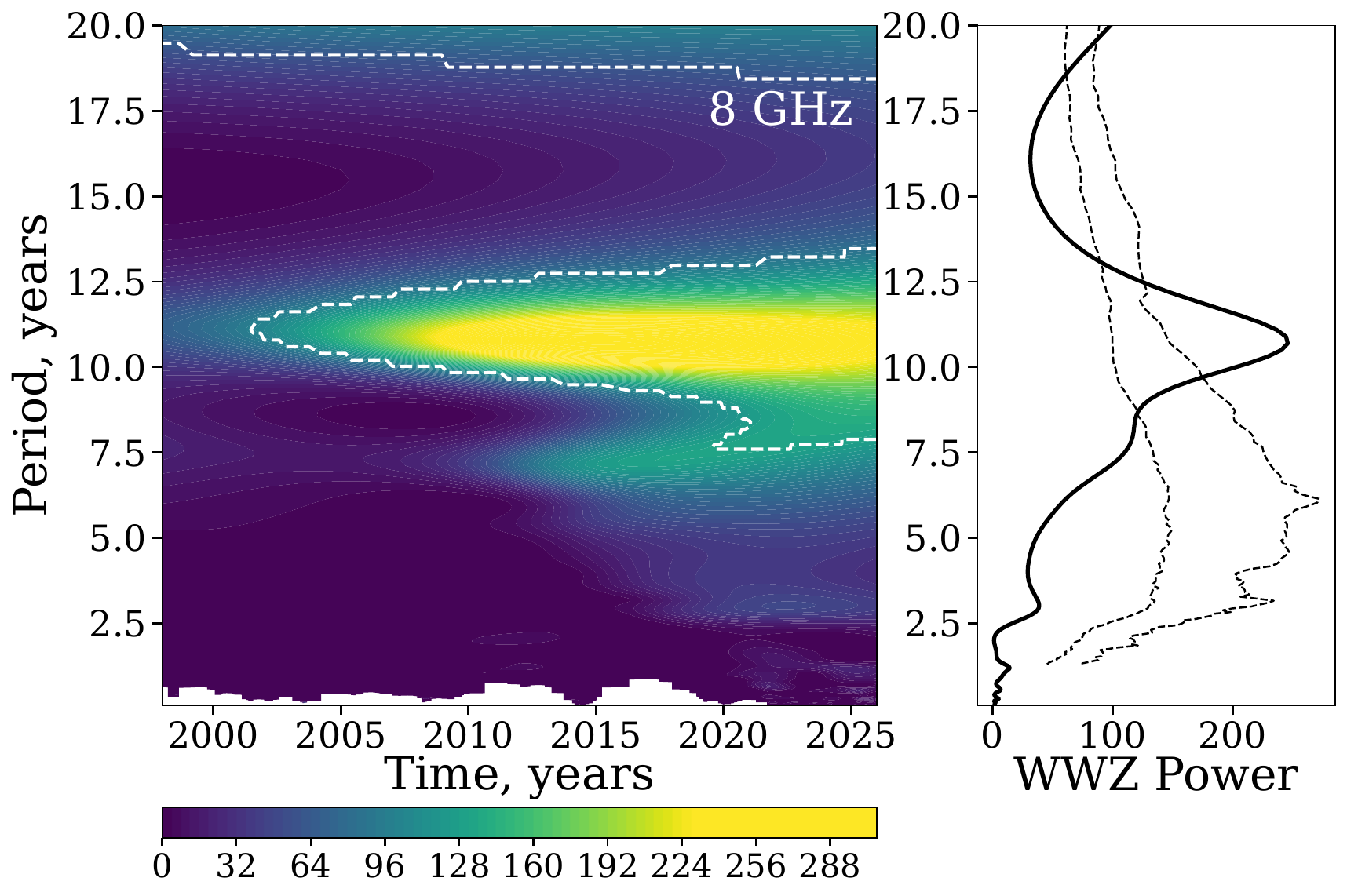}\includegraphics[height=5cm]{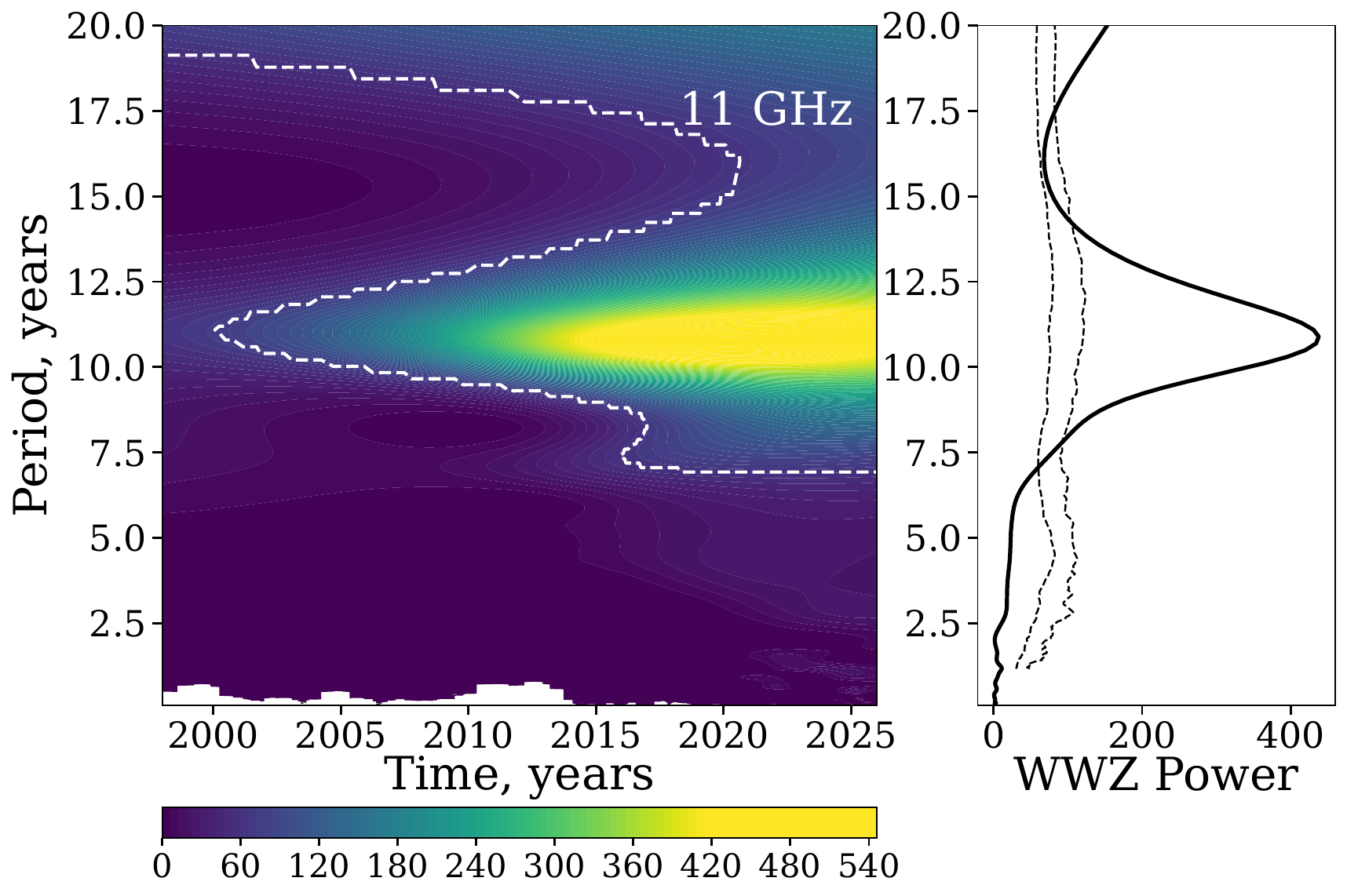}}
\centerline{\includegraphics[height=5cm]{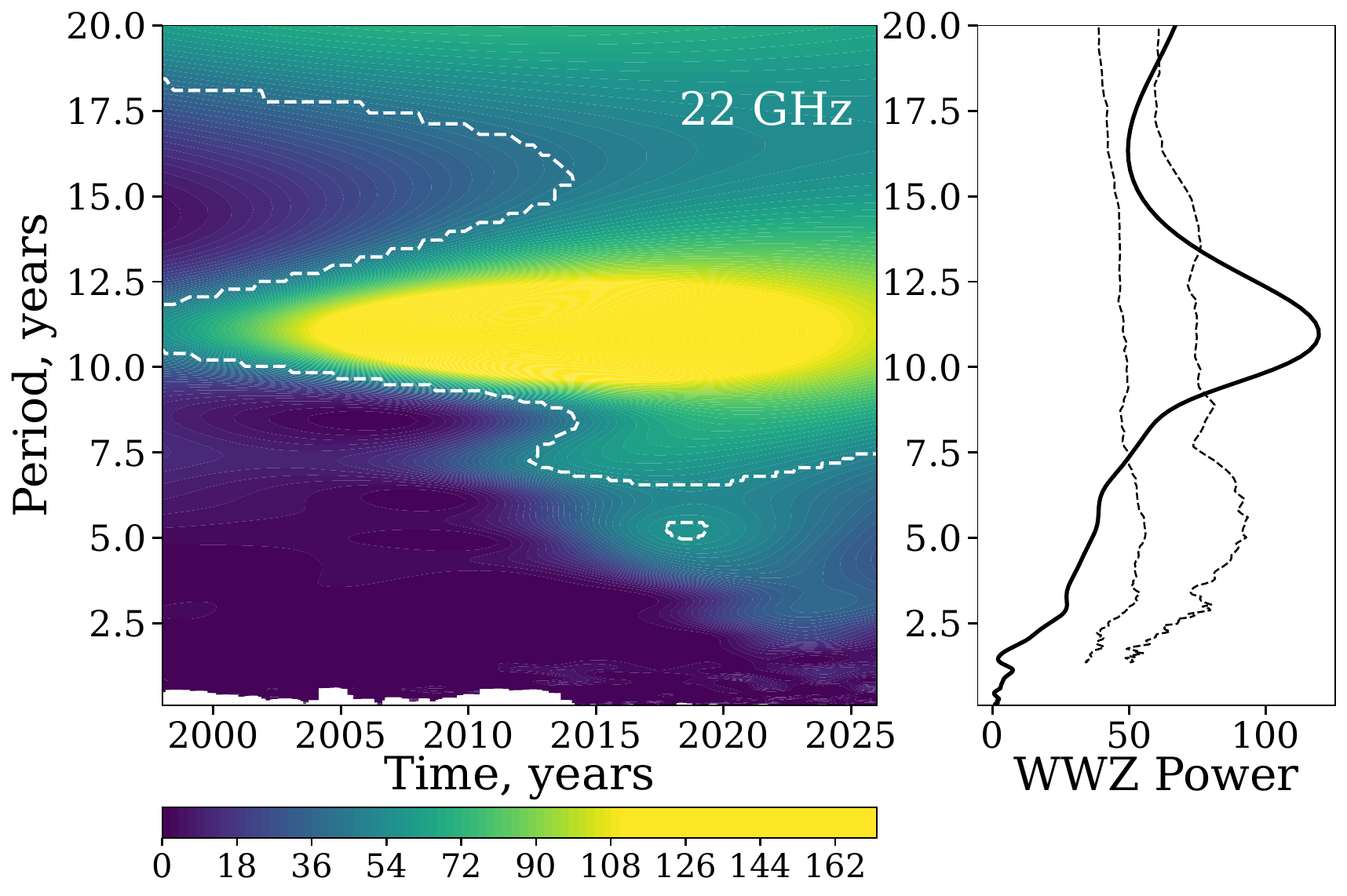}\includegraphics[height=5cm]{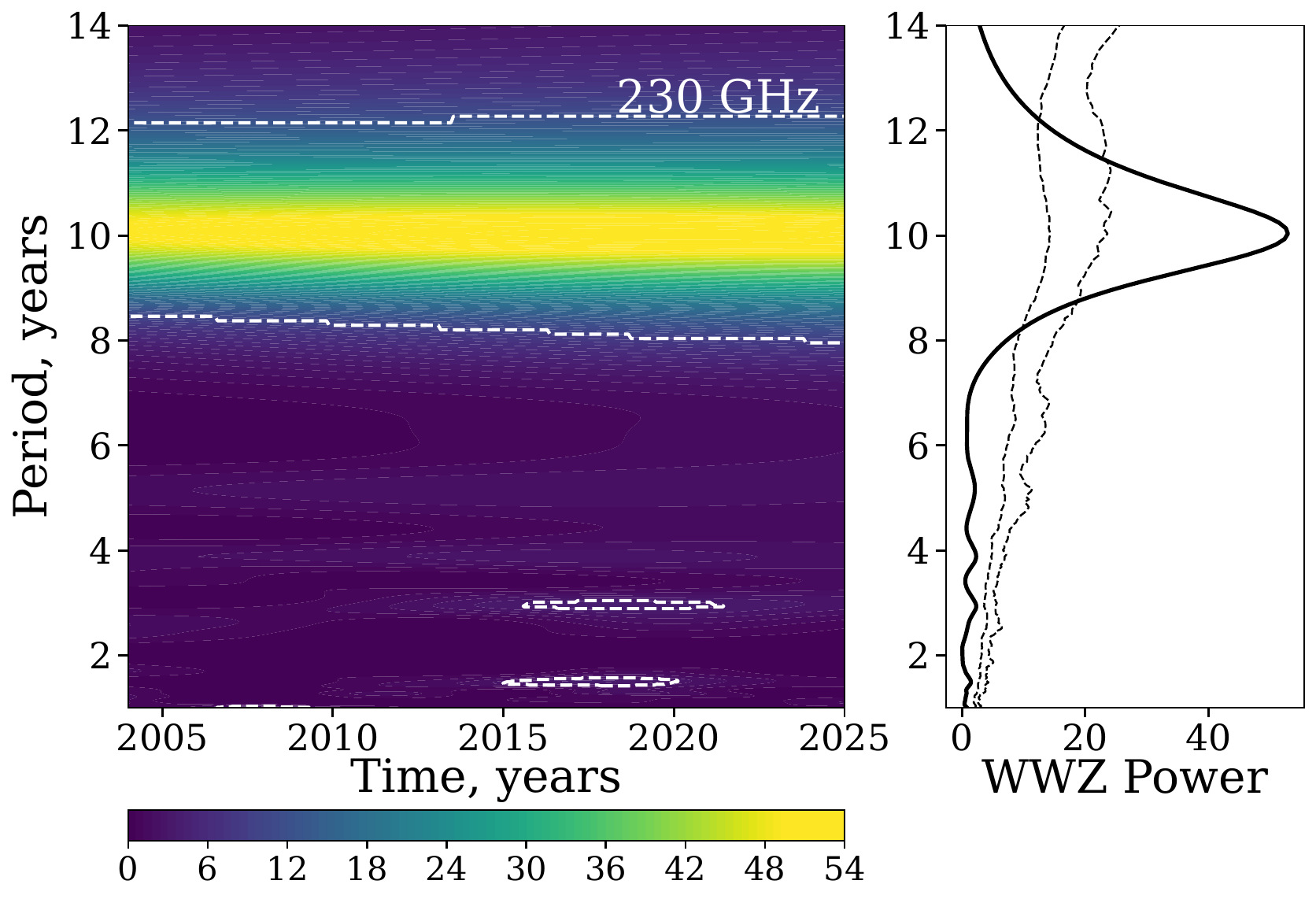}}
\centerline{\includegraphics[height=5cm]{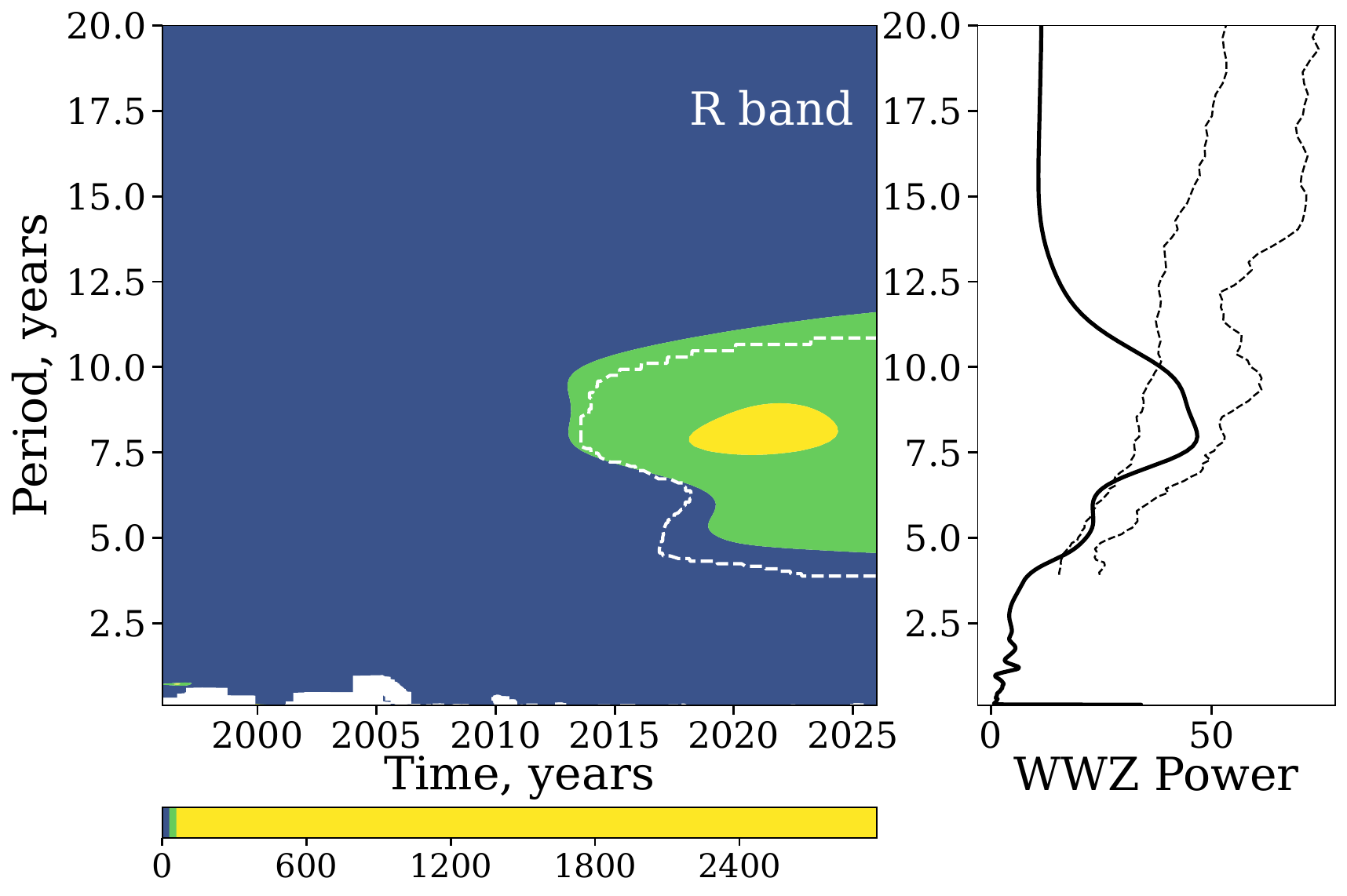}\includegraphics[height=5cm]{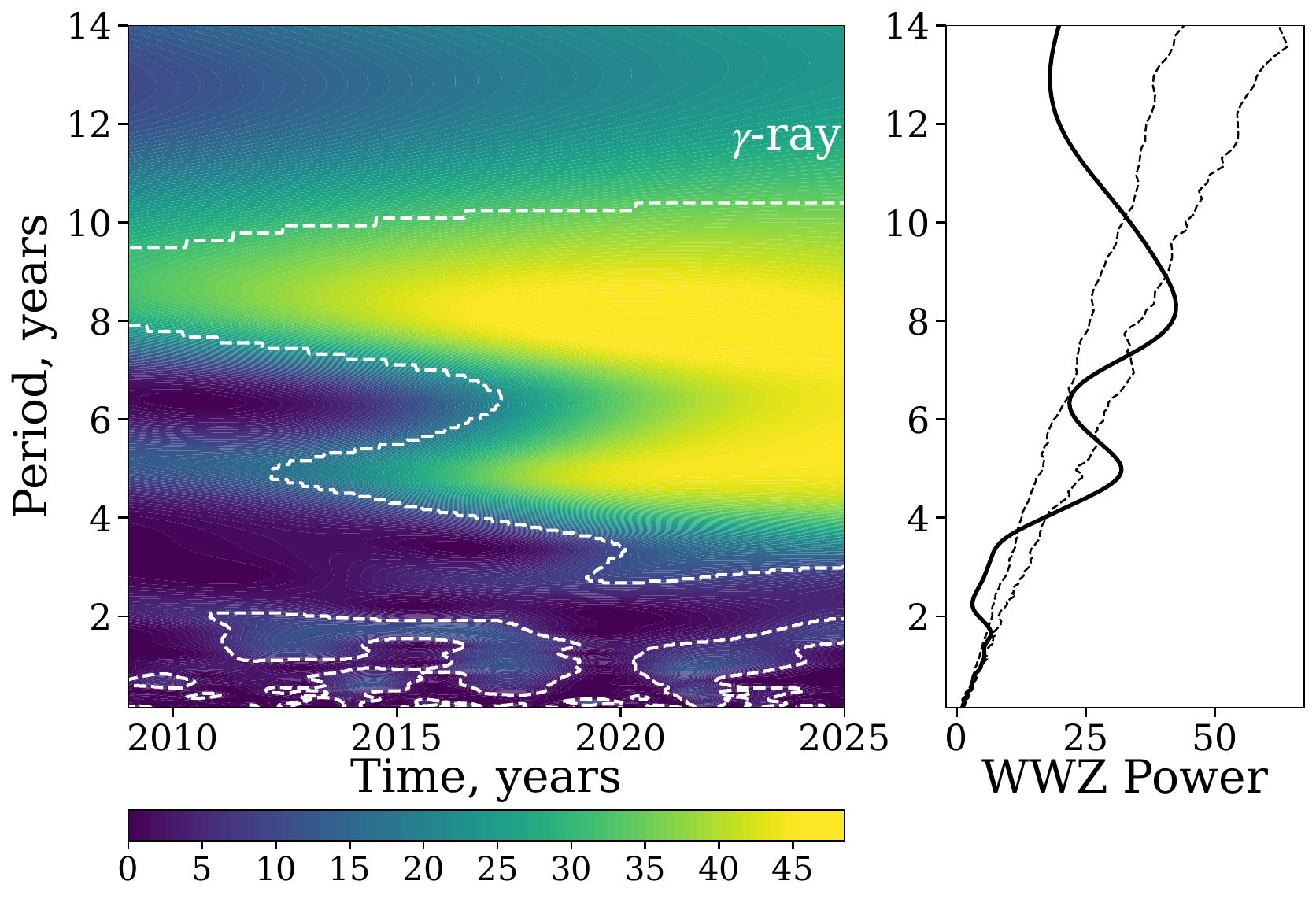}}
\caption{
WWZ power maps and global spectra for \pks. Each left-hand panel presents the WWZ power distribution as a function of time (horizontal axis, in years) and trial period (vertical axis, in years). The colour scale reflects the strength of the localised quasi-periodic signal, with brighter tones indicating higher WWZ power. White dashed contours outline regions where the WWZ power exceeds the period-dependent 95th percentile derived from 1000 Monte Carlo red-noise realisations, following the approach of \citet{2013MNRAS.433..907E}. These contours should be regarded as visual guides to enhanced WWZ power. The corresponding right-hand panels show the global WWZ spectra, obtained by averaging the WWZ power over the full time span at each trial period. Thick black solid curves represent the observed global spectra, while thin black dash-dotted and thin black solid lines mark the 95 per cent and 99 per cent significance levels, respectively, derived from the simulations. Note that transient but locally strong signals may not manifest themselves in the global spectrum if they are not sustained over the full observation window. The colour bars indicate the WWZ power scale applied in the left panels.}
\label{fig:wwz}
\end{figure*}

\section{VLBI core shift}
\label{sec:coreshift}

\begin{figure}
\centerline{\includegraphics[width=1.0\linewidth]{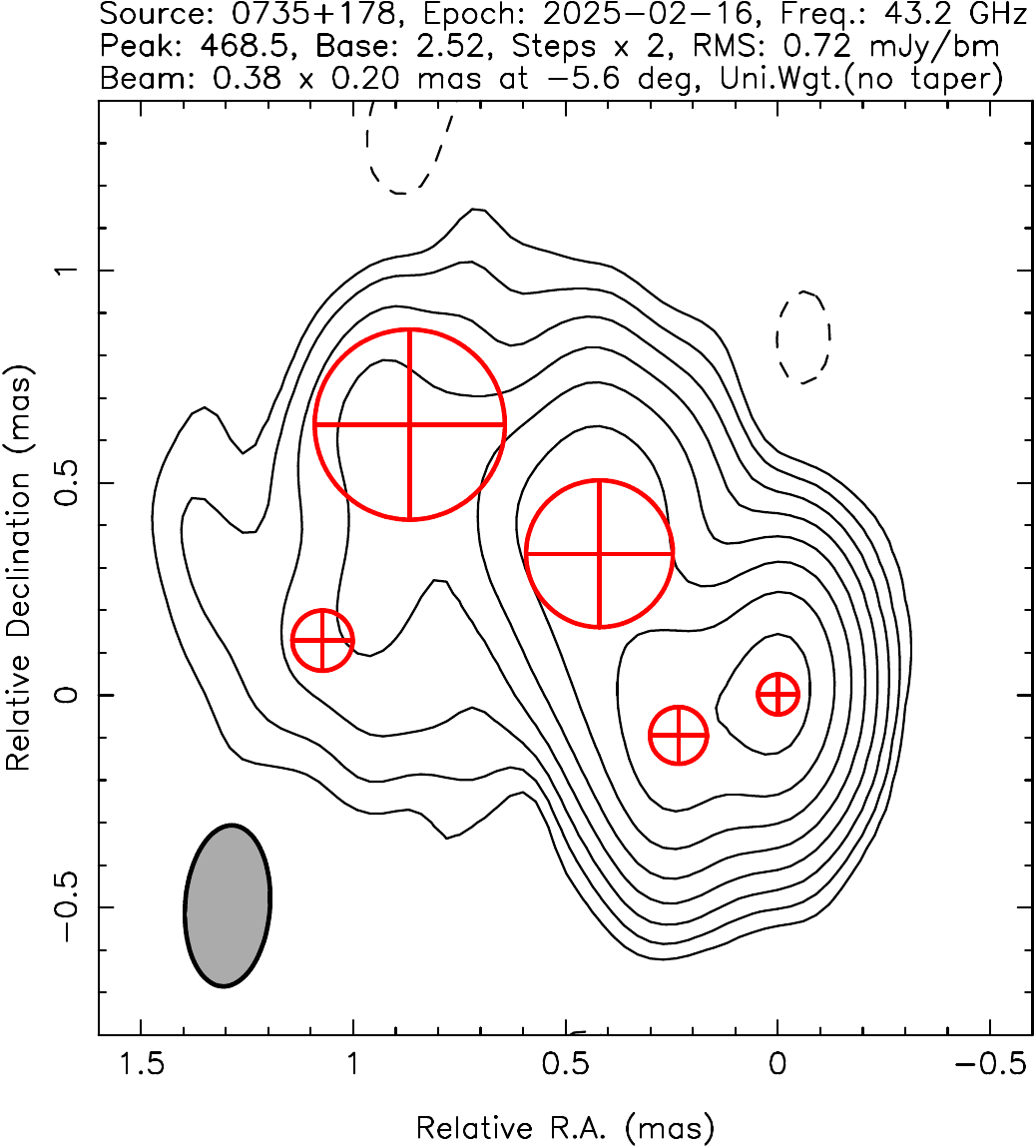}}\vspace{3mm}
\centerline{\includegraphics[width=1.0\linewidth]{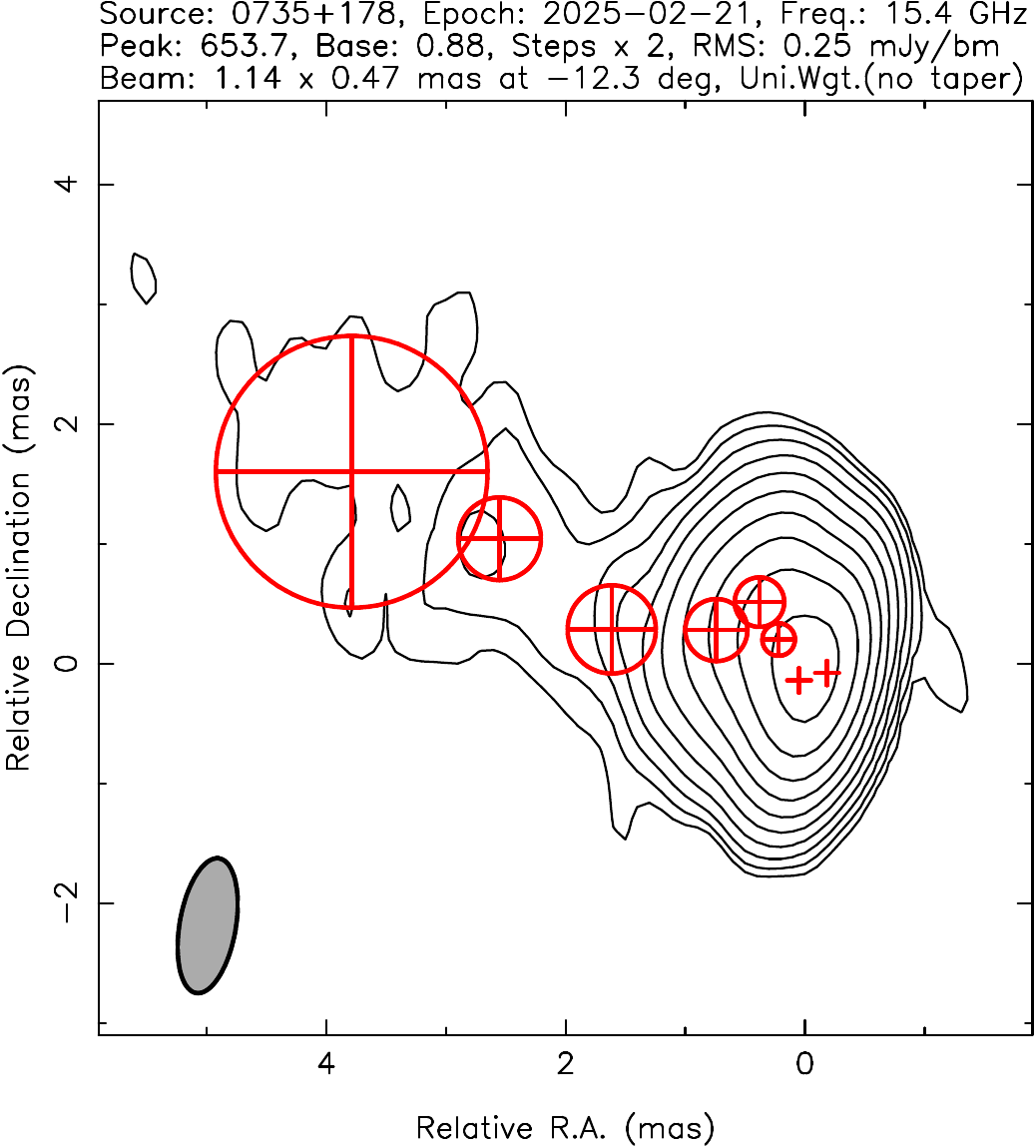}}
\caption{Uniformly weighted VLBA maps of \pks at 15 (left) and 43 GHz (right) at close epochs. 
         The contours are given at increasing powers of 2. The bottom level corresponds to 3.5 times image rms. 
         The grey ellipse in the left lower corner represents the full-width at half maximum (FWHM) of the Gaussian restoring beam. 
         The positions and FWHMs of the fitted components are shown in red. 
         } 
\label{fig:UQ_maps}
\end{figure}

\begin{figure}
\centerline{\includegraphics[width=1.0\linewidth]{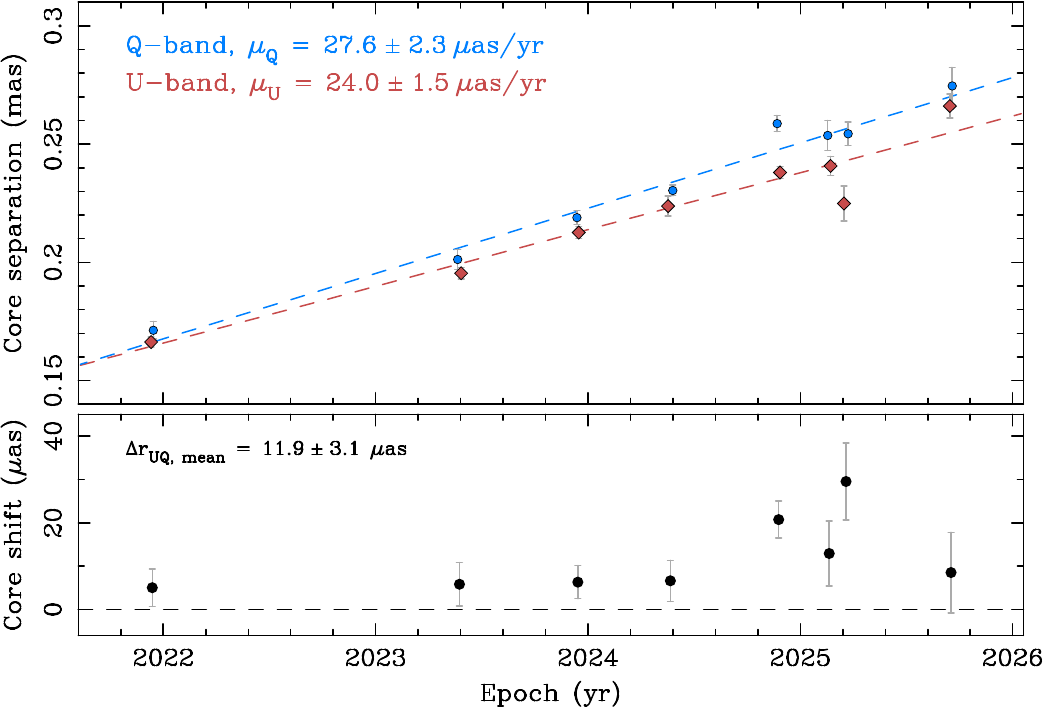}}
\caption{Angular separation from the core against time for the innermost jet component at 15 and 43~GHz (upper panel), 
         with linear kinematic fits obtained. The corresponding shifts of the core position (lower panel).}
\label{fig:coreshifts}
\end{figure}

The core shift effect, mainly caused by synchrotron self-absorption, is a powerful tool for probing physical conditions in the innermost regions of relativistic jets \citep{1998A&A...330...79L}. We measured the shift between the frequency-dependent core position at 15 (U band) and 43~GHz (Q band) using publicly available data from the MOJAVE \citep{MOJAVE} and VLBA-BU-BLAZAR monitoring programs \citep{Jorstad16,Jorstad17,2022ApJS..260...12W}, respectively. The high cadence of \pks observations, especially at 43~GHz, allowed us to select eight close epochs since 2021 with a difference of $3 <\Delta t < 8$ days. To derive the core shift, we applied the self-referencing method \citep[e.g.,][]{Kovalev08}, directly measuring the difference in the core position with respect to the same jet feature cross-identified at both frequencies. For this purpose we used the innermost component, as the jet is very curvy downstream (Fig.~\ref{fig:UQ_maps}). The component is compact and bright, resulting in high accuracy of its position. For each fully-calibrated $uv$ data set of the selected epochs, we performed a structure model fitting with the {\tt Difmap} package \citep{DIFMAP} using a limited number of circular Gaussian components that, being convolved with the restoring beam, adequately reproduce the observed source brightness distribution. In Fig.~\ref{fig:coreshifts}, we present the measured core shifts and component kinematics. The mean core shift $\Delta r_\text{UQ}=11.9\pm3.1$~$\upmu$as. Then, assuming that a disturbance propagating down the jet peaks in flux density at a given frequency when it reaches the VLBI core at that frequency \citep{Bach06,Kudryavtseva11,Pushkarev19,Kutkin19}, we can infer its proper motion as $\mu=\Delta r_\text{UQ}/\Delta t_\text{UQ}=28.8\pm14.0$~$\upmu$as/yr, where $\Delta t_\text{UQ}=151.1\pm12.8$~d 
is derived from the fitting
of Eq.~\ref{eq:opacity_fit}. This estimate is consistent with the jet component kinematics, $\mu_\text{Q}=27.6\pm2.3$~$\upmu$as/yr and $\mu_\text{U}=24.0\pm1.5$~$\upmu$as/yr (Fig.~\ref{fig:coreshifts}, upper panel).

Adopting the source redshift $z=0.5$, we obtain the apparent speed $\beta_\text{app,U}=0.77\pm0.04\,c$ and $\beta_\text{app,Q}=0.82\pm0.07\,c$, consistent with $\beta_\text{app,Q}=0.78\pm0.04$ \citep{2025A&A...699A.381K} taken from a larger number of Q-band epochs but somewhat narrower time range they span. We can also estimate the jet viewing angle
\begin{equation}
\theta = \text{arctan}\frac{2\beta_\text{app}}{\beta_\text{app}^2+\delta^2-1}=3.6^\circ\pm0.4^\circ\,,  
\end{equation}
\noindent
where $\delta = 4.8$ \citep{Homan21}. Introducing the core shift measure \citep{1998A&A...330...79L}
\begin{equation}
\Omega_{r\nu} = 4.85\frac{D_\text{L}\Delta r_\text{core,$\nu_1\nu_2$}}{(1+z)^2}\cdot\frac{\nu_1^{1/k_r}\nu_2^{1/k_r}}{\nu_2^{1/k_r}-\nu_1^{1/k_r}}=0.71\pm0.19\,\text{GHz\,pc}\,,
\end{equation}
where $D_\text{L}$ is the luminosity distance in Gpc, 
and
$\Delta r_\text{core,$\nu_1\nu_2$}$ is the core shift in mas, we can derive the absolute separation of the core from the jet apex
\begin{equation}
r_\text{core,U} = \Omega_\text{r$\nu$}/(\nu^{1/k_r}\sin\theta) = 2.8\pm2.2\,\text{pc.}
\end{equation}
Assuming a typical half intrinsic opening angle $\varphi = 1.3^\circ/2$ from \cite{Pushkarev17}, the magnetic field at 1 pc from the central engine can be estimated \citep{2009MNRAS.400...26O}
\begin{equation}
B_1\simeq0.025\left(\frac{\Omega_\text{r$\nu$}^3(1+z)^2}{\varphi\,\delta^2\sin^2\theta}\right)^{1/4} \approx 0.36\pm0.07~\text{G.}
\end{equation}

Then we estimate the jet magnetic flux from the relation \citep{2014Natur.510..126Z}:
\begin{equation}
    \Phi_{\rm jet}=1.2\times10^{34}f(a{_*})\Gamma_{\rm j} \theta_{\rm j}\frac{M_{\rm BH}}{10^9M_{\odot}}B_{1} [\rm G\,cm^2],
\end{equation}
where 
$f(a{_*})$ is a function which depends on the black hole spin $a{_*}$, $\Gamma_{\rm j}$ is the jet bulk Lorentz factor, and \mbox{$\theta_{\rm j}=2\varphi$} is the jet opening angle. Using the quantities calculated in this section and assuming $M_{\rm BH}=10^{9}M_{\odot}$, $a_{*}=0.7\div0.998$, we obtain $\Phi_{\rm jet}=(2.6\pm0.8)\times10^{32}\div(6.2\pm1.9)\times10^{32}$\,G\,cm$^{2}$. We can estimate the magnetic flux $\Phi_{\rm BH}\approx\Phi_{\rm jet}$ in the vicinity of the black hole using the assumption of a frozen-in magnetic field.

We have compared this value with the saturated magnetic flux $\Phi_{\rm MAD}$ when the magnetic pressure is equal to the ram pressure of the accreting matter \citep{2014Natur.510..126Z}: 
\begin{multline}
 \label{eq:PhiMAD}
    \Phi_{\rm MAD}\approx50 \sqrt{L_{\rm acc} r_g^2/\eta c}=\\
    =2.4\times10^{25} \left(\frac{\eta}{0.4}\right)^{-1/2}\frac{M_{\rm BH}}{M_{\odot}}  \left(\frac{L_{\rm acc}}{1.26\times10^{47} \rm erg\,s^{-1}}\right)^{1/2}  [\rm G\,cm^2],   
\end{multline}
where $\eta=\eta(a_*)$ is the radiative efficiency of the accretion flow, 
and
$L_{\rm acc}$ is the accretion luminosity. 
We assessed the ratio of the magnetic fluxes $\Phi_{\rm BH}/\Phi_{\rm MAD}\sim f(a_*)\,\eta^{1/2}\,\Gamma_{\rm j}\,\theta_{\rm j}\,B_1L_{\rm acc}^{-1/2}$
using our estimates for $\Gamma_{\rm j}$, $\theta_{\rm j}$, and $B_{1}$ from the core shift analysis. As above, we considered the spins of the central SMBH $a_{*}=0.7\div0.998$, motivated both by the current AGN SMBH spin constraints
(e.g., \citealt{2021ARA&A..59..117R}) and by the
theoretical predictions that a fast-spinning black hole is needed to generate powerful relativistic jets (e.g., \citealt{1977MNRAS.179..433B,2007MNRAS.377.1652N,2009ApJ...699..400G}). $\Phi_{\rm BH}/\Phi_{\rm MAD}$ varies within $\sim 25$~per~cent at $a_{*}=0.7\div0.998$ and a fixed $L_{\rm acc}$, i.e., we can neglect the spin dependence in our consideration. $\Phi_{\rm BH}/\Phi_{\rm MAD}\approx1$ at $L_{\rm acc}\approx1.5\times 10^{43}$~erg\,s$^{-1}$, meaning the magnetically arrested disk (MAD) state. The accretion disk of \pks is not in the MAD state at higher $L_{\rm acc}$. There are no detected emission lines in the
optical spectrum of the BL~Lac object \pks, and 
synchrotron radiation dominates in the optical continuum, 
we therefore do not have an opportunity to use common scaling relations to estimate $L_{\rm acc}$. Another way to estimate $L_{\rm acc}$ is to use the X-ray luminosity at \mbox{2--10~keV}: $L_{\rm acc}=kL_{2-10\,\rm keV}$, where $k\geq 1$ is the bolometric correction. We estimated $L_{2-10\,\rm keV}=5.8\times10^{44}$~erg\,s$^{-1}$ from the
published X-ray measurements with the ASCA satellite \citep{2001ApJS..133....1U}. If the X-ray emission is not subject to strong Doppler enhancement, we have a constraint
$L_{\rm acc}\geq5.8\times10^{44}$~erg\,s$^{-1}$. Then we can estimate $\Phi_{\rm BH}/\Phi_{\rm MAD}\leq 0.15\div0.19$.    

\section{Discussion}

\pks demonstrates a complex multiwavelength variability. Evidence of quasi-periodic emission has been found on different time-scales in many studies. \citet{2004PASP..116..161Q} found a quasi period of 13.8\,yr in the optical light curve measured in 1995--2000, which is very close to 14.2 yrs in the $R$ band, obtained by \citet{1997A&AS..125..525F}. \citet{2001MNRAS.328..873G} found a twisted jet structure at 5\,GHz and suggested that this could explain the periodicity within a geometric scenario. In many studies it was shown that a contribution to long-term periodic flux density variations in blazars may come from jet precession \citep{2023ApJ...951..106B}. 

\citealt{2025A&A...699A.381K} analysed archival VLBI data obtained during 2020--2024 (including the time of IC211208A) and found that the jet components of \pks showed high fractional linear polarization ($\geq$ 10 per cent), indicating low synchrotron opacity in the emitting regions. Polarized VLBI images revealed 
variation of the electric vector position angle (EVPA)
as shown in \citet{2025ApJ...989..208P}, tracing the orientation and evolution of the magnetic field within the jet. During the flare, the polarization signatures pointed to a propagating shock front originating from the jet core, compressing the plasma and ordering the magnetic field. This behaviour is consistent with shock-in-jet scenarios, in which shocks accelerate particles and enhance synchrotron emission.

These results support the interpretation that the flare observed in 2021 was driven by a shock propagating downstream in the jet, rather than by purely geometric effects such as variation of viewing angle. Within this framework, the delayed radio flare (peaking $\sim 300$--650\,days after the neutrino and multiwavelength peak) can be naturally explained as a consequence of the shock propagating through progressively optically thinner regions, where synchrotron emission at lower radio frequencies becomes observable.  

The strong and highly significant cross-correlation between the $\gamma$-ray and optical $R$ band emission (with a DCF amplitude $\sim 0.75$) as well as very short delays ($0\pm20$\, days for the full range and $-10\pm20$\,days for the last flaring epoch) between the flares in these bands provide an argument in favour of a multi-zone synchrotron self-Compton emission structure or contributions from additional processes, such as external Compton scattering or evolving shock conditions within the jet. The resulting delay between the $\gamma$-ray and optical $R$~band emission is consistent with the analysis of the short pre-flares at the neutrino event moment (Fig.~\ref{fig:short_flares_vv}). This interpretation is further supported by the presence of comparable intermediate-timescale quasi-periodic components ($\sim 5$--9 yr) in both the optical and $\gamma$-ray bands, suggesting that these emissions may originate from related regions in the inner jet.

The 12-day lag of the $\gamma$-ray emission relative to the optical flux is for the first time reliably determined for \pks. The previous studies gave only marginal indications on possible lags between these bands with large uncertainties \citep{2018MNRAS.480.5517L, 2023ApJ...954...70A, 2024MNRAS.529.3503B, 2025A&A...695A.266O}. In this context, these short flares provide an important diagnostic tool, as they probe the fastest variability time-scales and reveal deviations from simple co-spatial emission scenarios.

\subsection{Quantifying a coherent sinusoidal component}
\label{sec:sin_model}

We searched for quasi-periodic modulation or characteristic variability time-scales in the radio, optical, and $\gamma$-ray light curves using a generalized Lomb--Scargle (GLS) periodogram, which is equivalent to weighted least-squares fitting of a sinusoid with a floating mean to unevenly sampled data \citep{2009A&A...496..577Z}. To characterize the candidate modulation suggested by the WWZ analysis, we fitted a sinusoidal model following the likelihood-based approach of \citet{2022ApJ...926L..35O}. The model is 
\begin{equation}
S(t) = S_0 + A \sin \left[ \frac{2\pi (t - t_0)}{P} - \phi \right],
\end{equation}
where $S_0$ is the mean flux density, $A$ is the modulation amplitude, $P$ is the period, $\phi$ is the phase offset, and $t_0$ is the reference epoch. To account for intrinsic stochastic variability not captured by a single sinusoid, we included an additional intrinsic scatter term $\sigma_0$ added in quadrature to the measurement uncertainties in the Gaussian likelihood: 
\begin{equation}
\sigma_{\mathrm{tot}}^2 =
\sigma_{\mathrm{obs}}^2 + \sigma_0^2 ,
\end{equation}
as in \citet{2022ApJ...926L..35O}. To avoid degeneracies between $\sigma_0$ and $A$, the best-fitting sinusoid parameters were obtained by profiling the likelihood over the linear coefficients at each trial period and selecting the maximum-likelihood solution within the physically motivated window $P\in[9,11]$\,yr.

The statistical significance of the GLS peaks was evaluated against a red-noise null hypothesis using Monte Carlo simulations, adopting synthetic light curves that reproduce both the observed power spectral density (PSD) slope and the flux probability distribution function (PDF), following the method of \citet{2013MNRAS.433..907E}. For each band, we generated 5000 synthetic light curves with the same sampling as the data and computed the GLS periodogram over $P\in[2,30]$ yr. We then derived a global (look-elsewhere) p-value from the distribution of the maximum GLS power across the full search range, and a local p-value by restricting the maximum power to the targeted WWZ-motivated window $P\in[9,11]$ yr \citep[e.g.,][]{2022ApJ...926L..35O}. The results are summarized in Table~\ref{tab:sine_qpo}. The corresponding fitted curves and the joint likelihood curve are shown in Appendix~\ref{app:sin}.

The resulting fittings yield consistent modulation time-scales near $P \simeq 10$--11\,yr at 5--230\,GHz with local red-noise significances of $\sim2$--3.5$\sigma$, while no statistically significant modulation is detected at the lowest radio frequencies, likely due to strong opacity and blending effects in downstream jet regions (Fig.~\ref{fig:sin_model}). The ratio between the sinusoidal amplitude and intrinsic stochastic scatter ($A/\sigma_0 \gtrsim 2$) at mid--high radio frequencies indicates that the quasi-periodic component represents a measurable modulation superposed on stochastic jet variability rather than a strictly periodic process. 

The WWZ analysis suggests weaker and time-localised characteristic time-scales in the optical and $\gamma$-ray bands; however, these features are not recovered as statistically significant coherent sinusoidal modulations under the adopted red-noise model.

To test whether the $\sim 10$\,yr modulation detected in the individual bands is coherent across the radio spectrum, we performed a joint sinusoidal fitting to the 5--230 \,GHz light curves assuming a common period but allowing the amplitude and phase to vary for each band. The likelihood maximization yields a best-fitting period of $P = 10.9$\,yr (Fig.~\ref{fig:joint_period}). The statistical significance of the joint signal was evaluated using Monte Carlo simulations of 5000 synthetic light curves generated with a red-noise power spectral density $P(f)\propto f^{-2}$. The probability that a red-noise process produces a joint likelihood peak as strong as the observed one is $p_{\rm glob}=1.4\times10^{-3}$ over the full search range $P\in[2,30]$\,yr and $p_{\rm loc}=2\times10^{-4}$ within the WWZ-motivated interval $P\in[9,11]$\,yr. These results indicate that the $\sim$11 yr modulation is coherent across the radio bands.

The observed $\sim$10 yr modulation is most naturally interpreted as slow variation in energy release at the jet base that episodically triggers disturbances propagating downstream as shocks, thereby linking the long-term quasi-periodicity with the opacity-driven delays and flare evolution discussed below.

While the radio--mm bands show a coherent long-term modulation near 10--11\,yr, the optical and $\gamma$-ray bands exhibit weaker and less stable variability components on shorter time-scales of $\sim5$--9\,yr. The similarity of the optical and $\gamma$-ray characteristic time-scales suggests that the high-energy variability originates in compact inner-jet regions, whereas the longer radio modulation is associated with large-scale jet evolution and opacity-driven propagation effects.

This interpretation provides a natural framework for the following discussion of an opacity-stratified jet, in which the long-term quasi-periodic modulation regulates the rate of energy injection at the jet base, while the observed radio delays and spectral evolution arise from shocks propagating through frequency-dependent synchrotron photospheres along the jet.

\begin{table*}
\centering
\caption{Sinusoidal modelling results for the $\gamma$-ray, optical, and radio light curves of \pks. $P_{\rm GLS}$ is the best period from the generalized Lomb--Scargle periodogram. The sine model parameters are the maximum-likelihood estimates obtained within the WWZ-motivated period windows adopted for each band. $\sigma_0$ is the additional intrinsic scatter term. The red-noise significances are estimated from 5000 synthetic light curves; $p_{\rm glob}$ corresponds to the maximum GLS power anywhere in the search range $P\in[2,30]$\,yr (look-elsewhere test), while $p_{\rm loc}$ uses the maximum GLS power restricted to $P\in[9,11]$ yr. Note: the 9--11\,yr fitting in the $R$ band is a constraint motivated by the WWZ analysis, while the GLS peak occurs at a shorter period. The $\gamma$-ray fittings are shown for completeness using the WWZ-motivated windows near $\sim5$ and $\sim9$\,yr, but do not represent statistically significant long-term periodicities under the adopted red-noise model.}
\label{tab:sine_qpo}
\begin{tabular}{cccccccccc}
\hline
$\nu$ & $P_{\rm GLS}$ & $P$ & $A$ & $\phi$ & $S_0$ & $\sigma_0$ & $A/\sigma_0$ & $p_{\rm glob}$ ($\sigma$) & $p_{\rm loc}$ ($\sigma$) \\
(GHz) & (yr) & (yr) & (Jy) & (rad) & (Jy) & (Jy) &  &  &  \\
\hline
1  & 29.67 & 9.35  & 0.42 & 1.59 & 1.01 & 0.25 & 1.68 & 0.08 (1.4) & 0.403 (0.24) \\
2  & 27.45 & 10.34 & 0.21 & 0.71 & 1.13 & 0.18 & 1.17 & 0.09 (1.3) & 0.501 (0.00) \\
5  & 11.17 & 11.00 & 0.33 & $-0.42$& 1.11 & 0.12 & 2.75 & 0.19 (0.86) & 0.001 (3.04) \\
8  & 10.79 & 10.67 & 0.30 & $-0.02$ & 1.10 & 0.10 & 3.00 & 0.13 (1.12) & 0.002 (2.95) \\
11 & 10.81 & 10.87 & 0.37 & $-0.42$ & 0.97 & 0.104 & 3.56 & 0.11 (1.24) & 0.001 (3.54) \\
22 & 11.12 & 11.00 & 0.32 & $-0.66$ & 0.82 & 0.17 & 1.88 & 0.46 (0.11) & 0.021 (2.03) \\
230 & 10.06 & 10.15 & 0.23 & $-1.83$ & 0.51 & 0.09 & 2.56 & 0.14 (1.10) & 0.002 (2.95) \\
$R$ band & 7.67 & 9.60 & 0.009 & 0.23 & 0.0017 & 0.00099 & 1.68 &  0.92 (0.00)  & 0.453 (0.12) \\
$\gamma$-ray (P1) & 5.13 & 5.06 & 0.33 & 0.78 & 0.61 & 0.36 & 0.93 & 0.99 (0.00)  & 0.714 (0.00) \\
$\gamma$-ray (P2) & 9.00 & 8.79 & 0.32 & 2.46 & 0.59 & 0.35 & 0.93 & 0.99 (0.00)  & 0.820 (0.00) \\
\hline
\end{tabular}
\end{table*}

\subsection{Precession}

Periodic brightening in the AGN radio light curves can sometimes be explained by jet precession caused by a supermassive black hole binary or the Lense--Thirring effect (e.g., \citealt{2018MNRAS.478.3199B, 2023ApJ...951..106B}). The precessing jet produces variation of Doppler boosting, which alter the observed flux density at a particular frequency $\nu$ as
\begin{equation}
S_\nu(t) = S_{\nu,0}\,\delta(t)^{p-\alpha_\nu(t)}
\end{equation}
where $S_{\nu,0}$ is the intrinsic flux at the frequency $\nu$ excluding Doppler boosting, $\delta(t)$ is Doppler boosting, $\alpha_\nu(t)$ is the spectral index, and $p$ is the geometry factor: $p=2$ in the case of a uniform AGN jet.  
As seen from the above described observed data, for \pks $\alpha_\nu(t)$ is a function of frequency $\nu$ and time $t$: the shapes of spectra vary and cannot be described by a fixed index at later epochs (see Fig.~\ref{fig:spectra}). However, we simplify that the ``local'' spectral index at a particular frequency does not depend on frequency within the $\nu$ range corresponding to Doppler boosting blueshift, unlike the dependence of $\alpha_\nu$ on frequency in the total spectrum. In other words, for a particular light curve $\alpha_\nu(t)$ is taken as a function of time only.

The variation of $\delta(t)$ due to precession can be described \citep{2018MNRAS.478.3199B} as
\begin{equation}
\begin{split}
\delta(t) & = [\gamma\,(1-\beta\cos\Phi(t))]^{-1},\\
\beta & = \sqrt{1-1/\gamma^2},\\
\Phi(t) & = \arcsin\sqrt{X^2+Y^2},\\
X(t) & = \cos\Omega\sin\Phi_0+\sin\Omega\cos\Phi_0\sin(\omega(t-t_0)),\\
Y(t) & = \sin\Omega\cos(\omega(t-t_0)),\\
\omega & = 2\pi / P,
\end{split}
\end{equation}
where $\Phi(t)$ and $\Phi_0$ are the viewing angles of the jet and its precession axis, $\Omega$ is the precession cone half-opening angle,  $\gamma$ is the Lorentz factor, $\beta$ is the velocity of the ejected matter in units of the speed of light, $t_0$ is the time of the initial phase, $P$ is the period, and $\omega$ is angular velocity.

Although we tried to obtain the parameters from the light curves at particular frequencies using Bayesian statistics, a single-frequency light curve does not contain enough information to reliably derive all the free parameters of the precession model: the solutions were ambiguous, unstable, and too sensitive to the prior parameter distributions. To overcome this problem, the following model has been suggested.
\begin{itemize}
\item Instead of approximating each light curve separately, we take all the light curves as a single dataset.
\item Phase shifts between the light curves are set according to the DCF measurements of the lags relative to the light curve at 5~GHz (it has the largest number of measurements).  
\item The precession period can be a free parameter or be set based on other estimations; here we used $P=10.9$~yr derived from the sinusoidal fitting in Section~\ref{sec:sin_model}.
\item Spectral indices $\alpha_\nu(t)$ evolve with time. Their evolution is described as a linear regression derived from the measurements of the indices in quasi-simultaneous spectra. 
For this particular case we have used a time window of 135~days, which corresponds to the median cadence of the 230~GHz light curve with the most scarce measurements. The indices were calculated by the standard formula using flux densities at the neighboring frequencies. For the optical light curve, we adopted not evolving $\alpha=-1.25$ from \cite{2007A&A...467..465C}.
\item The viewing angle of the precession axis $\Phi_0$ and the precession cone half-opening angle $\Omega$ are free parameters of the model and are the same for all the light curves.
\item Other free parameters, individual for each light curve, are the Lorentz factors $\gamma$ and intrinsic flux densities $S_{\nu,0}$.
\end{itemize}
Thus, the model describes a jet as a combination of emission regions responsible for producing the light curves at corresponding frequencies, the jet precesses with a certain period within a single precession cone; the emissions from different regions are delayed relative to each other according to the observed time lags; the spectral indices vary in time.


To find the best fitting parameters, we used the \mbox{UltraNest} library \citep{2021JOSS....6.3001B}, which implements Bayesian optimization. For the optical $R$ light curve, we beforehand excluded supposed flare events (the spikes in the light curve that do not correspond to smooth flux density variation characteristic of the precession process) using the Bayesian blocks algorithm \citep{2013ApJ...764..167S} implemented in Astropy \citep{2022ApJ...935..167A}. 
The false alarm probability to detect a new block was set to $p_0=0.05$, and the flare detection threshold was 1~standard deviation from the median. The lag between the 5~GHz and $R$ light curves was then remeasured ($-290\pm40$~days).  

\begin{figure}
\includegraphics[width=\columnwidth]{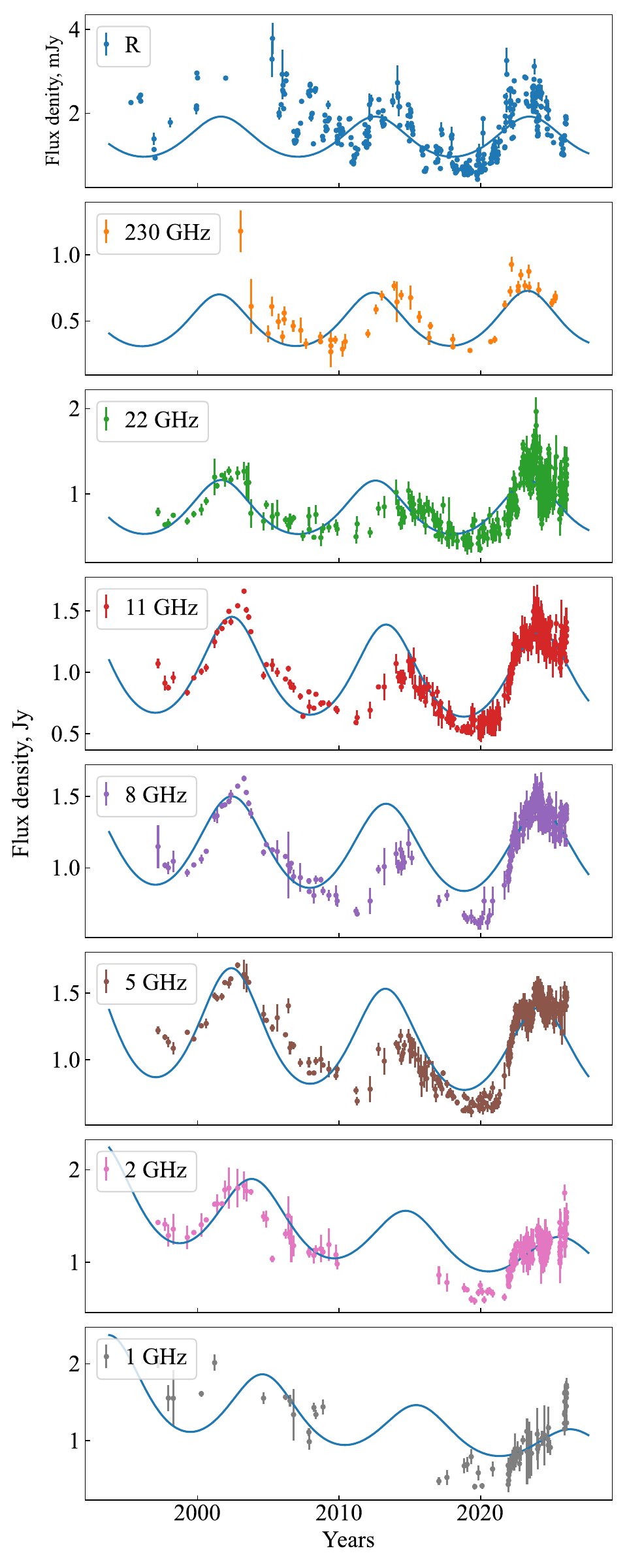}
\caption{Fitting the light curves by a precession model. Top to bottom: the optical $R$ band, 230, 22, 11, 8, 5, 2, and 1~GHz flux densities are shown by circles, the model curves are presented by solid lines.} 
\label{fig:precess}
\end{figure}
\begin{figure*}
\includegraphics[width=\textwidth]{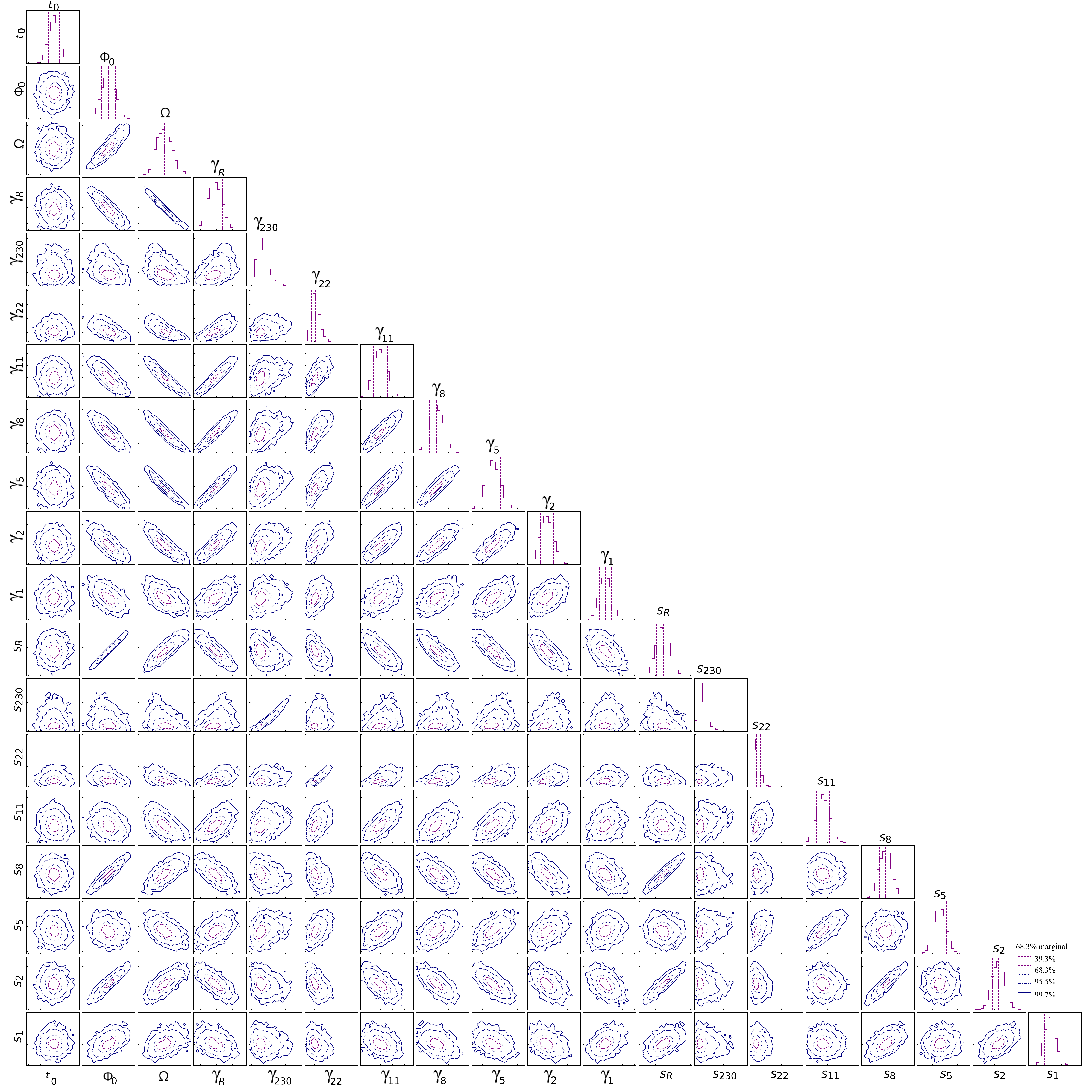}
\caption{Posterior distributions and parameter correlations (a corner diagram) for the jet-precession model. The diagonal panels show the marginalized one-dimensional posterior distributions, while the off-diagonal panels show the two-dimensional posterior distributions for each parameter pair. The plotted parameters are the reference epoch ($t_0$), the precession axis viewing angle ($\Phi_0$), the precession cone semi-opening angle~($\Omega$), the Lorentz factors $\gamma_\nu$ with the subscripts indicating the frequencies, and the corresponding intrinsic flux density normalizations $S_{\nu,0}$. The contours demonstrate the degree of correlation between the model parameters and enclose 39.3, 68.3, 95.5, and 99.7 per cent of the posterior probability in the two-dimensional projections, from inner to outer contours. The vertical dashed lines in the one-dimensional histograms mark the best-fitted parameters (the medians) and the 68.3 per cent marginal credible intervals, the corresponding numerical values are given in Table~\ref{tab:precess} and its caption.} 
\label{fig:precess_corner}
\end{figure*}

The results of fitting the precession model are given in Figs.~\ref{fig:precess}, ~\ref{fig:precess_corner} and Table~\ref{tab:precess}. We have not included the $\gamma$-ray curve in the modelling as it does not show the smooth flux density variation characteristic of precession. The model provides a satisfactory fitting of the light curves and even reproduces the overall decrease of intensity with time at 1--5 GHz thanks to allowing for spectral index evolution. Nevertheless, the derived combination of the precession axis viewing angle $\Phi_0=27\fdg4$ and the precession cone half-opening angle $\Omega=3\fdg1$ means that the actual jet viewing angle is always greater than $\sim24\deg$, which contradicts all other observations. 

\begin{table}
\centering
\caption{Parameters of the precession model: precession axis viewing angle $\Phi_0=27\fdg4\pm0\fdg4$, precession cone half-opening angle $\Omega=3\fdg1\pm0\fdg1$, initial phase $t_0(\text{5~GHz})=2005.176\pm0.002$~yr. The period $P=10.9$~yr is taken from the sinusoidal fitting. The lags with respect to the light curve at 5~GHz are taken from the DCF calculations.}
\label{tab:precess}
\begin{tabular}{c r@{$\,\pm\,$}l r@{$\,\pm\,$}l r@{$\,\pm\,$}l}
\hline
$\nu$ (GHz) &\multicolumn{2}{c}{$\gamma_\nu$}&\multicolumn{2}{c}{$S_{\nu, 0}$ (Jy)}&\multicolumn{2}{c}{Lag (days)}\\
\hline
1   &2.9&0.2 &0.28&0.01 &850&230 \\
2   &2.5&0.1 &0.27&0.01 &570&280 \\
5   &3.6&0.2 &0.29&0.01 &0&0 \\
8   &2.5&0.1 &0.21&0.01 &0&20 \\
11  &4.0&0.2 &0.25&0.01 &0&30 \\
22  &4.6&0.4 &0.27&0.03 &$-280$&80 \\
230 &5.8&1.0 &0.24&0.07 &$-340$&140 \\
$R$ &2.2&0.1 &0.11&0.01 &$-290$&40 \\
\hline
\end{tabular}
\end{table}

To interpret this obvious discrepancy, we first should mention that unlike in other investigations (e.g., \citealt{2018MNRAS.478.3199B, 2023ApJ...951..106B}), here we do not derive the precession geometry from direct VLBI monitoring of the jet but instead are trying to restore the angles solely from the light curves along with other free parameters. In turn, the model does not incorporate any other mechanisms of flux density variation other then precession. Any stochastic flux density variation may influence the computed likelihood and distort the derived parameters. Another reason for the discrepancy may be a jet structure more complex than a straight line supposed in the precession model. For instance, \cite{2010A&A...515A.105B} describe an appearing and disappearing ``staircase'' mode of the jet during a 1995--2008 VLBI monitoring and explain the effect by the non-ballistic model of \cite{2008MNRAS.389..315G}, where a precessing continuous jet produces discrete hotspots via interaction with the ambient medium.

We thus can conclude that the general variations in the observed light curves may potentially be explained by jet precession, but the derived parameters are likely far from their true values because of the influence of other effects that are not included in the precession model and/or due to a complex pattern of the jet emission regions.

\subsection{Shock propagation in an opacity-stratified jet}

While long-term geometric effects (e.g., jet bending or precession) may modulate the baseline emission level, the flare evolution in \pks appears to be dominated by disturbances propagating along the jet.

\begin{figure}
\centerline{\includegraphics[width=1.0\linewidth]{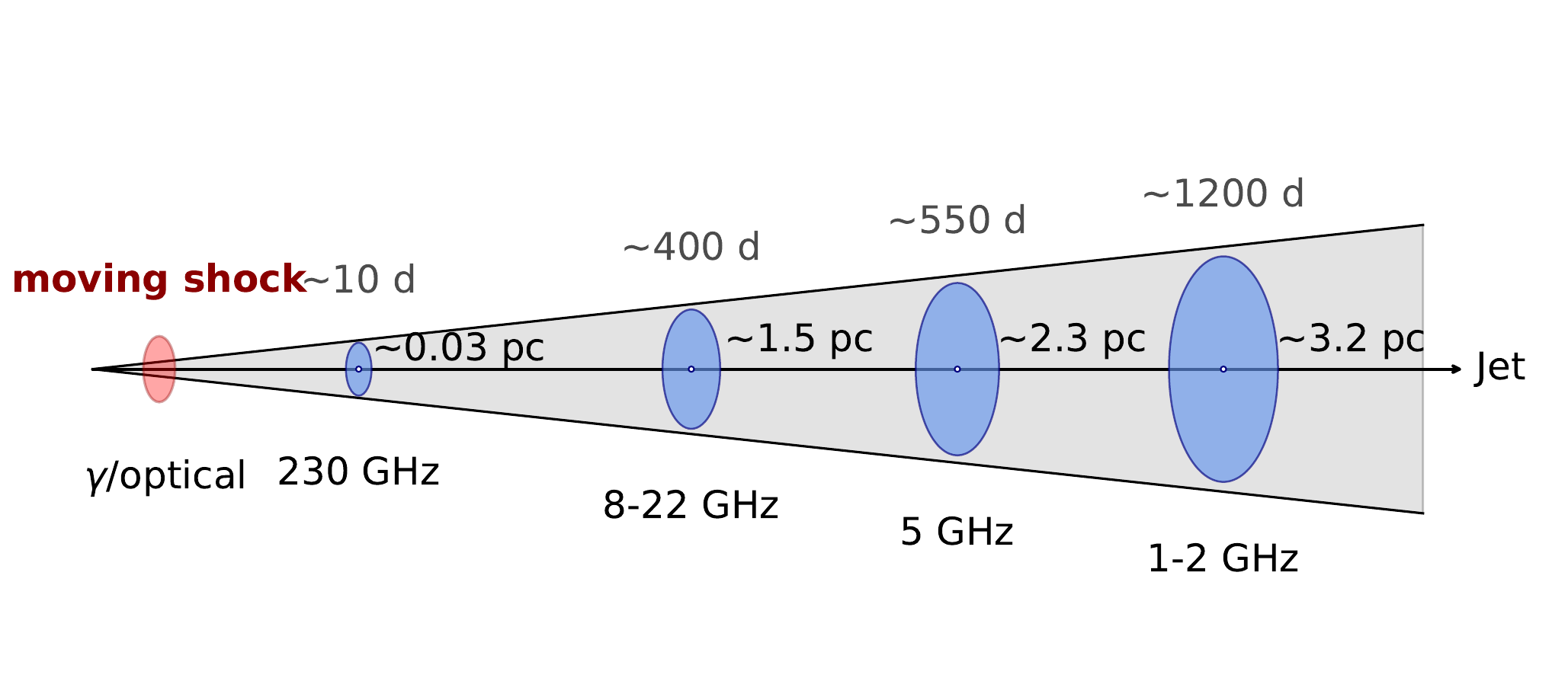}}
\caption{Schematic illustration of a propagating shock in the opacity-stratified parsec-scale jet of the blazar \pks. The shock (red circle) moves downstream along the jet axis (black cone). Blue circles mark the approximate locations of the synchrotron photospheres at different radio frequencies, separated by projected distances of $\sim 0.03$--3\,pc along the jet, with corresponding time delays 
(days after the shock--core interaction). The $\gamma$-ray/optical emission region is indicated upstream.} 
\label{fig:scheme}
\end{figure}

The observed frequency-dependent time delays (Section~\ref{sec:delays}) provide strong evidence for an opacity-stratified jet structure. In such a scenario, the radio ``core'' at each frequency corresponds to a $\tau_\nu \approx 1$ synchrotron photosphere, whose distance from the jet apex follows $r_{\rm core}(\nu) \propto \nu^{-1/k_r}$ \citep[e.g.,][]{1981ApJ...243..700K,1998A&A...330...79L}. A disturbance moving downstream becomes observable at progressively lower frequencies as it crosses these photospheres, naturally producing the observed delay--frequency relation. A schematic illustration of this scenario is shown in Fig.~\ref{fig:scheme}.

The obtained value $k_r = 1.96 \pm 0.24$ exceeds the canonical $k_r \sim 1$ expected for a conical equipartition jet, suggesting deviations from simple equipartition conditions. This may indicate steeper gradients in the magnetic field and particle density or additional stratification within the flow.

The spectral index evolution further supports this interpretation. The high-frequency spectral index remains in the range $\sim 0$ to $-0.5$, indicating that the emission is predominantly observed in the optically thin regime above a few GHz. This implies that the shocks are detected after their formation, when radiative cooling and particle evolution dominate the emission.

Within the framework of the shock-in-jet model \citep{1985ApJ...298..114M}, the flare evolution can be qualitatively divided into three stages: an initial high-frequency peak associated with particle injection, followed by a shift of the spectral peak toward lower frequencies, and finally a delayed low-frequency response dominated by adiabatic expansion.

The consistency between the delay-derived propagation speed and the VLBI kinematics provides additional support for this scenario, indicating that both the time delays and 
the core-shift measurements trace the same physical process, namely a disturbance propagating downstream through an opacity-stratified jet.

\section{Summary}

We have investigated the multiwavelength variability of \pks using $\sim$30 years of radio, optical, and $\gamma$-ray observations (1995--2026). The main results can be summarized as follows.
\begin{enumerate}

\item The source exhibits three major outbursts over the observed period. The third bright flare, coincided in time with IC211208A, is characterized by emission from a compact and optically thick region, which is consistent with enhanced activity and efficient particle acceleration. The $\gamma$-ray and optical light curves are closely correlated with no significant delay on long timescales, while the radio emission lags behind by $\sim$300--1200 days in a frequency-dependent manner. The increase of delay toward lower frequencies is consistent with synchrotron opacity and a stratified jet structure.

\item Short optical and $\gamma$-ray flares occurred within a few days of the IC211208A neutrino event. The $\sim$12-day lag of the $\gamma$-ray emission relative to the optical flare, reliably determined for the first time, suggests that the emission in these bands is not strictly co-spatial or instantaneous. This delay is consistent with a multi-zone emission structure or the contribution of additional processes, such as external Compton scattering or evolving shock conditions within the jet.

\item A characteristic variability time-scale of $\sim$10--11 yr is robustly detected across the radio--mm bands. In contrast, the optical and $\gamma$-ray variability show weaker, time-localised shorter-period signals that do not form a coherent long-term pattern. Statistical tests against red-noise variability indicate that the radio modulation is quasi-periodic rather than strictly periodic.

\item Application of a jet precession model reproduces the overall variability pattern; however, it requires unrealistically large viewing angles that are inconsistent with VLBI constraints, indicating that precession alone cannot explain the observed MW behaviour.

\item Independent estimates based on VLBI core-shift measurements and radio time delays provide consistent constraints on the jet parameters, including the propagation speed of disturbances and the location of emission regions. This agreement indicates that both the observed delays and the evolution of radio flares are governed by propagation effects along the jet.

\end{enumerate}
The strong temporal correlation found between IC211208A and the bright MW flare in \pks is an extra argument for the blazar--neutrino connection. Overall, the observational results indicate that the variability of \pks is governed by two distinct processes operating on different timescales: a long-term quasi-periodic modulation affecting the radio emission, and short-term flares associated with disturbances propagating through an opacity-stratified jet.

\section*{Acknowledgements}

This work is based on the data obtained with the RATAN-600 radio telescope, Zeiss-1000 and AS-500/2 optical reflectors at the Special Astrophysical Observatory of the Russian Academy of Sciences (SAO RAS). 
This work is supported by the Tianshan Talent Training Program (grant No.~2023TSYCCX0099) and the Urumqi Nanshan Astronomy and Deep Space Exploration Observation and Research Station of Xinjiang (XJYWZ2303).
The work of Y.V.S, Y.A.K., A.G.M., A.V.P., S.V.T., A.K.E., and P.G.T.\ is supported in the framework of the State project ``Science’’ by the Ministry of Science and Higher Education of the Russian Federation under the contract 075-15-2024-541. 
The work of A.B.P. is supported in the framework of the state assignment of the Federal State Budget Scientific Institution ``Crimean Astrophysical Observatory of RAS''.
YYK was supported by the MuSES project, which has received funding from the European Union (ERC grant agreement No 101142396). Views and opinions expressed are however those of the author(s) only and do not necessarily reflect those of the European Union or ERCEA. Neither the European Union nor the granting authority can be held responsible for them.

We thank M.~Gurwell for the Submillimeter Array (SMA) 230 GHz flux density time series for \pks. The SMA is a joint project between the Smithsonian Astrophysical Observatory and the Academia Sinica Institute of Astronomy and Astrophysics and is funded by the Smithsonian Institution and the Academia Sinica. We recognize that Maunakea, the site of the SMA, is culturally important for the indigenous Hawaiian people; we are privileged to study the cosmos from its summit.

This research has made use of the NASA/IPAC Extragalactic Database, which is funded by the National Aeronautics and Space Administration and operated by the California Institute of Technology.

The VLBA is a facility of the National Radio Astronomy Observatory, a facility of the National Science Foundation that is operated under cooperative agreement with Associated Universities, Inc. This research has made use of data from the \mbox{MOJAVE} database that is maintained by the \mbox{MOJAVE} team \citep{MOJAVE} and 43 GHz VLBA data from the VLBA-BU Blazar Monitoring Program (\mbox{VLBA-BU-BLAZAR}, \cite{Jorstad16}).

We thank V.~Makeev from the Max Planck Institute for Radio Astronomy for his valuable comments on the draft manuscript. The authors are grateful to the anonymous referee for constructive comments and suggestions, which helped to improve the presentation of the manuscript.

\section*{Data Availability}

The RATAN-600, Zeiss-1000, and AS-500/2 data underlying this article are available in the article and in its online supplementary material. The SMA data are available at \url{http://sma1.sma.hawaii.edu/callist/callist.html}.
The Catalina Sky Survey data are available at \url{http://nesssi.cacr.caltech.edu/DataRelease/}.
The KAIT Fermi AGN Light-Curves are available at \url {http://herculesii.astro.berkeley.edu/kait/agn/}.
The Zwicky Transient Facility data are available at  \url {https://irsa.ipac.caltech.edu/Missions/ztf.html}.
The raw \textit{Fermi} LAT data used to construct the light curve is available from the \textit{Fermi} Data Server at \url{http://heasarc.gsfc.nasa.gov/FTP/fermi/data/lat/weekly/diffuse}.

\textit{Facilities}: RATAN-600, Zeiss-1000, AS-500/2, CRTS, KAIT, ZTF, SMA, \textit{Fermi}-LAT, VLBA.

\bibliographystyle{mnras}
\bibliography{PKS0735+17}

@ARTICLE{1994AJ....108...56P,
       author = {{Perlman}, Eric S. and {Stocke}, John T.},
        title = "{New Very Large Array Maps of Highly Core-Dominated BL Lacs: testing Unified Schemes}",
      journal = {\aj},
         year = 1994,
        month = jul,
       volume = {108},
        pages = {56},
          doi = {10.1086/117044},
       adsurl = {https://ui.adsabs.harvard.edu/abs/1994AJ....108...56P}
}

@ARTICLE{2001ApJS..133....1U,
       author = {{Ueda}, Y. and {Ishisaki}, Y. and {Takahashi}, T. and {Makishima}, K. and {Ohashi}, T.},
        title = "{The ASCA Medium Sensitivity Survey (the GIS Catalog Project): Source Catalog}",
      journal = {\apjs},
         year = 2001,
        month = mar,
       volume = {133},
       number = {1},
        pages = {1-52},
          doi = {10.1086/319189},
archivePrefix = {arXiv},
       eprint = {astro-ph/9908128},
 primaryClass = {astro-ph},
       adsurl = {https://ui.adsabs.harvard.edu/abs/2001ApJS..133....1U}
}

@ARTICLE{1994ApJ...435..128G,
       author = {{Gabuzda}, D.~C. and {Wardle}, J.~F.~C. and {Roberts}, D.~H. and {Aller}, M.~F. and {Aller}, H.~D.},
        title = "{Unusual Evolution in the VLBI Structure of 0735+178}",
      journal = {\apj},
         year = 1994,
        month = nov,
       volume = {435},
        pages = {128},
          doi = {10.1086/174800},
       adsurl = {https://ui.adsabs.harvard.edu/abs/1994ApJ...435..128G}
}

@ARTICLE{2001MNRAS.328..873G,
       author = {{G{\'o}mez}, J.~L. and {Guirado}, J.~C. and {Agudo}, I. and {Marscher}, A.~P. and {Alberdi}, A. and {Marcaide}, J.~M. and {Gabuzda}, D.~C.},
        title = "{Changes in the trajectory of the radio jet in 0735+178?}",
      journal = {\mnras},
         year = 2001,
        month = dec,
       volume = {328},
       number = {3},
        pages = {873-881},
          doi = {10.1046/j.1365-8711.2001.04933.x},
archivePrefix = {arXiv},
       eprint = {astro-ph/0109179},
 primaryClass = {astro-ph},
       adsurl = {https://ui.adsabs.harvard.edu/abs/2001MNRAS.328..873G}
}

@ARTICLE{2001MNRAS.328..719G,
       author = {{Gabuzda}, D.~C. and {G{\'o}mez}, J.~L. and {Agudo}, I.},
        title = "{Evidence for parsec-scale absorption from VSOP observations of the BL Lacertae object 0735+178}",
      journal = {\mnras},
         year = 2001,
        month = dec,
       volume = {328},
       number = {3},
        pages = {719-725},
          doi = {10.1046/j.1365-8711.2001.04935.x},
       adsurl = {https://ui.adsabs.harvard.edu/abs/2001MNRAS.328..719G}
}

@ARTICLE{2021ARA&A..59..117R,
       author = {{Reynolds}, Christopher S.},
        title = "{Observational Constraints on Black Hole Spin}",
      journal = {\araa},
         year = 2021,
        month = sep,
       volume = {59},
        pages = {117-154},
          doi = {10.1146/annurev-astro-112420-035022},
archivePrefix = {arXiv},
       eprint = {2011.08948},
 primaryClass = {astro-ph.HE},
       adsurl = {https://ui.adsabs.harvard.edu/abs/2021ARA&A..59..117R}
}

@ARTICLE{1977MNRAS.179..433B,
       author = {{Blandford}, R.~D. and {Znajek}, R.~L.},
        title = "{Electromagnetic extraction of energy from Kerr black holes.}",
      journal = {\mnras},
         year = 1977,
        month = may,
       volume = {179},
        pages = {433-456},
          doi = {10.1093/mnras/179.3.433},
       adsurl = {https://ui.adsabs.harvard.edu/abs/1977MNRAS.179..433B}
}

@ARTICLE{2007MNRAS.377.1652N,
       author = {{Nemmen}, R.~S. and {Bower}, R.~G. and {Babul}, A. and {Storchi-Bergmann}, T.},
        title = "{Models for jet power in elliptical galaxies: a case for rapidly spinning black holes}",
      journal = {\mnras},
         year = 2007,
        month = jun,
       volume = {377},
       number = {4},
        pages = {1652-1662},
          doi = {10.1111/j.1365-2966.2007.11726.x},
archivePrefix = {arXiv},
       eprint = {astro-ph/0612354},
 primaryClass = {astro-ph},
       adsurl = {https://ui.adsabs.harvard.edu/abs/2007MNRAS.377.1652N}
}

@ARTICLE{2009ApJ...699..400G,
       author = {{Garofalo}, David},
        title = "{The Spin Dependence of the Blandford-Znajek Effect}",
      journal = {\apj},
         year = 2009,
        month = jul,
       volume = {699},
       number = {1},
        pages = {400-408},
          doi = {10.1088/0004-637X/699/1/400},
archivePrefix = {arXiv},
       eprint = {0904.3486},
 primaryClass = {astro-ph.HE},
       adsurl = {https://ui.adsabs.harvard.edu/abs/2009ApJ...699..400G}
}

@ARTICLE{2008MNRAS.389..315G,
       author = {{Gong}, Biping},
        title = "{The non-ballistic superluminal motion in the plane of the sky}",
      journal = {\mnras},
         year = 2008,
        month = sep,
       volume = {389},
       number = {1},
        pages = {315-320},
          doi = {10.1111/j.1365-2966.2008.13570.x},
archivePrefix = {arXiv},
       eprint = {0806.2708},
 primaryClass = {astro-ph},
       adsurl = {https://ui.adsabs.harvard.edu/abs/2008MNRAS.389..315G}
}

@ARTICLE{2010A&A...515A.105B,
       author = {{Britzen}, S. and {Witzel}, A. and {Gong}, B.~P. and {Zhang}, J.~W. and {Gopal-Krishna} and {Goyal}, A. and {Aller}, M.~F. and {Aller}, H.~D. and {Zensus}, J.~A.},
        title = "{Understanding BL Lacertae objects. Structural and kinematic mode changes in the BL Lac object PKS 0735+178}",
      journal = {\aap},
         year = 2010,
        month = jun,
       volume = {515},
          eid = {A105},
        pages = {A105},
          doi = {10.1051/0004-6361/200913685},
archivePrefix = {arXiv},
       eprint = {1002.3531},
 primaryClass = {astro-ph.CO},
       adsurl = {https://ui.adsabs.harvard.edu/abs/2010A&A...515A.105B}
}

@ARTICLE{2022ApJ...935..167A,
       author = {{Astropy Collaboration} and {Price-Whelan}, Adrian M. and {Lim}, Pey Lian and {Earl}, Nicholas and {Starkman}, Nathaniel and {Bradley}, Larry and {Shupe}, David L. and {Patil}, Aarya A. and {Corrales}, Lia and {Brasseur}, C.~E. and {N{\"o}the}, Maximilian and {Donath}, Axel and {Tollerud}, Erik and {Morris}, Brett M. and {Ginsburg}, Adam and {Vaher}, Eero and {Weaver}, Benjamin A. and {Tocknell}, James and {Jamieson}, William and {van Kerkwijk}, Marten H. and {Robitaille}, Thomas P. and {Merry}, Bruce and {Bachetti}, Matteo and {G{\"u}nther}, H. Moritz and {Aldcroft}, Thomas L. and {Alvarado-Montes}, Jaime A. and {Archibald}, Anne M. and {B{\'o}di}, Attila and {Bapat}, Shreyas and {Barentsen}, Geert and {Baz{\'a}n}, Juanjo and {Biswas}, Manish and {Boquien}, M{\'e}d{\'e}ric and {Burke}, D.~J. and {Cara}, Daria and {Cara}, Mihai and {Conroy}, Kyle E. and {Conseil}, Simon and {Craig}, Matthew W. and {Cross}, Robert M. and {Cruz}, Kelle L. and {D'Eugenio}, Francesco and {Dencheva}, Nadia and {Devillepoix}, Hadrien A.~R. and {Dietrich}, J{\"o}rg P. and {Eigenbrot}, Arthur Davis and {Erben}, Thomas and {Ferreira}, Leonardo and {Foreman-Mackey}, Daniel and {Fox}, Ryan and {Freij}, Nabil and {Garg}, Suyog and {Geda}, Robel and {Glattly}, Lauren and {Gondhalekar}, Yash and {Gordon}, Karl D. and {Grant}, David and {Greenfield}, Perry and {Groener}, Austen M. and {Guest}, Steve and {Gurovich}, Sebastian and {Handberg}, Rasmus and {Hart}, Akeem and {Hatfield-Dodds}, Zac and {Homeier}, Derek and {Hosseinzadeh}, Griffin and {Jenness}, Tim and {Jones}, Craig K. and {Joseph}, Prajwel and {Kalmbach}, J. Bryce and {Karamehmetoglu}, Emir and {Ka{\l}uszy{\'n}ski}, Miko{\l}aj and {Kelley}, Michael S.~P. and {Kern}, Nicholas and {Kerzendorf}, Wolfgang E. and {Koch}, Eric W. and {Kulumani}, Shankar and {Lee}, Antony and {Ly}, Chun and {Ma}, Zhiyuan and {MacBride}, Conor and {Maljaars}, Jakob M. and {Muna}, Demitri and {Murphy}, N.~A. and {Norman}, Henrik and {O'Steen}, Richard and {Oman}, Kyle A. and {Pacifici}, Camilla and {Pascual}, Sergio and {Pascual-Granado}, J. and {Patil}, Rohit R. and {Perren}, Gabriel I. and {Pickering}, Timothy E. and {Rastogi}, Tanuj and {Roulston}, Benjamin R. and {Ryan}, Daniel F. and {Rykoff}, Eli S. and {Sabater}, Jose and {Sakurikar}, Parikshit and {Salgado}, Jes{\'u}s and {Sanghi}, Aniket and {Saunders}, Nicholas and {Savchenko}, Volodymyr and {Schwardt}, Ludwig and {Seifert-Eckert}, Michael and {Shih}, Albert Y. and {Jain}, Anany Shrey and {Shukla}, Gyanendra and {Sick}, Jonathan and {Simpson}, Chris and {Singanamalla}, Sudheesh and {Singer}, Leo P. and {Singhal}, Jaladh and {Sinha}, Manodeep and {Sip{\H{o}}cz}, Brigitta M. and {Spitler}, Lee R. and {Stansby}, David and {Streicher}, Ole and {{\v{S}}umak}, Jani and {Swinbank}, John D. and {Taranu}, Dan S. and {Tewary}, Nikita and {Tremblay}, Grant R. and {de Val-Borro}, Miguel and {Van Kooten}, Samuel J. and {Vasovi{\'c}}, Zlatan and {Verma}, Shresth and {de Miranda Cardoso}, Jos{\'e} Vin{\'\i}cius and {Williams}, Peter K.~G. and {Wilson}, Tom J. and {Winkel}, Benjamin and {Wood-Vasey}, W.~M. and {Xue}, Rui and {Yoachim}, Peter and {Zhang}, Chen and {Zonca}, Andrea and {Astropy Project Contributors}},
        title = "{The Astropy Project: Sustaining and Growing a Community-oriented Open-source Project and the Latest Major Release (v5.0) of the Core Package}",
      journal = {\apj},
         year = 2022,
        month = aug,
       volume = {935},
       number = {2},
          eid = {167},
        pages = {167},
          doi = {10.3847/1538-4357/ac7c74},
archivePrefix = {arXiv},
       eprint = {2206.14220},
 primaryClass = {astro-ph.IM},
       adsurl = {https://ui.adsabs.harvard.edu/abs/2022ApJ...935..167A}
}

@ARTICLE{Pushkarev17,
       author = {{Pushkarev}, A.~B. and {Kovalev}, Y.~Y. and {Lister}, M.~L. and {Savolainen}, T.},
        title = "{MOJAVE - XIV. Shapes and opening angles of AGN jets}",
      journal = {\mnras},
         year = 2017,
        month = jul,
       volume = {468},
       number = {4},
        pages = {4992-5003},
          doi = {10.1093/mnras/stx854},
archivePrefix = {arXiv},
       eprint = {1705.02888},
 primaryClass = {astro-ph.HE},
       adsurl = {https://ui.adsabs.harvard.edu/abs/2017MNRAS.468.4992P}
}

@ARTICLE{2014Natur.510..126Z,
       author = {{Zamaninasab}, M. and {Clausen-Brown}, E. and {Savolainen}, T. and {Tchekhovskoy}, A.},
        title = "{Dynamically important magnetic fields near accreting supermassive black holes}",
      journal = {\nat},
         year = 2014,
        month = jun,
       volume = {510},
       number = {7503},
        pages = {126-128},
          doi = {10.1038/nature13399},
       adsurl = {https://ui.adsabs.harvard.edu/abs/2014Natur.510..126Z}
}

@ARTICLE{Homan21,
       author = {{Homan}, D.~C. and {Cohen}, M.~H. and {Hovatta}, T. and {Kellermann}, K.~I. and {Kovalev}, Y.~Y. and {Lister}, M.~L. and {Popkov}, A.~V. and {Pushkarev}, A.~B. and {Ros}, E. and {Savolainen}, T.},
        title = "{MOJAVE. XIX. Brightness Temperatures and Intrinsic Properties of Blazar Jets}",
      journal = {\apj},
         year = 2021,
        month = dec,
       volume = {923},
       number = {1},
          eid = {67},
        pages = {67},
          doi = {10.3847/1538-4357/ac27af},
archivePrefix = {arXiv},
       eprint = {2109.04977},
 primaryClass = {astro-ph.HE},
       adsurl = {https://ui.adsabs.harvard.edu/abs/2021ApJ...923...67H}
}

@ARTICLE{Kutkin19,
       author = {{Kutkin}, A.~M. and {Pashchenko}, I.~N. and {Sokolovsky}, K.~V. and {Kovalev}, Y.~Y. and {Aller}, M.~F. and {Aller}, H.~D.},
        title = "{Opacity, variability, and kinematics of AGN jets}",
      journal = {\mnras},
         year = 2019,
        month = jun,
       volume = {486},
       number = {1},
        pages = {430-439},
          doi = {10.1093/mnras/stz885},
archivePrefix = {arXiv},
       eprint = {1809.05536},
 primaryClass = {astro-ph.GA},
       adsurl = {https://ui.adsabs.harvard.edu/abs/2019MNRAS.486..430K}
}

@ARTICLE{Pushkarev19,
       author = {{Pushkarev}, A.~B. and {Butuzova}, M.~S. and {Kovalev}, Y.~Y. and {Hovatta}, T.},
        title = "{Multifrequency study of the gamma-ray flaring BL Lacertae object PKS 2233-148 in 2009-2012}",
      journal = {\mnras},
         year = 2019,
        month = jan,
       volume = {482},
       number = {2},
        pages = {2336-2353},
          doi = {10.1093/mnras/sty2724},
archivePrefix = {arXiv},
       eprint = {1808.06138},
 primaryClass = {astro-ph.HE},
       adsurl = {https://ui.adsabs.harvard.edu/abs/2019MNRAS.482.2336P}
}

@ARTICLE{Kudryavtseva11,
       author = {{Kudryavtseva}, N.~A. and {Gabuzda}, D.~C. and {Aller}, M.~F. and {Aller}, H.~D.},
        title = "{A new method for estimating frequency-dependent core shifts in active galactic nucleus jets}",
      journal = {\mnras},
         year = 2011,
        month = aug,
       volume = {415},
       number = {2},
        pages = {1631-1637},
          doi = {10.1111/j.1365-2966.2011.18808.x},
archivePrefix = {arXiv},
       eprint = {1106.0069},
 primaryClass = {astro-ph.HE},
       adsurl = {https://ui.adsabs.harvard.edu/abs/2011MNRAS.415.1631K}
}

@ARTICLE{Bach06,
       author = {{Bach}, U. and {Villata}, M. and {Raiteri}, C.~M. and {Agudo}, I. and {Aller}, H.~D. and {Aller}, M.~F. and {Denn}, G. and {G{\'o}mez}, J.~L. and {Jorstad}, S. and {Marscher}, A. and {Mutel}, R.~L. and {Ter{\"a}sranta}, H.},
        title = "{Structure and flux variability in the VLBI jet of BL Lacertae during the WEBT campaigns (1995-2004)}",
      journal = {\aap},
         year = 2006,
        month = sep,
       volume = {456},
       number = {1},
        pages = {105-115},
          doi = {10.1051/0004-6361:20065235},
archivePrefix = {arXiv},
       eprint = {astro-ph/0606050},
 primaryClass = {astro-ph},
       adsurl = {https://ui.adsabs.harvard.edu/abs/2006A&A...456..105B}
}

@ARTICLE{Jorstad17,
       author = {{Jorstad}, Svetlana G. and {Marscher}, Alan P. and {Morozova}, Daria A. and {Troitsky}, Ivan S. and {Agudo}, Iv{\'a}n and {Casadio}, Carolina and {Foord}, Adi and {G{\'o}mez}, Jos{\'e} L. and {MacDonald}, Nicholas R. and {Molina}, Sol N. and {L{\"a}hteenm{\"a}ki}, Anne and {Tammi}, Joni and {Tornikoski}, Merja},
        title = "{Kinematics of Parsec-scale Jets of Gamma-Ray Blazars at 43 GHz within the VLBA-BU-BLAZAR Program}",
      journal = {\apj},
         year = 2017,
        month = sep,
       volume = {846},
       number = {2},
          eid = {98},
        pages = {98},
          doi = {10.3847/1538-4357/aa8407},
archivePrefix = {arXiv},
       eprint = {1711.03983},
 primaryClass = {astro-ph.GA},
       adsurl = {https://ui.adsabs.harvard.edu/abs/2017ApJ...846...98J}
}

@ARTICLE{Jorstad16,
       author = {{Jorstad}, Svetlana and {Marscher}, Alan},
        title = "{The VLBA-BU-BLAZAR Multi-Wavelength Monitoring Program}",
      journal = {Galaxies},
         year = 2016,
        month = oct,
       volume = {4},
       number = {4},
          eid = {47},
        pages = {47},
          doi = {10.3390/galaxies4040047},
       adsurl = {https://ui.adsabs.harvard.edu/abs/2016Galax...4...47J}
}

@ARTICLE{MOJAVE,
       author = {{Lister}, M.~L. and {Aller}, M.~F. and {Aller}, H.~D. and {Hodge}, M.~A. and {Homan}, D.~C. and {Kovalev}, Y.~Y. and {Pushkarev}, A.~B. and {Savolainen}, T.},
        title = "{MOJAVE. XV. VLBA 15 GHz Total Intensity and Polarization Maps of 437 Parsec-scale AGN Jets from 1996 to 2017}",
      journal = {\apjs},
         year = 2018,
        month = jan,
       volume = {234},
       number = {1},
          eid = {12},
        pages = {12},
          doi = {10.3847/1538-4365/aa9c44},
archivePrefix = {arXiv},
       eprint = {1711.07802},
 primaryClass = {astro-ph.GA},
       adsurl = {https://ui.adsabs.harvard.edu/abs/2018ApJS..234...12L}
}

@ARTICLE{Kovalev08,
       author = {{Kovalev}, Y.~Y. and {Lobanov}, A.~P. and {Pushkarev}, A.~B. and {Zensus}, J.~A.},
        title = "{Opacity in compact extragalactic radio sources and its effect on astrophysical and astrometric studies}",
      journal = {\aap},
         year = 2008,
        month = jun,
       volume = {483},
       number = {3},
        pages = {759-768},
          doi = {10.1051/0004-6361:20078679},
archivePrefix = {arXiv},
       eprint = {0802.2970},
 primaryClass = {astro-ph},
       adsurl = {https://ui.adsabs.harvard.edu/abs/2008A&A...483..759K}
}

@INPROCEEDINGS{DIFMAP,
       author = {{Shepherd}, M.~C.},
        title = "{Difmap: an Interactive Program for Synthesis Imaging}",
    booktitle = {Astronomical Data Analysis Software and Systems VI},
         year = 1997,
       editor = {{Hunt}, Gareth and {Payne}, Harry},
       series = {Astronomical Society of the Pacific Conference Series},
       volume = {125},
        month = jan,
        pages = {77},
       adsurl = {https://ui.adsabs.harvard.edu/abs/1997ASPC..125...77S}
}

@ARTICLE{2021JOSS....6.3001B,
       author = {{Buchner}, Johannes},
        title = "{UltraNest - a robust, general purpose Bayesian inference engine}",
      journal = {The Journal of Open Source Software},
         year = 2021,
        month = apr,
       volume = {6},
       number = {60},
          eid = {3001},
        pages = {3001},
          doi = {10.21105/joss.03001},
archivePrefix = {arXiv},
       eprint = {2101.09604},
 primaryClass = {stat.CO},
       adsurl = {https://ui.adsabs.harvard.edu/abs/2021JOSS....6.3001B}
}

@ARTICLE{1998PASP..110..660P,
       author = {{Peterson}, Bradley M. and {Wanders}, Ignaz and {Horne}, Keith and {Collier}, Stefan and {Alexander}, Tal and {Kaspi}, Shai and {Maoz}, Dan},
        title = "{On Uncertainties in Cross-Correlation Lags and the Reality of Wavelength-dependent Continuum Lags in Active Galactic Nuclei}",
      journal = {\pasp},
         year = 1998,
        month = jun,
       volume = {110},
       number = {748},
        pages = {660-670},
          doi = {10.1086/316177},
archivePrefix = {arXiv},
       eprint = {astro-ph/9802103},
 primaryClass = {astro-ph},
       adsurl = {https://ui.adsabs.harvard.edu/abs/1998PASP..110..660P}
}

@ARTICLE{2009MNRAS.399.1622G,
       author = {{Goyal}, Arti and {Gopal-Krishna} and {Anupama}, G.~C. and {Sahu}, D.~K. and {Sagar}, R. and {Britzen}, S. and {Karouzos}, M. and {Aller}, M.~F. and {Aller}, H.~D.},
        title = "{Unusual optical quiescence of the classical BL Lac object PKS 0735+178 on intranight time-scale}",
      journal = {\mnras},
         year = 2009,
        month = nov,
       volume = {399},
       number = {3},
        pages = {1622-1632},
          doi = {10.1111/j.1365-2966.2009.15385.x},
archivePrefix = {arXiv},
       eprint = {0907.2285},
 primaryClass = {astro-ph.CO},
       adsurl = {https://ui.adsabs.harvard.edu/abs/2009MNRAS.399.1622G}
}

@ARTICLE{1985ApJ...298..296A,
       author = {{Aller}, H.~D. and {Aller}, M.~F. and {Hughes}, P.~A.},
        title = "{Polarized radio outbursts in BL Lacertae. I. Polarized emission from a compact jet.}",
      journal = {\apj},
         year = 1985,
        month = nov,
       volume = {298},
        pages = {296-300},
          doi = {10.1086/163610},
       adsurl = {https://ui.adsabs.harvard.edu/abs/1985ApJ...298..296A}
}

@ARTICLE{4FGL-DR3,
       author = {{Abdollahi}, S. and {Acero}, F. and {Baldini}, L. and {Ballet}, J. and {Bastieri}, D. and {Bellazzini}, R. and {Berenji}, B. and {Berretta}, A. and {Bissaldi}, E. and {Blandford}, R.~D. and {Bloom}, E. and {Bonino}, R. and {Brill}, A. and {Britto}, R.~J. and {Bruel}, P. and {Burnett}, T.~H. and {Buson}, S. and {Cameron}, R.~A. and {Caputo}, R. and {Caraveo}, P.~A. and {Castro}, D. and {Chaty}, S. and {Cheung}, C.~C. and {Chiaro}, G. and {Cibrario}, N. and {Ciprini}, S. and {Coronado-Bl{\'a}zquez}, J. and {Crnogorcevic}, M. and {Cutini}, S. and {D'Ammando}, F. and {De Gaetano}, S. and {Digel}, S.~W. and {Di Lalla}, N. and {Dirirsa}, F. and {Di Venere}, L. and {Dom{\'\i}nguez}, A. and {Fallah Ramazani}, V. and {Fegan}, S.~J. and {Ferrara}, E.~C. and {Fiori}, A. and {Fleischhack}, H. and {Franckowiak}, A. and {Fukazawa}, Y. and {Funk}, S. and {Fusco}, P. and {Galanti}, G. and {Gammaldi}, V. and {Gargano}, F. and {Garrappa}, S. and {Gasparrini}, D. and {Giacchino}, F. and {Giglietto}, N. and {Giordano}, F. and {Giroletti}, M. and {Glanzman}, T. and {Green}, D. and {Grenier}, I.~A. and {Grondin}, M.-H. and {Guillemot}, L. and {Guiriec}, S. and {Gustafsson}, M. and {Harding}, A.~K. and {Hays}, E. and {Hewitt}, J.~W. and {Horan}, D. and {Hou}, X. and {J{\'o}hannesson}, G. and {Karwin}, C. and {Kayanoki}, T. and {Kerr}, M. and {Kuss}, M. and {Landriu}, D. and {Larsson}, S. and {Latronico}, L. and {Lemoine-Goumard}, M. and {Li}, J. and {Liodakis}, I. and {Longo}, F. and {Loparco}, F. and {Lott}, B. and {Lubrano}, P. and {Maldera}, S. and {Malyshev}, D. and {Manfreda}, A. and {Mart{\'\i}-Devesa}, G. and {Mazziotta}, M.~N. and {Mereu}, I. and {Meyer}, M. and {Michelson}, P.~F. and {Mirabal}, N. and {Mitthumsiri}, W. and {Mizuno}, T. and {Moiseev}, A.~A. and {Monzani}, M.~E. and {Morselli}, A. and {Moskalenko}, I.~V. and {Negro}, M. and {Nuss}, E. and {Omodei}, N. and {Orienti}, M. and {Orlando}, E. and {Paneque}, D. and {Pei}, Z. and {Perkins}, J.~S. and {Persic}, M. and {Pesce-Rollins}, M. and {Petrosian}, V. and {Pillera}, R. and {Poon}, H. and {Porter}, T.~A. and {Principe}, G. and {Rain{\`o}}, S. and {Rando}, R. and {Rani}, B. and {Razzano}, M. and {Razzaque}, S. and {Reimer}, A. and {Reimer}, O. and {Reposeur}, T. and {S{\'a}nchez-Conde}, M. and {Saz Parkinson}, P.~M. and {Scotton}, L. and {Serini}, D. and {Sgr{\`o}}, C. and {Siskind}, E.~J. and {Smith}, D.~A. and {Spandre}, G. and {Spinelli}, P. and {Sueoka}, K. and {Suson}, D.~J. and {Tajima}, H. and {Tak}, D. and {Thayer}, J.~B. and {Thompson}, D.~J. and {Torres}, D.~F. and {Troja}, E. and {Valverde}, J. and {Wood}, K. and {Zaharijas}, G.},
        title = "{Incremental Fermi Large Area Telescope Fourth Source Catalog}",
      journal = {\apjs},
         year = 2022,
        month = jun,
       volume = {260},
       number = {2},
          eid = {53},
        pages = {53},
          doi = {10.3847/1538-4365/ac6751},
archivePrefix = {arXiv},
       eprint = {2201.11184},
 primaryClass = {astro-ph.HE},
       adsurl = {https://ui.adsabs.harvard.edu/abs/2022ApJS..260...53A}
}

@ARTICLE{RFC,
       author = {{Petrov}, L.~Y. and {Kovalev}, Y.~Y.},
        title = "{The Radio Fundamental Catalog. I. Astrometry}",
      journal = {\apjs},
         year = 2025,
        month = feb,
       volume = {276},
       number = {2},
          eid = {38},
        pages = {38},
          doi = {10.3847/1538-4365/ad8c36},
archivePrefix = {arXiv},
       eprint = {2410.11794},
 primaryClass = {astro-ph.IM},
       adsurl = {https://ui.adsabs.harvard.edu/abs/2025ApJS..276...38P}
}

@ARTICLE{1981ApJ...243..700K,
       author = {{Konigl}, A.},
        title = "{Relativistic jets as X-ray and gamma-ray sources.}",
      journal = {\apj},
         year = 1981,
        month = feb,
       volume = {243},
        pages = {700-709},
          doi = {10.1086/158638},
       adsurl = {https://ui.adsabs.harvard.edu/abs/1981ApJ...243..700K}
}

@ARTICLE{1998A&A...330...79L,
       author = {{Lobanov}, A.~P.},
        title = "{Ultracompact jets in active galactic nuclei}",
      journal = {\aap},
         year = 1998,
        month = feb,
       volume = {330},
        pages = {79-89},
          doi = {10.48550/arXiv.astro-ph/9712132},
archivePrefix = {arXiv},
       eprint = {astro-ph/9712132},
 primaryClass = {astro-ph},
       adsurl = {https://ui.adsabs.harvard.edu/abs/1998A&A...330...79L}
}

@ARTICLE{2018MNRAS.478.3199B,
       author = {{Britzen}, S. and {Fendt}, C. and {Witzel}, G. and {Qian}, S. -J. and {Pashchenko}, I.~N. and {Kurtanidze}, O. and {Zajacek}, M. and {Martinez}, G. and {Karas}, V. and {Aller}, M. and {Aller}, H. and {Eckart}, A. and {Nilsson}, K. and {Ar{\'e}valo}, P. and {Cuadra}, J. and {Subroweit}, M. and {Witzel}, A.},
        title = "{OJ287: deciphering the `Rosetta stone of blazars}",
      journal = {\mnras},
         year = 2018,
        month = aug,
       volume = {478},
       number = {3},
        pages = {3199-3219},
          doi = {10.1093/mnras/sty1026},
       adsurl = {https://ui.adsabs.harvard.edu/abs/2018MNRAS.478.3199B}
}

@ARTICLE{2023ApJ...951..106B,
       author = {{Britzen}, Silke and {Zaja{\v{c}}ek}, Michal and {Gopal-Krishna} and {Fendt}, Christian and {Kun}, Emma and {Jaron}, Fr{\'e}d{\'e}ric and {Sillanp{\"a}{\"a}}, Aimo and {Eckart}, Andreas},
        title = "{Precession-induced Variability in AGN Jets and OJ 287}",
      journal = {\apj},
         year = 2023,
        month = jul,
       volume = {951},
       number = {2},
          eid = {106},
        pages = {106},
          doi = {10.3847/1538-4357/accbbc},
archivePrefix = {arXiv},
       eprint = {2307.05838},
 primaryClass = {astro-ph.HE},
       adsurl = {https://ui.adsabs.harvard.edu/abs/2023ApJ...951..106B}
}

@ARTICLE{2021ATel15105....1K,
       author = {{Kadler}, Matthias and {Benke}, Petra and {Gokus}, Andrea and {Hessdoerfer}, Jonas and {Sinapius}, Jonas and {Amp} and {Weber}, Philip and {TELAMON Team} and {Tornikoski}, Merja and {Righini}, Simona and {Marchili}, Nicola and {Hovatta}, Talvikki and {Readhead}, Anthony C. and {Kiehlmann}, Sebastian and {Kovalev}, Yuri A. and {Popkov}, Alexander V. and {Kovalev}, Yuri Y.},
        title = "{TELAMON, Metsahovi, Medicina, OVRO and RATAN-600 programs find a long-term radio flare in PKS0735+17 coincident with IceCube-211208A}",
      journal = {The Astronomer's Telegram},
         year = 2021,
        month = dec,
       volume = {15105},
        pages = {1},
       adsurl = {https://ui.adsabs.harvard.edu/abs/2021ATel15105....1K}
}

@ARTICLE{1997A&AS..125..525F,
       author = {{Fan}, J.~H. and {Xie}, G.~Z. and {Lin}, R.~G. and {Qin}, Y.~P. and {Li}, K.~H. and {Zhang}, X.},
        title = "{The long-term variability of BL Lac object PKS 0735+178}",
      journal = {\aaps},
         year = 1997,
        month = nov,
       volume = {125},
        pages = {525-528},
          doi = {10.1051/aas:1997240},
       adsurl = {https://ui.adsabs.harvard.edu/abs/1997A&AS..125..525F}
}

@ARTICLE{2004PASP..116..161Q,
       author = {{Qian}, Bochen and {Tao}, Jun},
        title = "{Optical Monitoring of PKS 0735+178 from 1995 to 2001 and Its Historical Periodic Light Curve}",
      journal = {\pasp},
         year = 2004,
        month = feb,
       volume = {116},
       number = {816},
        pages = {161-169},
          doi = {10.1086/381536},
       adsurl = {https://ui.adsabs.harvard.edu/abs/2004PASP..116..161Q}
}

@ARTICLE{Lott12,
       author = {{Lott}, B. and {Escande}, L. and {Larsson}, S. and {Ballet}, J.},
        title = "{An adaptive-binning method for generating constant-uncertainty/constant-significance light curves with Fermi-LAT data}",
      journal = {\aap},
         year = 2012,
        month = aug,
       volume = {544},
          eid = {A6},
        pages = {A6},
          doi = {10.1051/0004-6361/201218873},
archivePrefix = {arXiv},
       eprint = {1201.4851},
 primaryClass = {astro-ph.HE},
       adsurl = {https://ui.adsabs.harvard.edu/abs/2012A&A...544A...6L}
}

@ARTICLE{2025A&A...699A.381K,
       author = {{Kim}, Yu-Sik and {Kim}, Jae-Young},
        title = "{The dynamics of the parsec-scale jet in the neutrino blazar PKS 0735+178}",
      journal = {\aap},
         year = 2025,
        month = jul,
       volume = {699},
          eid = {A381},
        pages = {A381},
          doi = {10.1051/0004-6361/202452111},
archivePrefix = {arXiv},
       eprint = {2505.13876},
 primaryClass = {astro-ph.HE},
       adsurl = {https://ui.adsabs.harvard.edu/abs/2025A&A...699A.381K}
}

@ARTICLE{2025ApJ...989..208P,
       author = {{Paraschos}, G.~F. and {Traianou}, E. and {Debbrecht}, L.~C. and {Liodakis}, I. and {Ros}, E.},
        title = "{Polarization as a Probe of Neutrino Emission from Blazars}",
      journal = {\apj},
         year = 2025,
        month = aug,
       volume = {989},
       number = {2},
          eid = {208},
        pages = {208},
          doi = {10.3847/1538-4357/adf110},
archivePrefix = {arXiv},
       eprint = {2507.16929},
 primaryClass = {astro-ph.HE},
       adsurl = {https://ui.adsabs.harvard.edu/abs/2025ApJ...989..208P}
}

@ARTICLE{2022ApJS..260...12W,
       author = {{Weaver}, Zachary R. and {Jorstad}, Svetlana G. and {Marscher}, Alan P. and {Morozova}, Daria A. and {Troitsky}, Ivan S. and {Agudo}, Iv{\'a}n and {G{\'o}mez}, Jos{\'e} L. and {L{\"a}hteenm{\"a}ki}, Anne and {Tammi}, Joni and {Tornikoski}, Merja},
        title = "{Kinematics of Parsec-scale Jets of Gamma-Ray Blazars at 43 GHz during 10 yr of the VLBA-BU-BLAZAR Program}",
      journal = {\apjs},
         year = 2022,
        month = may,
       volume = {260},
       number = {1},
          eid = {12},
        pages = {12},
          doi = {10.3847/1538-4365/ac589c},
archivePrefix = {arXiv},
       eprint = {2202.12290},
 primaryClass = {astro-ph.HE},
       adsurl = {https://ui.adsabs.harvard.edu/abs/2022ApJS..260...12W}
}

@INPROCEEDINGS{2020gbar.conf...32S,
       author = {{Sotnikova}, Yu. V.},
        title = "{RATAN-600 Radio Telescope: Observing Programs and Outlook}",
    booktitle = {Ground-Based Astronomy in Russia. 21st Century},
         year = 2020,
       editor = {{Romanyuk}, I.~I. and {Yakunin}, I.~A. and {Valeev}, A.~F. and {Kudryavtsev}, D.~O.},
        month = dec,
        pages = {32-40},
          doi = {10.26119/978-5-6045062-0-2_2020_32},
       adsurl = {https://ui.adsabs.harvard.edu/abs/2020gbar.conf...32S}
}

@ARTICLE{2009A&A...494..527H,
       author = {{Hovatta}, T. and {Valtaoja}, E. and {Tornikoski}, M. and {L{\"a}hteenm{\"a}ki}, A.},
        title = "{Doppler factors, Lorentz factors and viewing angles for quasars, BL Lacertae objects and radio galaxies}",
      journal = {\aap},
         year = 2009,
        month = feb,
       volume = {494},
       number = {2},
        pages = {527-537},
          doi = {10.1051/0004-6361:200811150},
archivePrefix = {arXiv},
       eprint = {0811.4278},
 primaryClass = {astro-ph},
       adsurl = {https://ui.adsabs.harvard.edu/abs/2009A&A...494..527H}
}

@ARTICLE{2018AstBu..73..494T,
       author = {{Tsybulev}, P.~G. and {Nizhelskii}, N.~A. and {Dugin}, M.~V. and {Borisov}, A.~N. and {Kratov}, D.~V. and {Udovitskii}, R. Yu.},
        title = "{C-Band Radiometer for Continuum Observations at RATAN-600 Radio Telescope}",
      journal = {Astrophysical Bulletin},
         year = 2018,
        month = oct,
       volume = {73},
       number = {4},
        pages = {494-500},
          doi = {10.1134/S1990341318040132},
       adsurl = {https://ui.adsabs.harvard.edu/abs/2018AstBu..73..494T}
}

@STRING(pasp="PASP")

@ARTICLE{2008A&A...488..897H,
   author = {{Hovatta}, T. and {Lehto}, H.~J. and {Tornikoski}, M.},
    title = "{Wavelet analysis of a large sample of AGN at high radio frequencies}",
  journal = {\aap},
   eprint = {0807.1796},
     year = 2008,
    month = sep,
   volume = 488,
    pages = {897-903},
      doi ={10.1051/0004-6361:200810200},
   adsurl = {http://adsabs.harvard.edu/abs/2008A%26A...488..897H}
}

@ARTICLE{1994A&A...284..331O,
   author = {{Ott}, M. and {Witzel}, A. and {Quirrenbach}, A. and {Krichbaum}, T.~P. and
	{Standke}, K.~J. and {Schalinski}, C.~J. and {Hummel}, C.~A.},
    title = "{An updated list of radio flux density calibrators}",
  journal = {\aap},
     year = 1994,
    month = apr,
   volume = 284,
    pages = {331-339},
   adsurl = {http://adsabs.harvard.edu/abs/1994A%26A...284..331O}
}

@ARTICLE{1980A&AS...39..379T,
   author = {{Tabara}, H. and {Inoue}, M.},
    title = "{A catalogue of linear polarization of radio sources}",
  journal = {\aaps},
     year = 1980,
    month = mar,
   volume = 39,
    pages = {379-393},
   adsurl = {http://adsabs.harvard.edu/abs/980A%26AS...39..379T}
}

@ARTICLE{1993IAPM...35....7P,
   author = {{Parijskij}, Y.~N.},
    title = "{RATAN-600 - The world's biggest reflector at the 'cross roads'}",
  journal = {IEEE Antennas and Propagation Magazine},
     year = 1993,
    month = aug,
   volume = 35,
    pages = {7-12},
      doi = {10.1109/74.229840},
   adsurl = {http://adsabs.harvard.edu/abs/1993IAPM...35....7P}
}

@ARTICLE{2011AstBu..66..109T,
   author = {{Tsybulev}, P.~G.},
    title = "{New-generation data acquisition and control system for continuum radio-astronomic observations with RATAN-600 radio telescope: Development, observations, and measurements}",
  journal = {Astrophysical Bulletin},
     year = 2011,
    month = jan,
   volume = 66,
    pages = {109-122},
      doi = {10.1134/S199034131101010X},
   adsurl = {http://adsabs.harvard.edu/abs/2011AstBu..66..109T}
}

@ARTICLE{1979S&T....57..324K,
   author = {{Korolkov}, D.~V. and {Pariiskii}, I.~N.},
    title = "{The Soviet RATAN-600 radio telescope}",
  journal = {\skytel},
     year = 1979,
    month = apr,
   volume = 57,
    pages = {324-329},
   adsurl = {http://adsabs.harvard.edu/abs/1979S%26T....57..324K}
}

@ARTICLE{1977A&A....61...99B,
   author = {{Baars}, J.~W.~M. and {Genzel}, R. and {Pauliny-Toth}, I.~I.~K. and
    {Witzel}, A.},
    title = "{The absolute spectrum of CAS A - an accurate flux density scale and a set of secondary calibrators}",
  journal = {\aap},
     year = 1977,
    month = oct,
   volume = 61,
    pages = {99-106},
   adsurl = {http://adsabs.harvard.edu/abs/1977A%26A....61...99B}
}

@ARTICLE{2016AstBu..71..496U,
       author = {{Udovitskiy}, R. Yu. and {Sotnikova}, Yu. V. and {Mingaliev}, M.~G. and
         {Tsybulev}, P.~G. and {Zhekanis}, G.~V. and {Nizhelskij}, N.~A.},
        title = "{Automated system for reduction of observational data on RATAN-600 radio telescope}",
      journal = {Astrophysical Bulletin},
         year = 2016,
        month = oct,
       volume = {71},
       number = {4},
        pages = {496-505},
          doi = {10.1134/S1990341316040131},
       adsurl = {https://ui.adsabs.harvard.edu/abs/2016AstBu..71..496U}
}

@ARTICLE{2019AstBu..74..497S,
       author = {{Sotnikova}, Yu. V. and {Kovalev}, Yu. A. and {Erkenov}, A.~K.},
        title = "{The Synchronous Calibration Method for the RATAN-600 using Its Two Sectors}",
      journal = {Astrophysical Bulletin},
         year = 2019,
        month = dec,
       volume = {74},
       number = {4},
        pages = {497-505},
          doi = {10.1134/S1990341319040151},
       adsurl = {https://ui.adsabs.harvard.edu/abs/2019AstBu..74..497S}
}

@ARTICLE{2017ApJS..230....7P,
       author = {{Perley}, R.~A. and {Butler}, B.~J.},
        title = "{An Accurate Flux Density Scale from 50 MHz to 50 GHz}",
      journal = {\apjs},
         year = 2017,
        month = may,
       volume = {230},
       number = {1},
          eid = {7},
        pages = {7},
          doi = {10.3847/1538-4365/aa6df9},
archivePrefix = {arXiv},
       eprint = {1609.05940},
 primaryClass = {astro-ph.IM},
       adsurl = {https://ui.adsabs.harvard.edu/abs/2017ApJS..230....7P}
}

@ARTICLE{2013ApJS..204...19P,
       author = {{Perley}, R.~A. and {Butler}, B.~J.},
        title = "{An Accurate Flux Density Scale from 1 to 50 GHz}",
      journal = {\apjs},
         year = 2013,
        month = feb,
       volume = {204},
       number = {2},
          eid = {19},
        pages = {19},
          doi = {10.1088/0067-0049/204/2/19},
archivePrefix = {arXiv},
       eprint = {1211.1300},
 primaryClass = {astro-ph.IM},
       adsurl = {https://ui.adsabs.harvard.edu/abs/2013ApJS..204...19P}
}

@ARTICLE{1985ApJ...298..114M,
       author = {{Marscher}, A.~P. and {Gear}, W.~K.},
        title = "{Models for high-frequency radio outbursts in extragalactic sources, with application to the early 1983 millimeter-to-infrared flare of 3C 273.}",
      journal = {\apj},
         year = 1985,
        month = nov,
       volume = {298},
        pages = {114-127},
          doi = {10.1086/163592},
       adsurl = {https://ui.adsabs.harvard.edu/abs/1985ApJ...298..114M}
}

@ARTICLE{1999A&AS..139..545K,
       author = {{Kovalev}, Y.~Y. and {Nizhelsky}, N.~A. and {Kovalev}, Yu. A. and {Berlin}, A.~B. and {Zhekanis}, G.~V. and {Mingaliev}, M.~G. and {Bogdantsov}, A.~V.},
        title = "{Survey of instantaneous 1-22 GHz spectra of 550 compact extragalactic objects with declinations from -30$^{deg}$ to +43$^{deg}$}",
      journal = {\aaps},
         year = 1999,
        month = nov,
       volume = {139},
        pages = {545-554},
          doi = {10.1051/aas:1999406},
archivePrefix = {arXiv},
       eprint = {astro-ph/0408264},
 primaryClass = {astro-ph},
       adsurl = {https://ui.adsabs.harvard.edu/abs/1999A&AS..139..545K}
}

@ARTICLE{1988ApJ...333..646E,
   author = {{Edelson}, R.~A. and {Krolik}, J.~H.},
    title = "{The Discrete Correlation Function: A New Method for Analyzing Unevenly Sampled Variability Data}",
  journal = {\apj},
     year = 1988,
    month = october,
   volume = 333,
    pages = {646},
      doi = {10.1086/166773},
   adsurl = {https://ui.adsabs.harvard.edu/abs/1988ApJ...333..646E/abstract}
}

@ARTICLE{2015MNRAS.453.3455R,
       author = {{Robertson}, D.~R.~S. and {Gallo}, L.~C. and {Zoghbi}, A. and
         {Fabian}, A.~C.},
        title = "{Searching for correlations in simultaneous X-ray and UV emission in the narrow-line Seyfert 1 galaxy 1H 0707-495}",
      journal = {\mnras},
         year = 2015,
        month = nov,
       volume = {453},
       number = {4},
        pages = {3455-3460},
          doi = {10.1093/mnras/stv1575},
archivePrefix = {arXiv},
       eprint = {1507.05201},
 primaryClass = {astro-ph.HE},
       adsurl = {https://ui.adsabs.harvard.edu/abs/2015MNRAS.453.3455R}
}

@ARTICLE{2013MNRAS.433..907E,
       author = {{Emmanoulopoulos}, D. and {McHardy}, I.~M. and {Papadakis}, I.~E.},
        title = "{Generating artificial light curves: revisited and updated}",
      journal = {\mnras},
         year = 2013,
        month = aug,
       volume = {433},
       number = {2},
        pages = {907-927},
          doi = {10.1093/mnras/stt764},
archivePrefix = {arXiv},
       eprint = {1305.0304},
 primaryClass = {astro-ph.IM},
       adsurl = {https://ui.adsabs.harvard.edu/abs/2013MNRAS.433..907E}
}

@ARTICLE{1990A&AS...83..183M,
       author = {{Mead}, A.~R.~G. and {Ballard}, K.~R. and {Brand}, P.~W.~J.~L. and {Hough}, J.~H. and {Brindle}, C. and {Bailey}, J.~A.},
        title = "{Optical and infrared polarimetry and photometry of blazars.}",
      journal = {\aaps},
         year = 1990,
        month = apr,
       volume = {83},
        pages = {183-204},
       adsurl = {https://ui.adsabs.harvard.edu/abs/1990A&AS...83..183M}
}

@ARTICLE{1999ApJS..120...95V,
       author = {{Valtaoja}, E. and {L{\"a}hteenm{\"a}ki}, A. and {Ter{\"a}sranta}, H. and {Lainela}, M.},
        title = "{Total Flux Density Variations in Extragalactic Radio Sources. I. Decomposition of Variations into Exponential Flares}",
      journal = {\apjs},
         year = 1999,
        month = jan,
       volume = {120},
       number = {1},
        pages = {95-99},
          doi = {10.1086/313170},
       adsurl = {https://ui.adsabs.harvard.edu/abs/1999ApJS..120...95V}
}

@ARTICLE{2018MNRAS.480.5517L,
       author = {{Liodakis}, I. and {Romani}, R.~W. and {Filippenko}, A.~V. and {Kiehlmann}, S. and {Max-Moerbeck}, W. and {Readhead}, A.~C.~S. and {Zheng}, W.},
        title = "{Multiwavelength cross-correlations and flaring activity in bright blazars}",
      journal = {\mnras},
         year = 2018,
        month = nov,
       volume = {480},
       number = {4},
        pages = {5517-5528},
          doi = {10.1093/mnras/sty2264},
archivePrefix = {arXiv},
       eprint = {1808.05625},
 primaryClass = {astro-ph.HE},
       adsurl = {https://ui.adsabs.harvard.edu/abs/2018MNRAS.480.5517L}
}

@ARTICLE{2013ApJ...764..167S,
       author = {{Scargle}, Jeffrey D. and {Norris}, Jay P. and {Jackson}, Brad and {Chiang}, James},
        title = "{Studies in Astronomical Time Series Analysis. VI. Bayesian Block Representations}",
      journal = {\apj},
         year = 2013,
        month = feb,
       volume = {764},
       number = {2},
          eid = {167},
        pages = {167},
          doi = {10.1088/0004-637X/764/2/167},
archivePrefix = {arXiv},
       eprint = {1207.5578},
 primaryClass = {astro-ph.IM},
       adsurl = {https://ui.adsabs.harvard.edu/abs/2013ApJ...764..167S}
}

@ARTICLE{2015arXiv150306676C,
       author = {{Connolly}, S D},
        title = "{A Python Code for the Emmanoulopoulos et al. [arXiv:1305.0304] Light Curve Simulation Algorithm}",
      journal = {arXiv e-prints},
         year = 2015,
        month = mar,
          eid = {arXiv:1503.06676},
        pages = {arXiv:1503.06676},
          doi = {10.48550/arXiv.1503.06676},
archivePrefix = {arXiv},
       eprint = {1503.06676},
 primaryClass = {astro-ph.IM},
       adsurl = {https://ui.adsabs.harvard.edu/abs/2015arXiv150306676C}
}

@ARTICLE{2019PASP..131a8002B,
       author = {{Bellm}, Eric C. and {Kulkarni}, Shrinivas R. and {Graham}, Matthew J. and {Dekany}, Richard and {Smith}, Roger M. and {Riddle}, Reed and {Masci}, Frank J. and {Helou}, George and {Prince}, Thomas A. and {Adams}, Scott M. and {Barbarino}, C. and {Barlow}, Tom and {Bauer}, James and {Beck}, Ron and {Belicki}, Justin and {Biswas}, Rahul and {Blagorodnova}, Nadejda and {Bodewits}, Dennis and {Bolin}, Bryce and {Brinnel}, Valery and {Brooke}, Tim and {Bue}, Brian and {Bulla}, Mattia and {Burruss}, Rick and {Cenko}, S. Bradley and {Chang}, Chan-Kao and {Connolly}, Andrew and {Coughlin}, Michael and {Cromer}, John and {Cunningham}, Virginia and {De}, Kishalay and {Delacroix}, Alex and {Desai}, Vandana and {Duev}, Dmitry A. and {Eadie}, Gwendolyn and {Farnham}, Tony L. and {Feeney}, Michael and {Feindt}, Ulrich and {Flynn}, David and {Franckowiak}, Anna and {Frederick}, S. and {Fremling}, C. and {Gal-Yam}, Avishay and {Gezari}, Suvi and {Giomi}, Matteo and {Goldstein}, Daniel A. and {Golkhou}, V. Zach and {Goobar}, Ariel and {Groom}, Steven and {Hacopians}, Eugean and {Hale}, David and {Henning}, John and {Ho}, Anna Y.~Q. and {Hover}, David and {Howell}, Justin and {Hung}, Tiara and {Huppenkothen}, Daniela and {Imel}, David and {Ip}, Wing-Huen and {Ivezi{\'c}}, {\v{Z}}eljko and {Jackson}, Edward and {Jones}, Lynne and {Juric}, Mario and {Kasliwal}, Mansi M. and {Kaspi}, S. and {Kaye}, Stephen and {Kelley}, Michael S.~P. and {Kowalski}, Marek and {Kramer}, Emily and {Kupfer}, Thomas and {Landry}, Walter and {Laher}, Russ R. and {Lee}, Chien-De and {Lin}, Hsing Wen and {Lin}, Zhong-Yi and {Lunnan}, Ragnhild and {Giomi}, Matteo and {Mahabal}, Ashish and {Mao}, Peter and {Miller}, Adam A. and {Monkewitz}, Serge and {Murphy}, Patrick and {Ngeow}, Chow-Choong and {Nordin}, Jakob and {Nugent}, Peter and {Ofek}, Eran and {Patterson}, Maria T. and {Penprase}, Bryan and {Porter}, Michael and {Rauch}, Ludwig and {Rebbapragada}, Umaa and {Reiley}, Dan and {Rigault}, Mickael and {Rodriguez}, Hector and {van Roestel}, Jan and {Rusholme}, Ben and {van Santen}, Jakob and {Schulze}, S. and {Shupe}, David L. and {Singer}, Leo P. and {Soumagnac}, Maayane T. and {Stein}, Robert and {Surace}, Jason and {Sollerman}, Jesper and {Szkody}, Paula and {Taddia}, F. and {Terek}, Scott and {Van Sistine}, Angela and {van Velzen}, Sjoert and {Vestrand}, W. Thomas and {Walters}, Richard and {Ward}, Charlotte and {Ye}, Quan-Zhi and {Yu}, Po-Chieh and {Yan}, Lin and {Zolkower}, Jeffry},
        title = "{The Zwicky Transient Facility: System Overview, Performance, and First Results}",
      journal = {\pasp},
         year = 2019,
        month = jan,
       volume = {131},
       number = {995},
        pages = {018002},
          doi = {10.1088/1538-3873/aaecbe},
archivePrefix = {arXiv},
       eprint = {1902.01932},
 primaryClass = {astro-ph.IM},
       adsurl = {https://ui.adsabs.harvard.edu/abs/2019PASP..131a8002B}
}

@INPROCEEDINGS{2007ASPC..375..234G,
       author = {{Gurwell}, M.~A. and {Peck}, A.~B. and {Hostler}, S.~R. and {Darrah}, M.~R. and {Katz}, C.~A.},
        title = "{Monitoring Phase Calibrators at Submillimeter Wavelengths}",
    booktitle = {From Z-Machines to ALMA: (Sub)Millimeter Spectroscopy of Galaxies},
         year = 2007,
       editor = {{Baker}, A.~J. and {Glenn}, J. and {Harris}, A.~I. and {Mangum}, J.~G. and {Yun}, M.~S.},
       series = {Astronomical Society of the Pacific Conference Series},
       volume = {375},
        month = oct,
        pages = {234},
       adsurl = {https://ui.adsabs.harvard.edu/abs/2007ASPC..375..234G}
}

@ARTICLE{1985BSAO...19...59A,
       author = {{Aliakberov}, K.~D. and {Mingaliev}, M.~G. and {Naugol'n aya}, M.~N. and {Trushkin}, S.~A. and {Sharipova}, L.~M. and {Yusupova}, S.~N.},
        title = "{Determination of the flux densities of radio sources on the set of broadband continuous-spectrum radiometers for the RATAN-600 radio telescope.}",
      journal = {Bulletin of the Special Astrophysics Observatory},
         year = 1985,
        month = jan,
       volume = {19},
        pages = {59-65},
       adsurl = {https://ui.adsabs.harvard.edu/abs/1985BSAO...19...59A}
}

@ARTICLE{2006A&A...453..477A,
       author = {{Agudo}, I. and {G{\'o}mez}, J.~L. and {Gabuzda}, D.~C. and {Marscher}, A.~P. and {Jorstad}, S.~G. and {Alberdi}, A.},
        title = "{The milliarcsecond-scale jet of PKS 0735+178 during quiescence}",
      journal = {\aap},
         year = 2006,
        month = jul,
       volume = {453},
       number = {2},
        pages = {477-486},
          doi = {10.1051/0004-6361:20054517},
archivePrefix = {arXiv},
       eprint = {astro-ph/0604543},
 primaryClass = {astro-ph},
       adsurl = {https://ui.adsabs.harvard.edu/abs/2006A&A...453..477A}
}

@ARTICLE{2022ApJ...926L..35O,
       author = {{O'Neill}, S. and {Kiehlmann}, S. and {Readhead}, A.~C.~S. and {Aller}, M.~F. and {Blandford}, R.~D. and {Liodakis}, I. and {Lister}, M.~L. and {Mr{\'o}z}, P. and {O'Dea}, C.~P. and {Pearson}, T.~J. and {Ravi}, V. and {Vallisneri}, M. and {Cleary}, K.~A. and {Graham}, M.~J. and {Grainge}, K.~J.~B. and {Hodges}, M.~W. and {Hovatta}, T. and {L{\"a}hteenm{\"a}ki}, A. and {Lamb}, J.~W. and {Lazio}, T.~J.~W. and {Max-Moerbeck}, W. and {Pavlidou}, V. and {Prince}, T.~A. and {Reeves}, R.~A. and {Tornikoski}, M. and {Vergara de la Parra}, P. and {Zensus}, J.~A.},
        title = "{The Unanticipated Phenomenology of the Blazar PKS 2131-021: A Unique Supermassive Black Hole Binary Candidate}",
      journal = {\apjl},
         year = 2022,
        month = feb,
       volume = {926},
       number = {2},
          eid = {L35},
        pages = {L35},
          doi = {10.3847/2041-8213/ac504b},
archivePrefix = {arXiv},
       eprint = {2111.02436},
 primaryClass = {astro-ph.HE},
       adsurl = {https://ui.adsabs.harvard.edu/abs/2022ApJ...926L..35O}
}

@ARTICLE{2009A&A...496..577Z,
       author = {{Zechmeister}, M. and {K{\"u}rster}, M.},
        title = "{The generalised Lomb-Scargle periodogram. A new formalism for the floating-mean and Keplerian periodograms}",
      journal = {\aap},
         year = 2009,
        month = mar,
       volume = {496},
       number = {2},
        pages = {577-584},
          doi = {10.1051/0004-6361:200811296},
archivePrefix = {arXiv},
       eprint = {0901.2573},
 primaryClass = {astro-ph.IM},
       adsurl = {https://ui.adsabs.harvard.edu/abs/2009A&A...496..577Z}
}

@ARTICLE{2012A&A...547A...1N,
       author = {{Nilsson}, K. and {Pursimo}, T. and {Villforth}, C. and {Lindfors}, E. and {Takalo}, L.~O. and {Sillanp{\"a}{\"a}}, A.},
        title = "{Redshift constraints for RGB 0136+391 and PKS 0735+178 from deep optical imaging}",
      journal = {\aap},
         year = 2012,
        month = nov,
       volume = {547},
          eid = {A1},
        pages = {A1},
          doi = {10.1051/0004-6361/201219848},
archivePrefix = {arXiv},
       eprint = {1209.4755},
 primaryClass = {astro-ph.CO},
       adsurl = {https://ui.adsabs.harvard.edu/abs/2012A&A...547A...1N}
}

@ARTICLE{2009MNRAS.400...26O,
       author = {{O'Sullivan}, S.~P. and {Gabuzda}, D.~C.},
        title = "{Magnetic field strength and spectral distribution of six parsec-scale active galactic nuclei jets}",
      journal = {\mnras},
         year = 2009,
        month = nov,
       volume = {400},
       number = {1},
        pages = {26-42},
          doi = {10.1111/j.1365-2966.2009.15428.x},
archivePrefix = {arXiv},
       eprint = {0907.5211},
 primaryClass = {astro-ph.CO},
       adsurl = {https://ui.adsabs.harvard.edu/abs/2009MNRAS.400...26O}
}

@ARTICLE{2021ATel15132....1F,
       author = {{Falomo}, Renato and {Treves}, Aldo and {Paiano}, Simona},
        title = "{Optical view of neutrino emitter candidate PKS 0735 +178}",
      journal = {The Astronomer's Telegram},
         year = 2021,
        month = dec,
       volume = {15132},
        pages = {1},
       adsurl = {https://ui.adsabs.harvard.edu/abs/2021ATel15132....1F}
}

@ARTICLE{2007A&A...467..465C,
       author = {{Ciprini}, S. and {Takalo}, L.~O. and {Tosti}, G. and {Raiteri}, C.~M. and {Fiorucci}, M. and {Villata}, M. and {Nucciarelli}, G. and {Lanteri}, L. and {Nilsson}, K. and {Ros}, J.~A.},
        title = "{Ten-year optical monitoring of PKS 0735+178: historical comparison, multiband behavior, and variability timescales}",
      journal = {\aap},
         year = 2007,
        month = may,
       volume = {467},
       number = {2},
        pages = {465-483},
          doi = {10.1051/0004-6361:20052646},
archivePrefix = {arXiv},
       eprint = {astro-ph/0701420},
 primaryClass = {astro-ph},
       adsurl = {https://ui.adsabs.harvard.edu/abs/2007A&A...467..465C}
}

@ARTICLE{2023ApJ...954...70A,
       author = {{Acharyya}, A. and {Adams}, C.~B. and {Archer}, A. and {Bangale}, P. and {Bartkoske}, J.~T. and {Batista}, P. and {Benbow}, W. and {Brill}, A. and {Buckley}, J.~H. and {Christiansen}, J.~L. and {Chromey}, A.~J. and {Errando}, M. and {Falcone}, A. and {Feng}, Q. and {Foote}, Juniper and {Fortson}, L. and {Furniss}, A. and {Gallagher}, G. and {Hanlon}, W. and {Hanna}, D. and {Hervet}, O. and {Hinrichs}, C.~E. and {Hoang}, J. and {Holder}, J. and {Humensky}, T.~B. and {Jin}, W. and {Kaaret}, P. and {Kertzman}, M. and {Kherlakian}, M. and {Kieda}, D. and {Kleiner}, T.~K. and {Korzoun}, N. and {Kumar}, S. and {Lang}, M.~J. and {Lundy}, M. and {Maier}, G. and {McGrath}, C.~E. and {Millard}, M.~J. and {Millis}, J. and {Mooney}, C.~L. and {Moriarty}, P. and {Mukherjee}, R. and {O'Brien}, S. and {Ong}, R.~A. and {Pohl}, M. and {Pueschel}, E. and {Quinn}, J. and {Ragan}, K. and {Reynolds}, P.~T. and {Ribeiro}, D. and {Roache}, E. and {Sadeh}, I. and {Sadun}, A.~C. and {Saha}, L. and {Santander}, M. and {Sembroski}, G.~H. and {Shang}, R. and {Splettstoesser}, M. and {Talluri}, A. Kaushik and {Tucci}, J.~V. and {Vassiliev}, V.~V. and {Weinstein}, A. and {Williams}, D.~A. and {Wong}, S.~L. and {Woo}, J. and {Aharonian}, F. and {Aschersleben}, J. and {Backes}, M. and {Martins}, V. Barbosa and {Batzofin}, R. and {Becherini}, Y. and {Berge}, D. and {Bernl{\"o}hr}, K. and {Bi}, B. and {B{\"o}ttcher}, M. and {Boisson}, C. and {Bolmont}, J. and {de Bony de Lavergne}, M. and {Borowska}, J. and {Bouyahiaoui}, M. and {Bradascio}, F. and {Breuhaus}, M. and {Brose}, R. and {Brun}, F. and {Bruno}, B. and {Bulik}, T. and {Burger-Scheidlin}, C. and {Caroff}, S. and {Casanova}, S. and {Cecil}, R. and {Celic}, J. and {Cerruti}, M. and {Chand}, T. and {Chandra}, S. and {Chen}, A. and {Chibueze}, J. and {Chibueze}, O. and {Cotter}, G. and {Dai}, S. and {Mbarubucyeye}, J. Damascene and {Djannati-Ata{\"\i}}, A. and {Dmytriiev}, A. and {Doroshenko}, V. and {Einecke}, S. and {Ernenwein}, J.-P. and {de Clairfontaine}, G. Fichet and {Filipovic}, M. and {Fontaine}, G. and {F{\"u}{\ss}ling}, M. and {Funk}, S. and {Gabici}, S. and {Ghafourizadeh}, S. and {Giavitto}, G. and {Glawion}, D. and {Glicenstein}, J.~F. and {Goswami}, P. and {Grolleron}, G. and {Haerer}, L. and {Hinton}, J.~A. and {Holch}, T.~L. and {Holler}, M. and {Horns}, D. and {Jamrozy}, M. and {Jankowsky}, F. and {Joshi}, V. and {Jung-Richardt}, I. and {Kasai}, E. and {Katarzy{\'n}ski}, K. and {Khatoon}, R. and {Kh{\'e}lifi}, B. and {Klepser}, S. and {Klu{\'z}niak}, W. and {Kosack}, K. and {Kostunin}, D. and {Lang}, R.~G. and {Le Stum}, S. and {Lemi{\`e}re}, A. and {Lenain}, J.-P. and {Leuschner}, F. and {Lohse}, T. and {Luashvili}, A. and {Lypova}, I. and {Mackey}, J. and {Malyshev}, D. and {Marandon}, V. and {Marchegiani}, P. and {Marcowith}, A. and {Mart{\'\i}-Devesa}, G. and {Marx}, R. and {Mitchell}, A. and {Moderski}, R. and {Mohrmann}, L. and {Montanari}, A. and {Moulin}, E. and {Murach}, T. and {Nakashima}, K. and {Niemiec}, J. and {Noel}, A. Priyana and {O'Brien}, P. and {Olivera-Nieto}, L. and {de Ona Wilhelmi}, E. and {Ostrowski}, M. and {Panny}, S. and {Panter}, M. and {Peron}, G. and {Prokhorov}, D.~A. and {P{\"u}hlhofer}, G. and {Punch}, M. and {Quirrenbach}, A. and {Reichherzer}, P. and {Reimer}, A. and {Reimer}, O. and {Ren}, H. and {Renaud}, M. and {Rieger}, F. and {Rudak}, B. and {Ruiz-Velasco}, E. and {Sahakian}, V. and {Santangelo}, A. and {Sasaki}, M. and {Sch{\"a}fer}, J. and {Sch{\"u}ssler}, F. and {Schutte}, H.~M. and {Schwanke}, U. and {Shapopi}, J.~N.~S. and {Specovius}, A. and {Spencer}, S. and {Stawarz}, {\L}. and {Steenkamp}, R. and {Steinmassl}, S. and {Sushch}, I. and {Suzuki}, H. and {Takahashi}, T. and {Tanaka}, T. and {Terrier}, R. and {van Eldik}, C. and {Vecchi}, M. and {Veh}, J. and {Venter}, C. and {Vink}, J.},
        title = "{Multiwavelength Observations of the Blazar PKS 0735+178 in Spatial and Temporal Coincidence with an Astrophysical Neutrino Candidate IceCube-211208A}",
      journal = {\apj},
         year = 2023,
        month = sep,
       volume = {954},
       number = {1},
          eid = {70},
        pages = {70},
          doi = {10.3847/1538-4357/ace327},
archivePrefix = {arXiv},
       eprint = {2306.17819},
 primaryClass = {astro-ph.HE},
       adsurl = {https://ui.adsabs.harvard.edu/abs/2023ApJ...954...70A}
}

@ARTICLE{2023MNRAS.519.1396S,
       author = {{Sahakyan}, N. and {Giommi}, P. and {Padovani}, P. and {Petropoulou}, M. and {B{\'e}gu{\'e}}, D. and {Boccardi}, B. and {Gasparyan}, S.},
        title = "{A multimessenger study of the blazar PKS 0735+178: a new major neutrino source candidate}",
      journal = {\mnras},
         year = 2023,
        month = feb,
       volume = {519},
       number = {1},
        pages = {1396-1408},
          doi = {10.1093/mnras/stac3607},
archivePrefix = {arXiv},
       eprint = {2204.05060},
 primaryClass = {astro-ph.HE},
       adsurl = {https://ui.adsabs.harvard.edu/abs/2023MNRAS.519.1396S}
}

@ARTICLE{1988AJ.....95..374W,
       author = {{Webb}, James R. and {Smith}, Alex G. and {Leacock}, Robert J. and {Fitzgibbons}, Gregory L. and {Gombola}, Paul P. and {Shepherd}, David W.},
        title = "{Optical Observations of 22 Violently Variable Extragalactic Sources: 1968-1986}",
      journal = {\aj},
         year = 1988,
        month = feb,
       volume = {95},
        pages = {374},
          doi = {10.1086/114641},
       adsurl = {https://ui.adsabs.harvard.edu/abs/1988AJ.....95..374W}
}

@ARTICLE{2025A&A...695A.266O,
       author = {{Omeliukh}, A. and {Garrappa}, S. and {Fallah Ramazani}, V. and {Franckowiak}, A. and {Winter}, W. and {Lindfors}, E. and {Nilsson}, K. and {Jormanainen}, J. and {Wierda}, F. and {Filippenko}, A.~V. and {Zheng}, W. and {Tornikoski}, M. and {L{\"a}hteenm{\"a}ki}, A. and {Kankkunen}, S. and {Tammi}, J.},
        title = "{Multi-epoch leptohadronic modeling of neutrino source candidate blazar PKS 0735+178}",
      journal = {\aap},
         year = 2025,
        month = mar,
       volume = {695},
          eid = {A266},
        pages = {A266},
          doi = {10.1051/0004-6361/202452143},
archivePrefix = {arXiv},
       eprint = {2409.04165},
 primaryClass = {astro-ph.HE},
       adsurl = {https://ui.adsabs.harvard.edu/abs/2025A&A...695A.266O}
}

@ARTICLE{2021GCN.31191....1I,
       author = {{IceCube Collaboration}},
        title = "{IceCube-211208A - IceCube observation of a high-energy neutrino candidate track-like event}",
      journal = {GRB Coordinates Network},
         year = 2021,
        month = dec,
       volume = {31191},
        pages = {1},
       adsurl = {https://ui.adsabs.harvard.edu/abs/2021GCN.31191....1I}
}

@ARTICLE{2021ATel15112....1D,
       author = {{Dzhilkibaev}, Zh.-A. and {Suvorova}, O. and {Baikal-GVD Collaboration}},
        title = "{Baikal-GVD observation of a high-energy neutrino candidate event from the blazar PKS 0735+17 at the day of the IceCube-211208A neutrino alert from the same direction}",
      journal = {The Astronomer's Telegram},
         year = 2021,
        month = dec,
       volume = {15112},
        pages = {1},
       adsurl = {https://ui.adsabs.harvard.edu/abs/2021ATel15112....1D}
}

@ARTICLE{2021ATel15143....1P,
       author = {{Petkov}, V.~B. and {Novoseltsev}, Yu. F. and {Novoseltseva}, R.~V. and {Baksan Underground Scintillation Telescope Group}},
        title = "{Baksan Underground Scintillation Telescope observation of a GeV neutrino candidate event at the time of a gamma-ray flare of the blazar PKS 0735+17, a possible source of coinciding IceCube and Baikal high-energy neutrinos}",
      journal = {The Astronomer's Telegram},
         year = 2021,
        month = dec,
       volume = {15143},
        pages = {1},
       adsurl = {https://ui.adsabs.harvard.edu/abs/2021ATel15143....1P}
}

@ARTICLE{2022ATel15290....1F,
       author = {{Filippini}, F. and {Illuminati}, G. and {Heijboer}, A. and {Gatius}, C. and {Muller}, R. and {Dornic}, D. and {Huang}, F. and {Le Stum}, S. and {Palacios Gonz{\'a}lez}, J. and {Celli}, S. and {Zegarelli}, A. and {Coniglione}, R. and {Samtleben}, D. and {Kovalev}, Y.~Y. and {Plavin}, A.},
        title = "{Search for neutrino counterpart to the blazar PKS0735+178 potentially associated with IceCube-211208A and Baikal-GVD-211208A with the KM3NeT neutrino detectors.}",
      journal = {The Astronomer's Telegram},
         year = 2022,
        month = mar,
       volume = {15290},
        pages = {1},
       adsurl = {https://ui.adsabs.harvard.edu/abs/2022ATel15290....1F}
}

@ARTICLE{2024MNRAS.529.3503B,
       author = {{Bharathan}, Athira M. and {Stalin}, C.~S. and {Sahayanathan}, S. and {Bhattacharyya}, Subir and {Mathew}, Blesson},
        title = "{Multiwavelength spectral modelling of the candidate neutrino blazar PKS 0735+178}",
      journal = {\mnras},
         year = 2024,
        month = apr,
       volume = {529},
       number = {4},
        pages = {3503-3510},
          doi = {10.1093/mnras/stae296},
archivePrefix = {arXiv},
       eprint = {2401.12680},
 primaryClass = {astro-ph.HE},
       adsurl = {https://ui.adsabs.harvard.edu/abs/2024MNRAS.529.3503B}
}

@ARTICLE{2024MNRAS.527.8746P,
       author = {{Prince}, Raj and {Das}, Saikat and {Gupta}, Nayantara and {Majumdar}, Pratik and {Czerny}, Bo{\.z}ena},
        title = "{Dissecting the broad-band emission from {\ensuremath{\gamma}}-ray blazar PKS 0735+178 in search of neutrinos}",
      journal = {\mnras},
         year = 2024,
        month = jan,
       volume = {527},
       number = {3},
        pages = {8746-8754},
          doi = {10.1093/mnras/stad3804},
archivePrefix = {arXiv},
       eprint = {2301.06565},
 primaryClass = {astro-ph.HE},
       adsurl = {https://ui.adsabs.harvard.edu/abs/2024MNRAS.527.8746P}
}

@ARTICLE{2022ApJ...933..224F,
       author = {{Fang}, Yue and {Chen}, Qihang and {Zhang}, Yan and {Wu}, Jianghua},
        title = "{Multiwavelength Variation Phenomena of PKS 0735+178 on Diverse Timescales}",
      journal = {\apj},
         year = 2022,
        month = jul,
       volume = {933},
       number = {2},
          eid = {224},
        pages = {224},
          doi = {10.3847/1538-4357/ac7647},
archivePrefix = {arXiv},
       eprint = {2206.03296},
 primaryClass = {astro-ph.GA},
       adsurl = {https://ui.adsabs.harvard.edu/abs/2022ApJ...933..224F}
}

@ARTICLE{2004IJMPD..13..771D,
       author = {{Ding}, S.~X. and {Xie}, G.~Z. and {Liang}, E.~W. and {Zhou}, S.~B. and {Ma}, L.},
        title = "{The Periodicity Analysis of the Light Curve of PKS 0735+178 and Implications for its Central Structure}",
      journal = {International Journal of Modern Physics D},
         year = 2004,
        month = jan,
       volume = {13},
       number = {4},
        pages = {771-782},
          doi = {10.1142/S0218271804004694},
       adsurl = {https://ui.adsabs.harvard.edu/abs/2004IJMPD..13..771D}
}

@INPROCEEDINGS{2001ASPC..246..121F,
       author = {{Filippenko}, Alexei V. and {Li}, W.~D. and {Treffers}, R.~R. and {Modjaz}, Maryam},
        title = "{The Lick Observatory Supernova Search with the Katzman Automatic Imaging Telescope}",
    booktitle = {IAU Colloquium 183: Small Telescope Astronomy on Global Scales},
         year = 2001,
       editor = {{Paczynski}, Bohdan and {Chen}, Wen-Ping and {Lemme}, Claudia},
       series = {Astronomical Society of the Pacific Conference Series},
       volume = {246},
        month = jan,
        pages = {121},
       adsurl = {https://ui.adsabs.harvard.edu/abs/2001ASPC..246..121F}
}

@ARTICLE{2009ApJ...696..870D,
       author = {{Drake}, A.~J. and {Djorgovski}, S.~G. and {Mahabal}, A. and {Beshore}, E. and {Larson}, S. and {Graham}, M.~J. and {Williams}, R. and {Christensen}, E. and {Catelan}, M. and {Boattini}, A. and {Gibbs}, A. and {Hill}, R. and {Kowalski}, R.},
        title = "{First Results from the Catalina Real-Time Transient Survey}",
      journal = {\apj},
         year = 2009,
        month = may,
       volume = {696},
       number = {1},
        pages = {870-884},
          doi = {10.1088/0004-637X/696/1/870},
archivePrefix = {arXiv},
       eprint = {0809.1394},
 primaryClass = {astro-ph},
       adsurl = {https://ui.adsabs.harvard.edu/abs/2009ApJ...696..870D}
}

@ARTICLE{2022Photo...9..950V,
       author = {{Valyavin}, Gennady and {Beskin}, Grigory and {Valeev}, Azamat and {Galazutdinov}, Gazinur and {Fabrika}, Sergei and {Romanyuk}, Iosif and {Aitov}, Vitaly and {Yakovlev}, Oleg and {Ivanova}, Anastasia and {Baluev}, Roman and {Vlasyuk}, Valery and {Han}, Inwoo and {Karpov}, Sergei and {Sasyuk}, Vyacheslav and {Perkov}, Alexei and {Bondar}, Sergei and {Musaev}, Faig and {Emelianov}, Eduard and {Fatkhullin}, Timur and {Drabek}, Sergei and {Shergin}, Vladimir and {Lee}, Byeong-Cheol and {Mitiani}, Guram and {Burlakova}, Tatiana and {Yushkin}, Maksim and {Sendzikas}, Eugene and {Gadelshin}, Damir and {Chmyreva}, Lisa and {Beskakotov}, Anatoly and {Dyachenko}, Vladimir and {Rastegaev}, Denis and {Mitrofanova}, Arina and {Yakunin}, Ilia and {Antonyuk}, Kirill and {Plokhotnichenko}, Vladimir and {Gutaev}, Alexei and {Lyapsina}, Nadezhda and {Chernenkov}, Vladimir and {Biryukov}, Anton and {Ivanov}, Evgenij and {Katkova}, Elena and {Belinski}, Alexander and {Sokov}, Eugene and {Tavrov}, Alexander and {Korablev}, Oleg and {Park}, Myeong-Gu and {Stolyarov}, Vladislav and {Bychkov}, Victor and {Gorda}, Stanislav and {Popov}, A.~A. and {Sobolev}, A.~M.},
        title = "{EXPLANATION: Exoplanet and Transient Event Investigation Project{\textemdash}Optical Facilities and Solutions}",
      journal = {Photonics},
         year = 2022,
        month = dec,
       volume = {9},
       number = {12},
          eid = {950},
        pages = {950},
          doi = {10.3390/photonics9120950},
       adsurl = {https://ui.adsabs.harvard.edu/abs/2022Photo...9..950V}
}

@ARTICLE{2024MNRAS.535.2775V,
       author = {{Vlasyuk}, V.~V. and {Sotnikova}, Y.~V. and {Volvach}, A.~E. and {Mufakharov}, T.~V. and {Kovalev}, Y.~A. and {Spiridonova}, O.~I. and {Khabibullina}, M.~L. and {Kovalev}, Y.~Y. and {Mikhailov}, A.~G. and {Stolyarov}, V.~A. and {Kudryavtsev}, D.~O. and {Mingaliev}, M.~G. and {Razzaque}, S. and {Semenova}, T.~A. and {Kudryashova}, A.~K. and {Bursov}, N.~N. and {Trushkin}, S.~A. and {Popkov}, A.~V. and {Erkenov}, A.~K. and {Rakhimov}, I.~A. and {Kharinov}, M.~A. and {Gurwell}, M.~A. and {Tsybulev}, P.~G. and {Moskvitin}, A.~S. and {Fatkhullin}, T.~A. and {Emelianov}, E.~V. and {Arshinova}, A. and {Iuzhanina}, K.~V. and {Andreeva}, T.~S. and {Volvach}, L.~N. and {Ghosh}, A.},
        title = "{Multiwavelength variability of the blazar AO 0235+164}",
      journal = {\mnras},
         year = 2024,
        month = dec,
       volume = {535},
       number = {3},
        pages = {2775-2799},
          doi = {10.1093/mnras/stae2491},
archivePrefix = {arXiv},
       eprint = {2411.01497},
 primaryClass = {astro-ph.HE},
       adsurl = {https://ui.adsabs.harvard.edu/abs/2024MNRAS.535.2775V}
}

@ARTICLE{1996AJ....112.1709F,
       author = {{Foster}, Grant},
        title = "{Wavelets for period analysis of unevenly sampled time series}",
      journal = {\aj},
         year = 1996,
        month = oct,
       volume = {112},
        pages = {1709-1729},
          doi = {10.1086/118137},
       adsurl = {https://ui.adsabs.harvard.edu/abs/1996AJ....112.1709F}
      }

@ARTICLE{2026MNRAS.tmp..392S,
       author = {{Sotnikova}, Yu V. and {Mufakharov}, T.~V. and {Volvach}, A.~E. and {Vlasyuk}, V.~V. and {Khabibullina}, M.~L. and {Mikhailov}, A.~G. and {An}, T. and {Kudryavtsev}, D.~O. and {Kovalev}, Yu A. and {Kovalev}, Y.~Y. and {Popkov}, A.~V. and {Savchenko}, S.~S. and {Erkenov}, A.~K. and {Morozova}, D.~A. and {Semenova}, T.~A. and {Spiridonova}, O.~I. and {Kharinov}, M.~A. and {Rakhimov}, I.~A. and {Andreeva}, T.~S. and {Cui}, L. and {Wang}, X. and {Chang}, N. and {Udovitskiy}, R. Yu and {Zhekanis}, P.~G. and {Borman}, G.~A. and {Grishina}, T.~S. and {Kopatskaya}, E.~N. and {Larionova}, E.~G. and {Troitskiy}, I.~S. and {Troitskaya}, Yu V. and {Vasilyev}, A.~A. and {Zhovtan}, A.~V. and {Kratov}, D.~V. and {Volvach}, L.~N. and {Shishkina}, E.~V. and {Dmytrotsa}, A.~I. and {Zharov}, V.~I.},
        title = "{Multiwavelength quasi-periodic variability of the blazar Ton 599}",
      journal = {\mnras},
         year = 2026,
        month = mar,
          doi = {10.1093/mnras/stag333},
archivePrefix = {arXiv},
       eprint = {2603.05894},
 primaryClass = {astro-ph.HE},
       adsurl = {https://ui.adsabs.harvard.edu/abs/2026MNRAS.tmp..392S}
}

\clearpage
\appendix

\section{Discrete correlation function measurements}
\label{app:dcf_figs}

Fig.~\ref{fig:dcf} presents the DCF measurements for the entire observed time interval. The figures correspond to the measured lags in Table~\ref{tab:ccf-lags}. For each frequency pair, the measured lag between the light curves is indicated by the orange vertical line in the lower panels, and the lag uncertainty is shown by the orange area. Both are derived by the method of flux redistribution and random subset selection (FR/RSS), introduced by \cite{1998PASP..110..660P}. The uncertainties of the lags are measured as the 16th/84th percentiles in the FR/RSS lag distribution.

\begin{figure*}
\centerline{
\includegraphics[width=0.7\columnwidth]{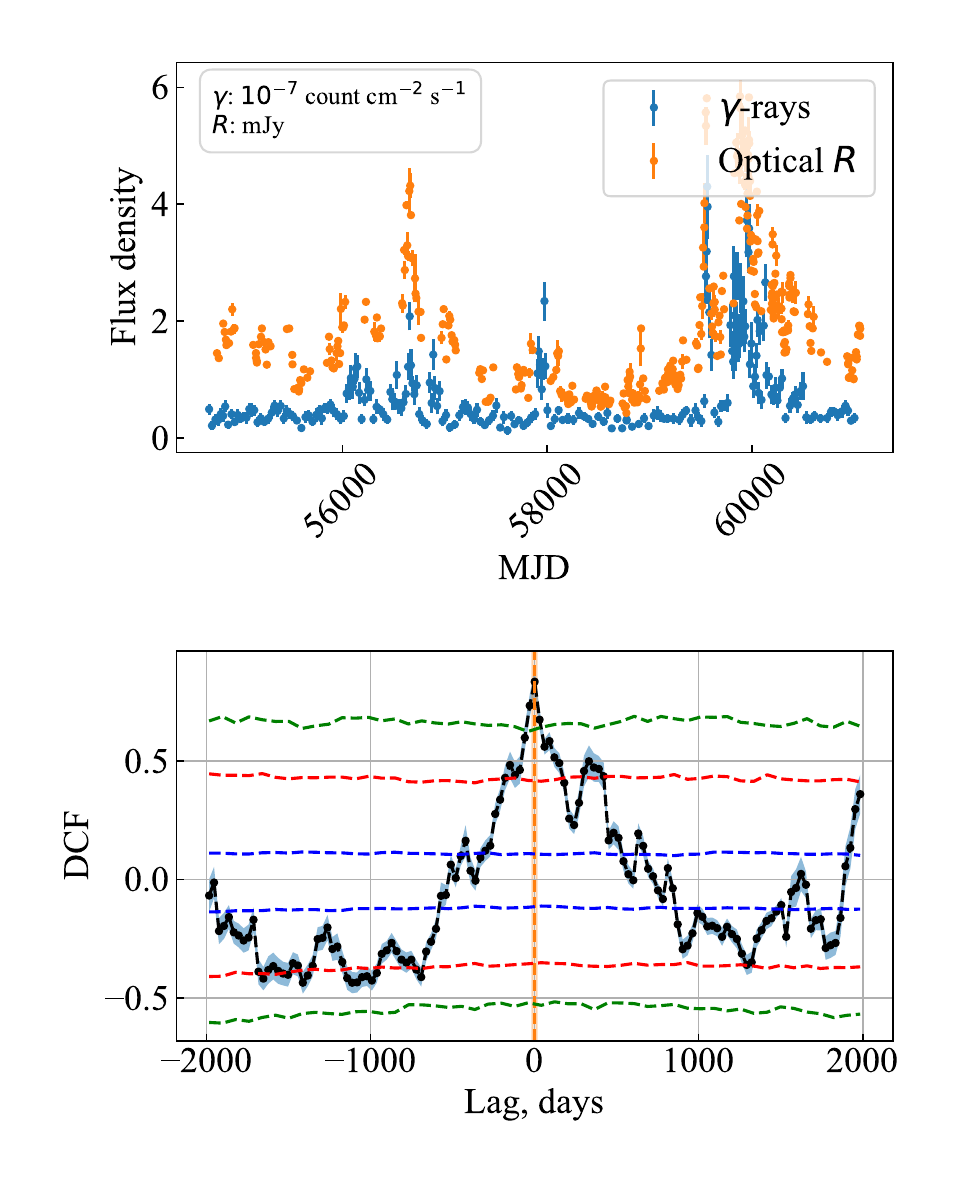}
\includegraphics[width=0.7\columnwidth]{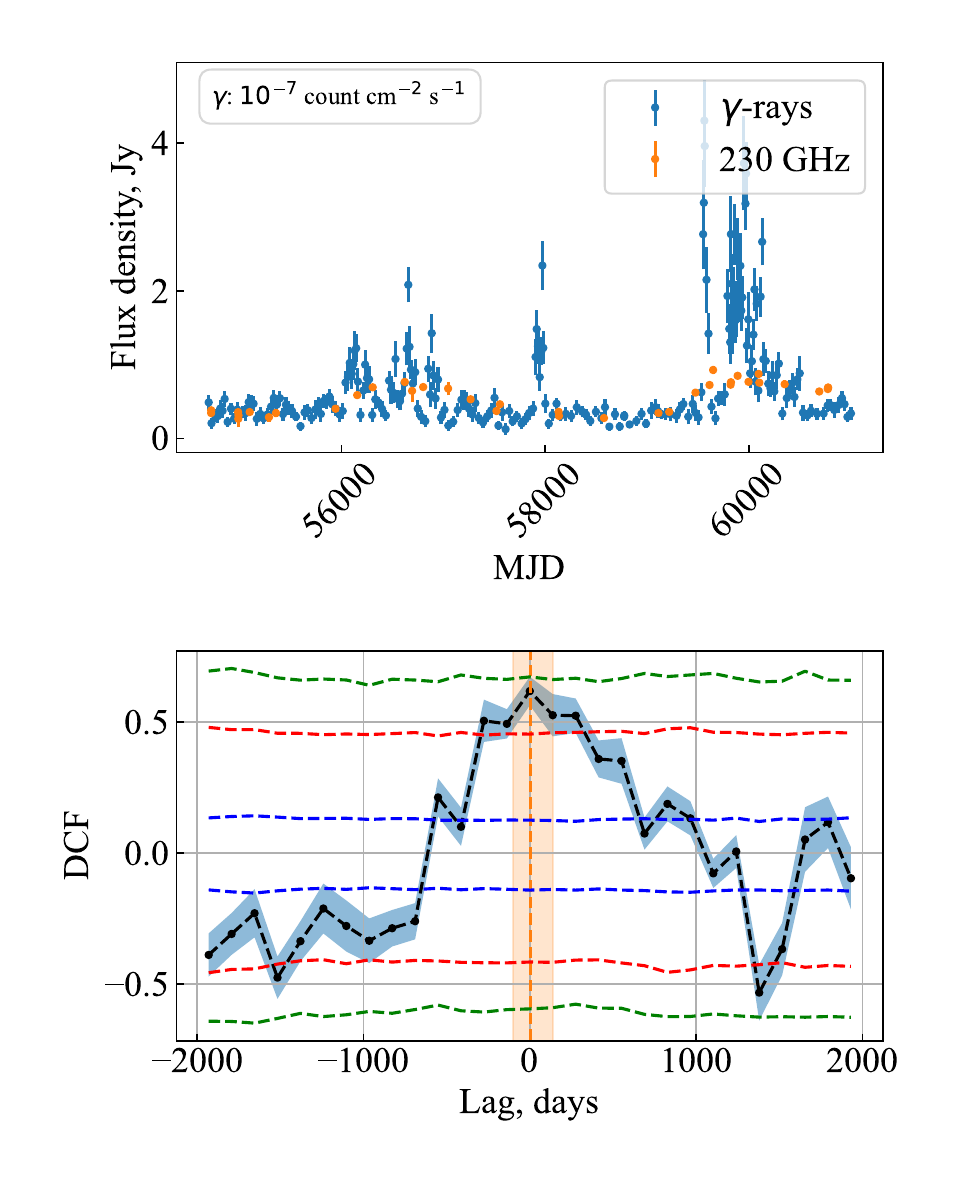}
\includegraphics[width=0.7\columnwidth]{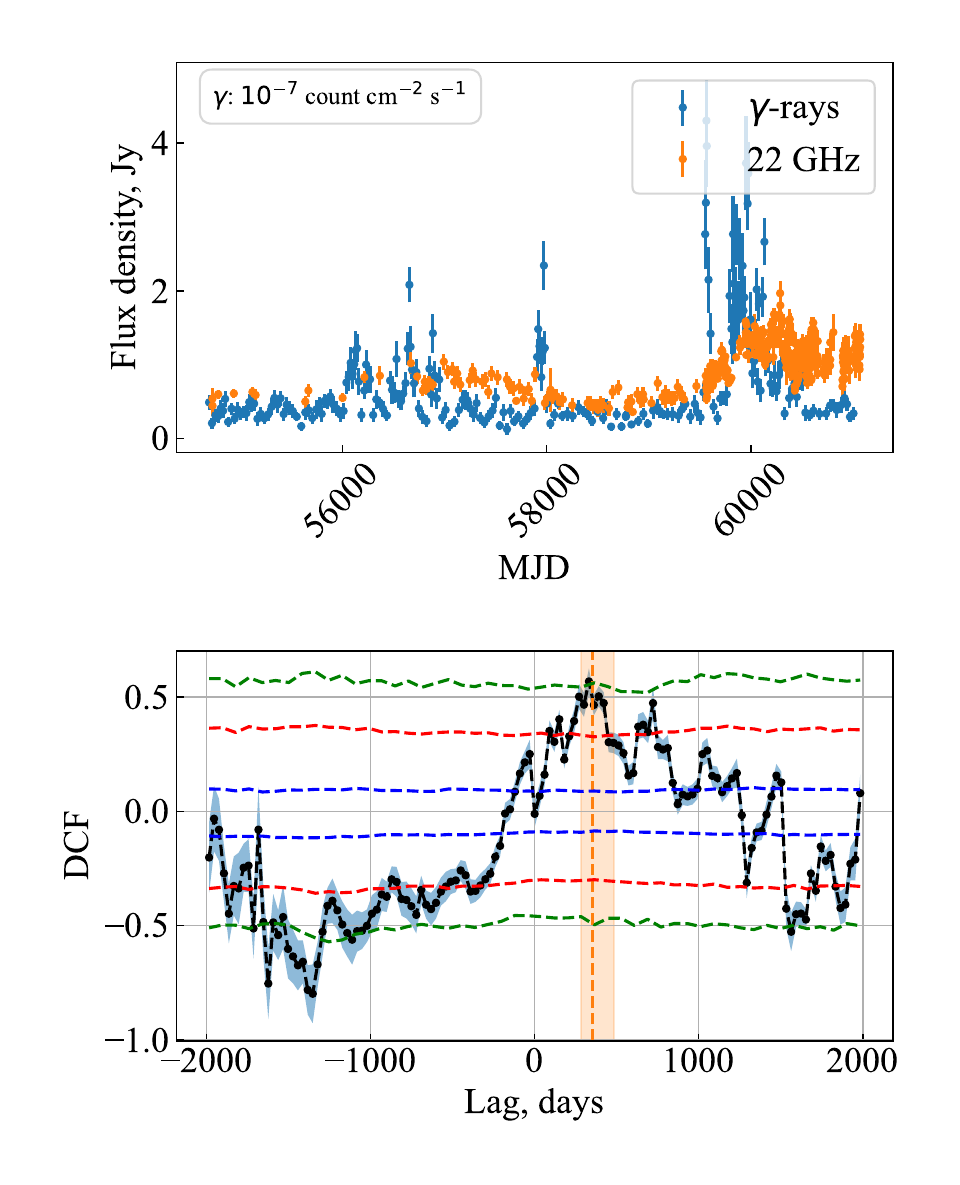}
}
\centerline{
\includegraphics[width=0.7\columnwidth]{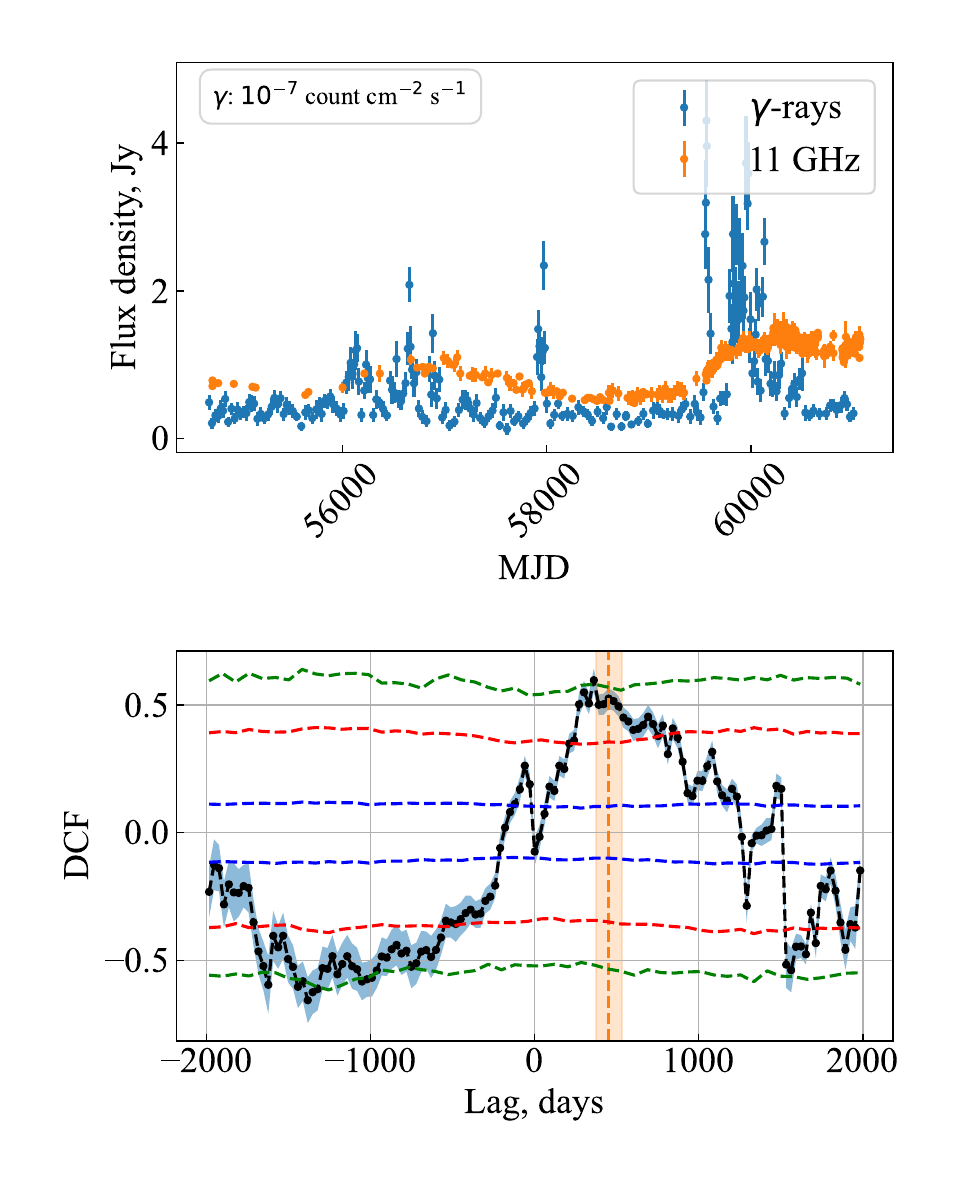}
\includegraphics[width=0.7\columnwidth]{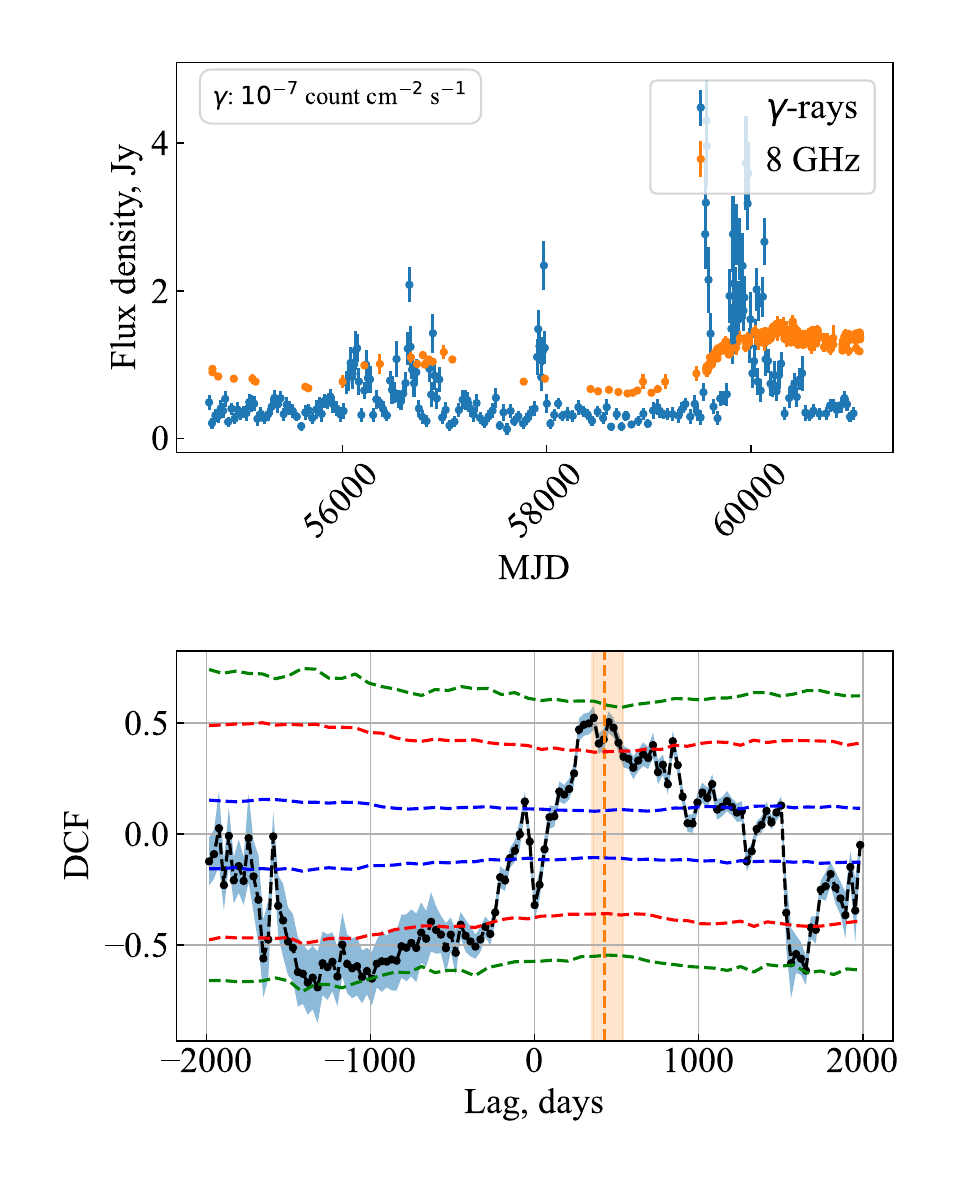}
\includegraphics[width=0.7\columnwidth]{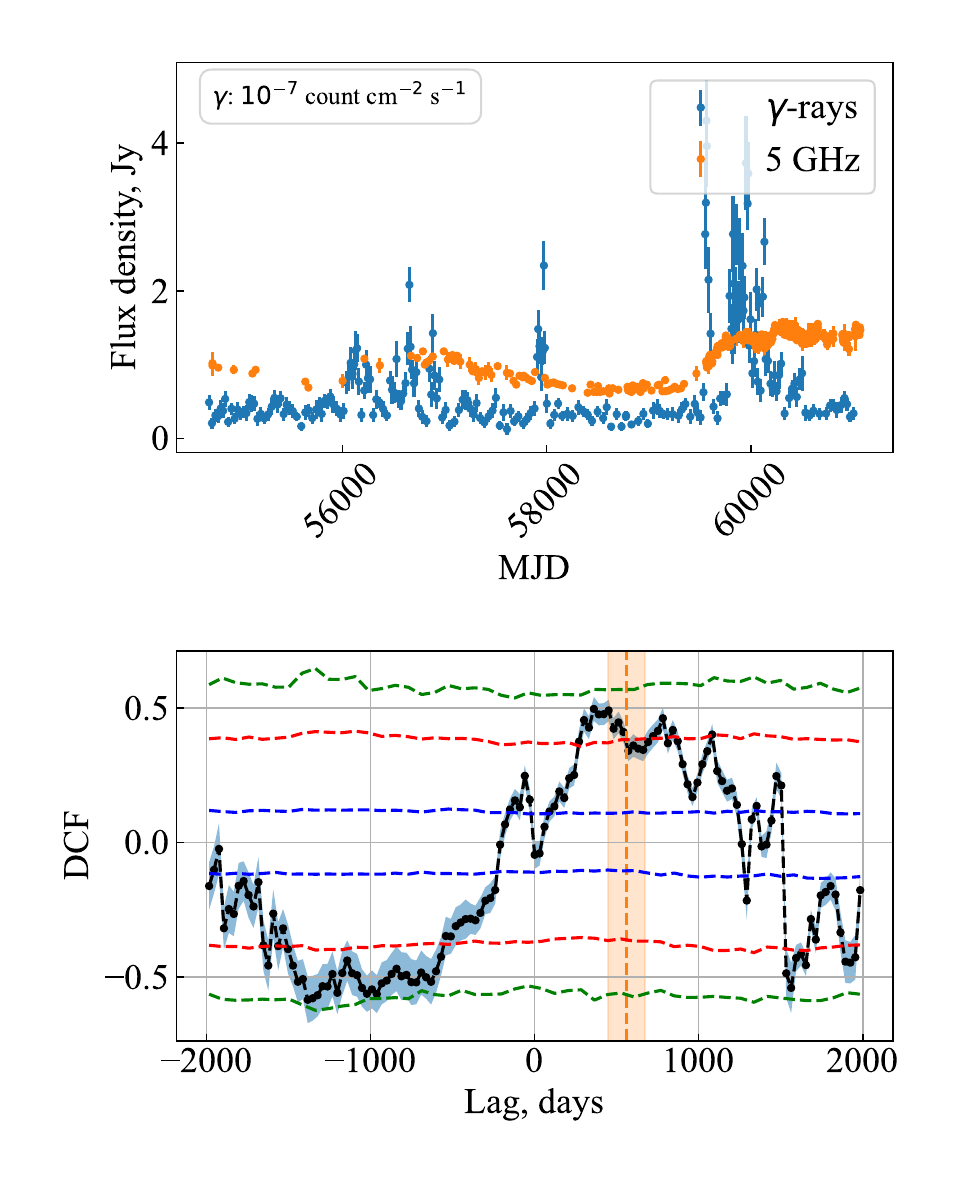}
}
\centerline{
\includegraphics[width=0.7\columnwidth]{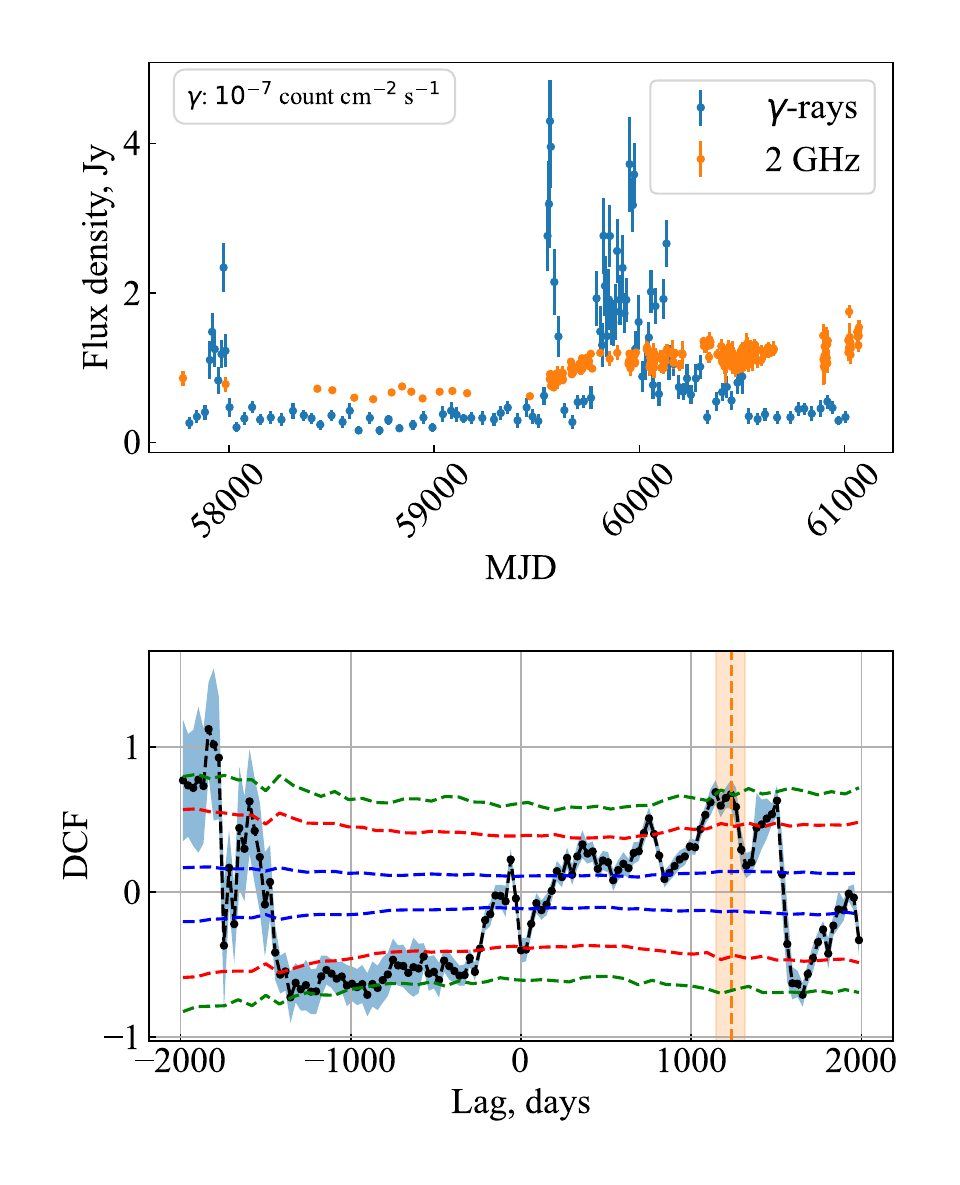}
\includegraphics[width=0.7\columnwidth]{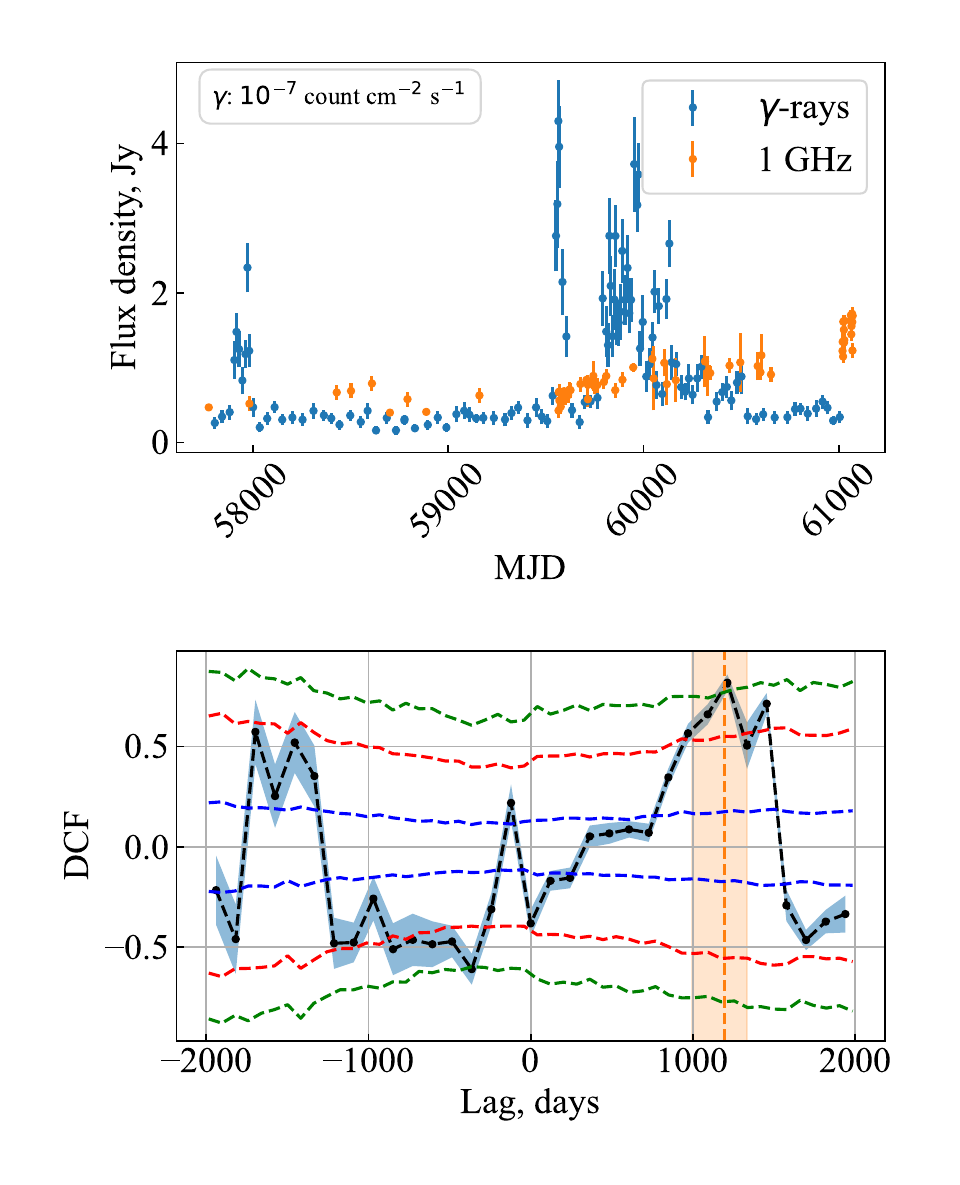}
\includegraphics[width=0.7\columnwidth]{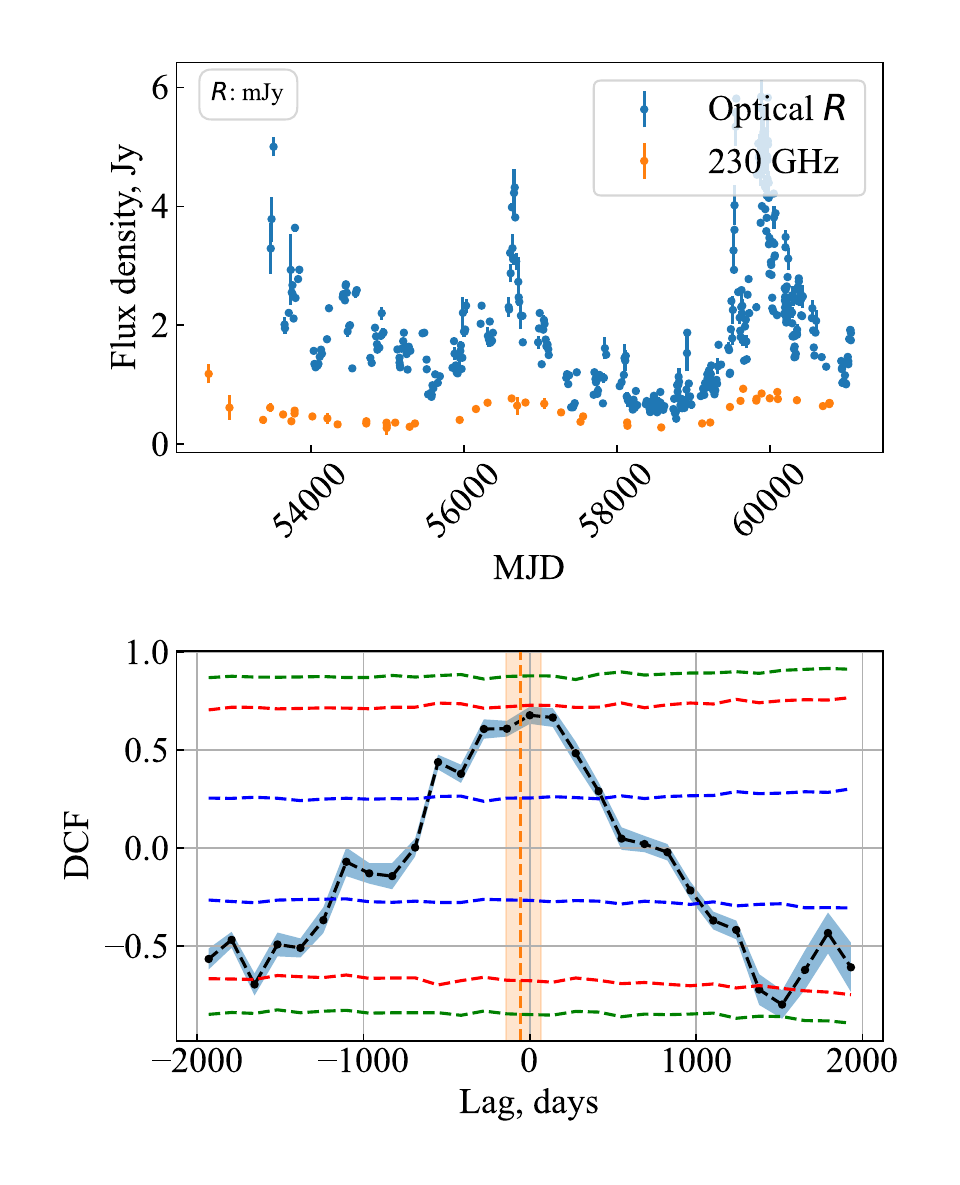}
}
\caption{The light curves for the entire observed time interval (upper panels) and corresponding DCFs (lower panels). The pairs of frequencies are designated in the legends. The black lines with the blue areas are the DCF values with their uncertainties. The dashed blue, red, and green horizontal lines are the 1, 2, and 3$\sigma$ significance levels, respectively. The orange vertical lines are the lags measured by the FR/RSS method. The orange areas are the lag uncertainties (16th and 84th percentiles).}
\label{fig:dcf}
\end{figure*}

\begin{figure*}
\centerline{
\includegraphics[width=0.7\columnwidth]{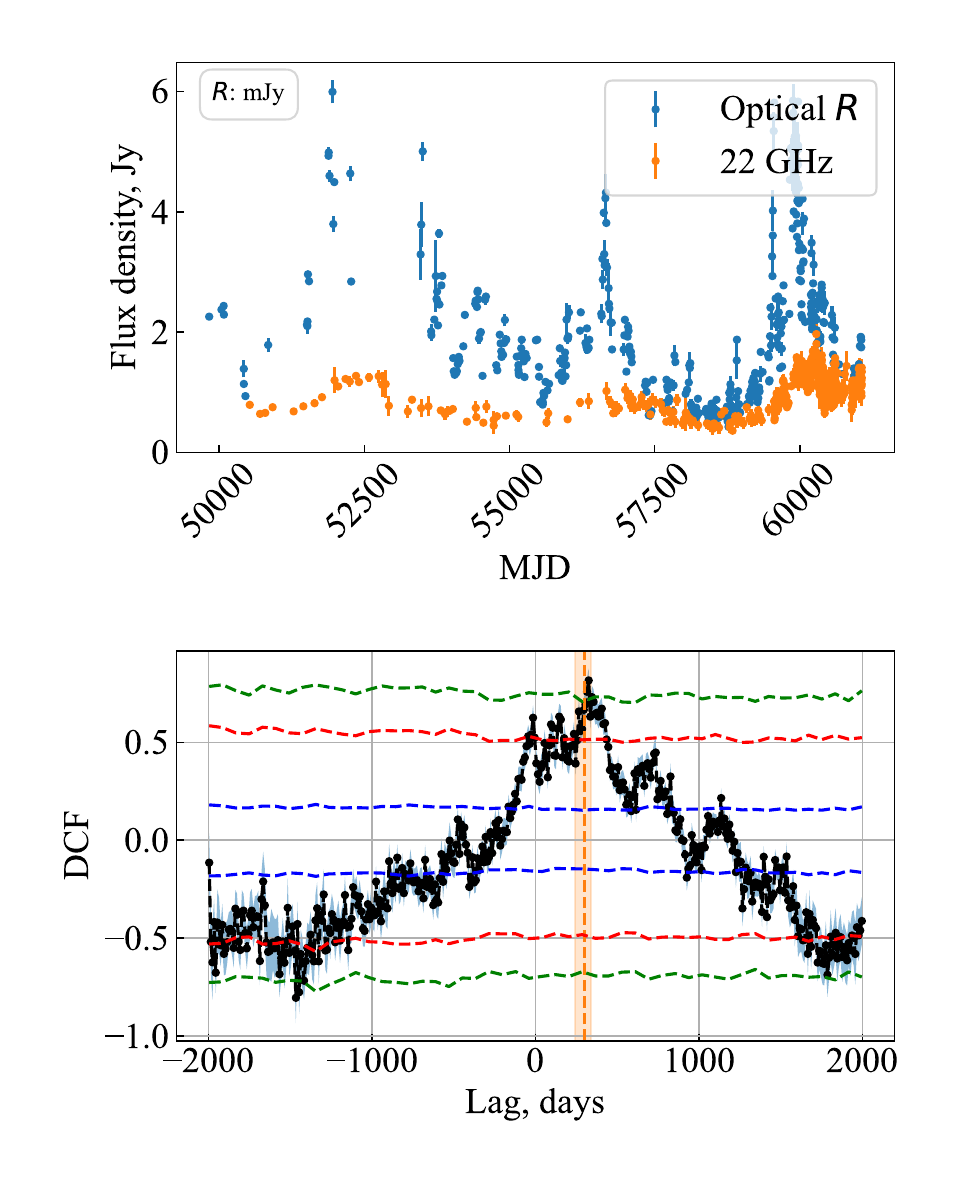}
\includegraphics[width=0.7\columnwidth]{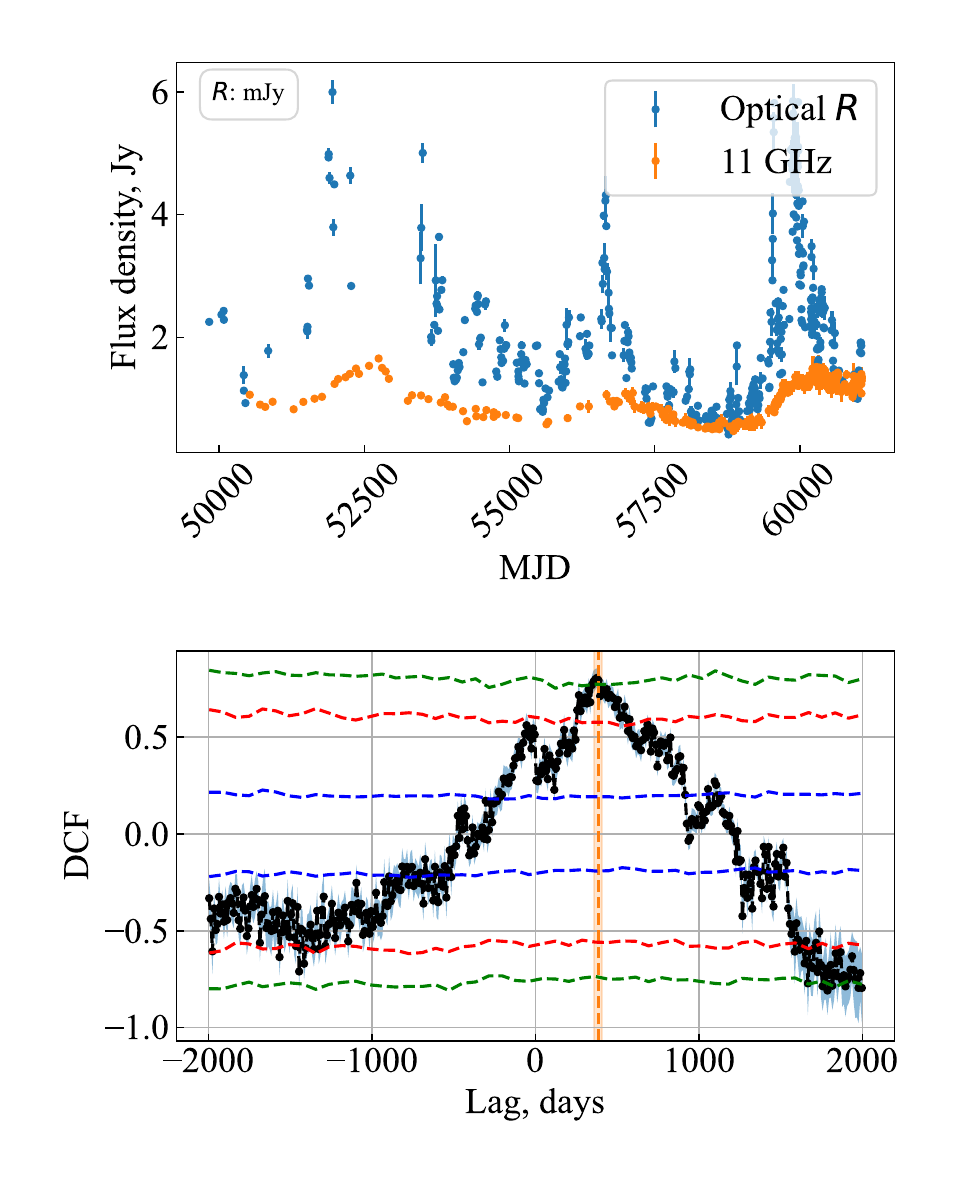}
\includegraphics[width=0.7\columnwidth]{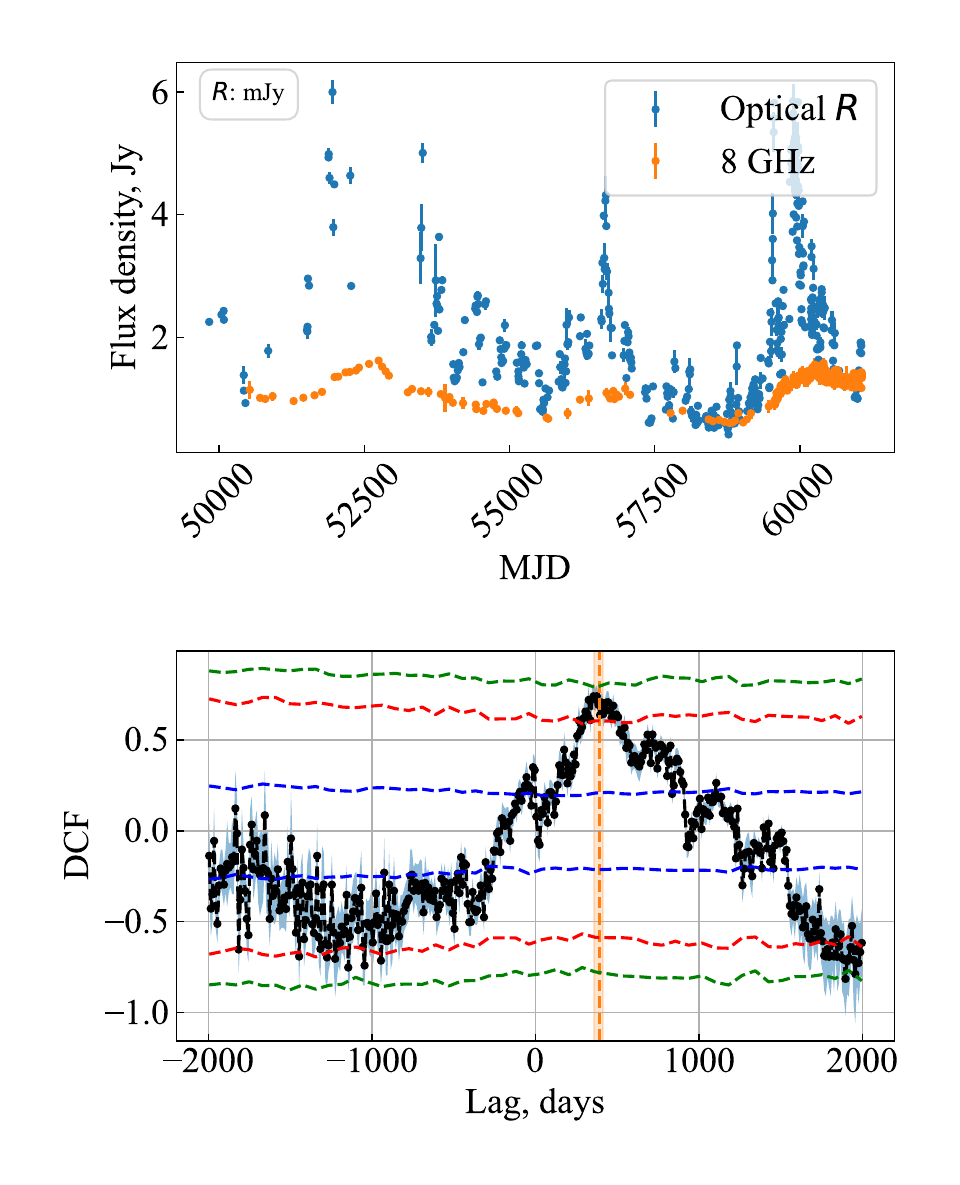}
}
\centerline{
\includegraphics[width=0.7\columnwidth]{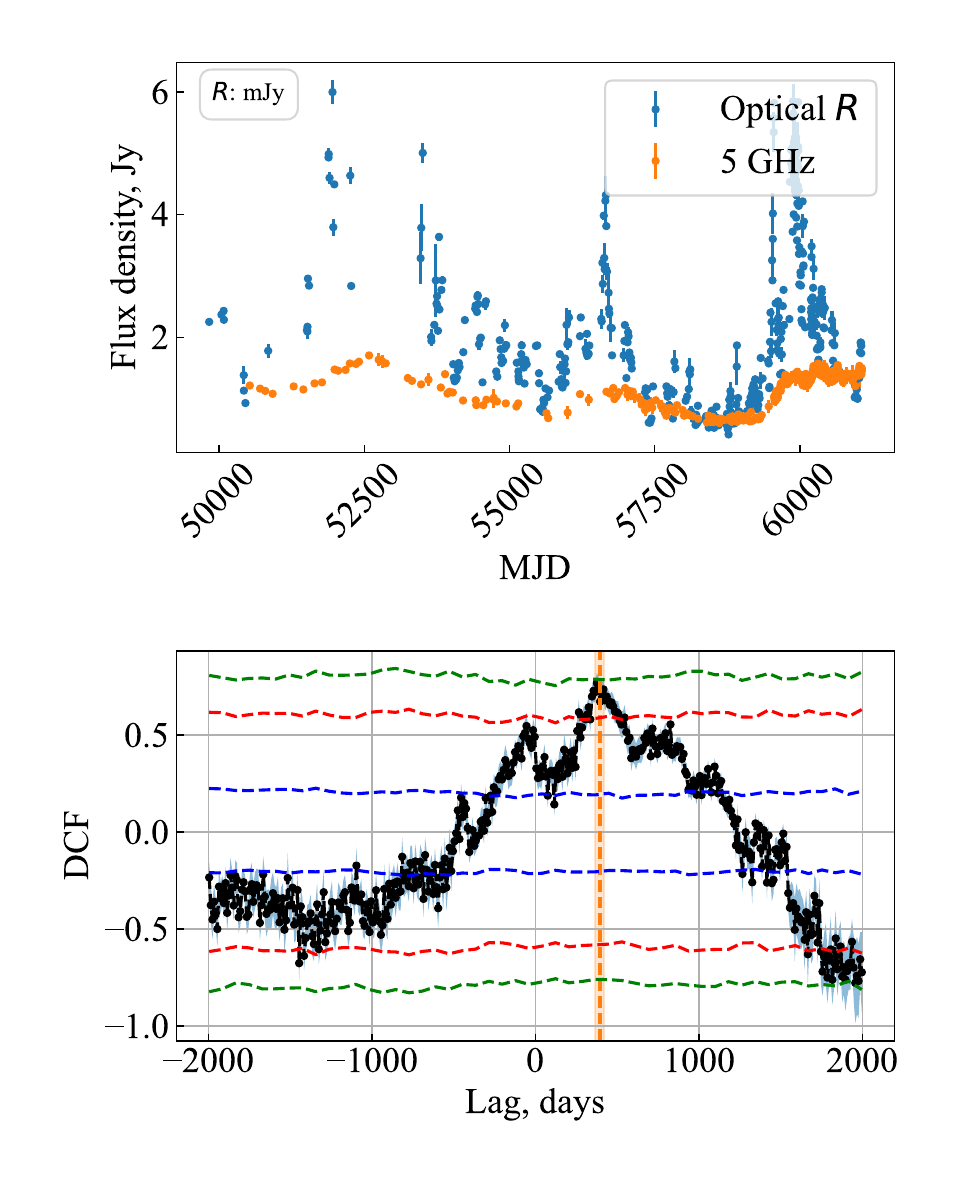}
\includegraphics[width=0.7\columnwidth]{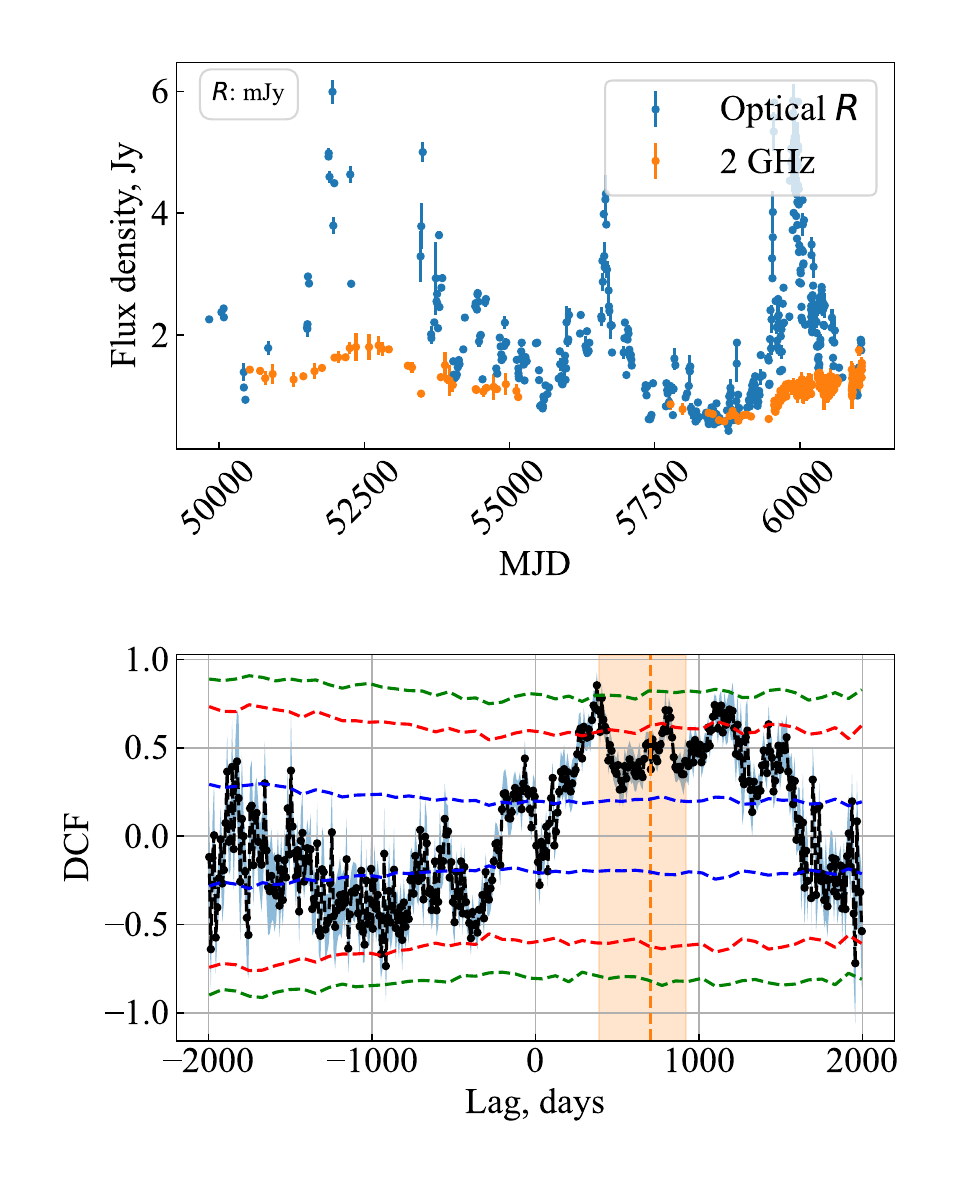}
\includegraphics[width=0.7\columnwidth]{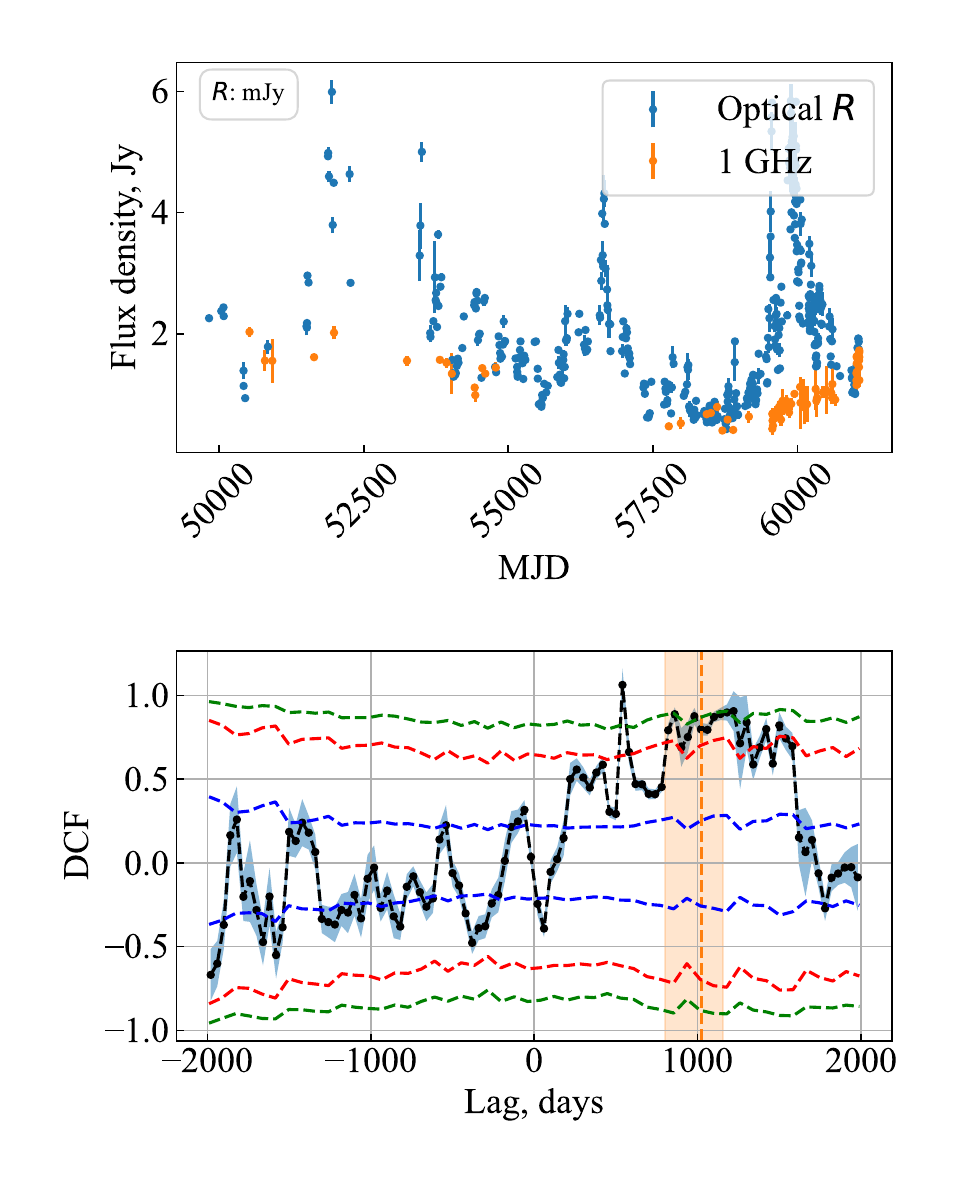}
}
\centerline{
\includegraphics[width=0.7\columnwidth]{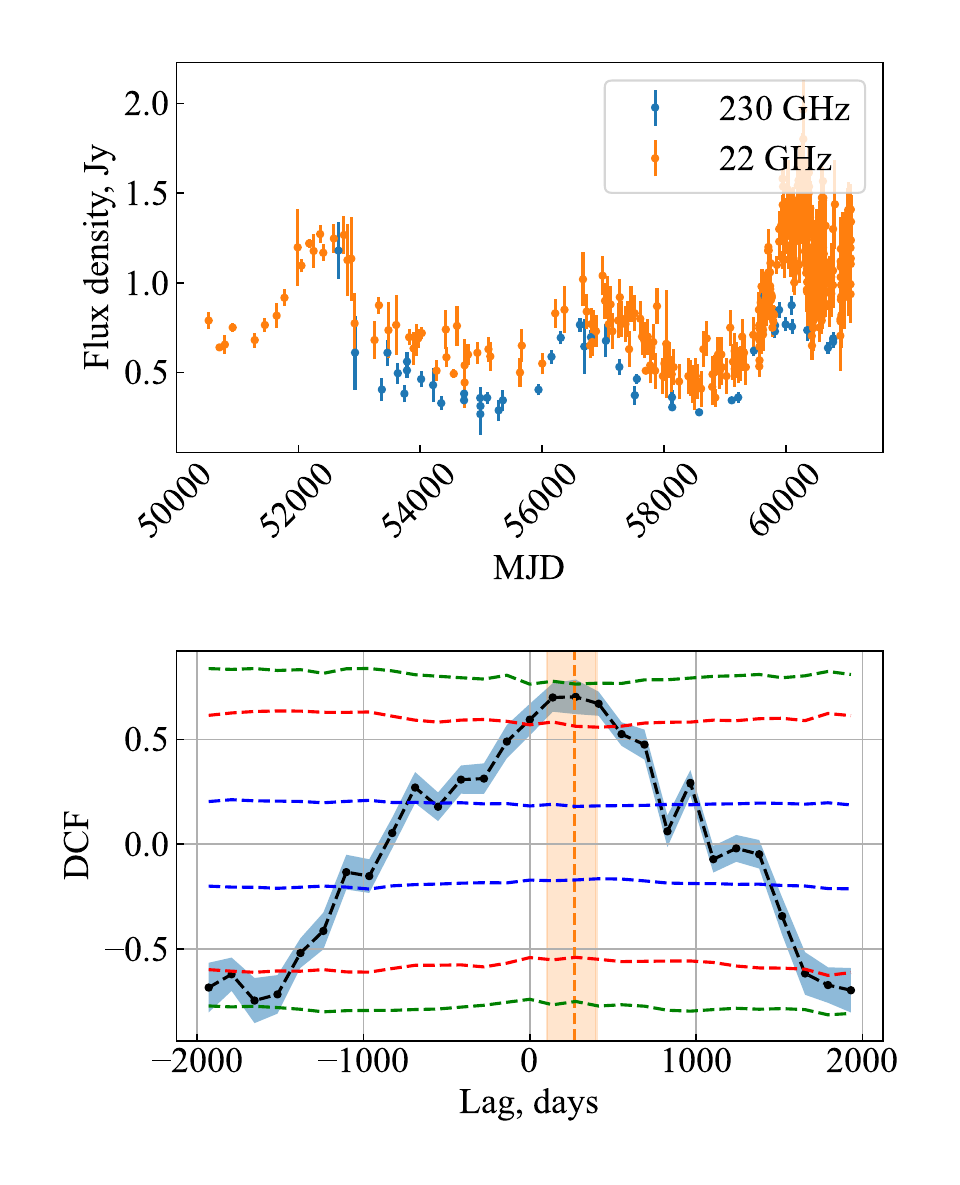}
\includegraphics[width=0.7\columnwidth]{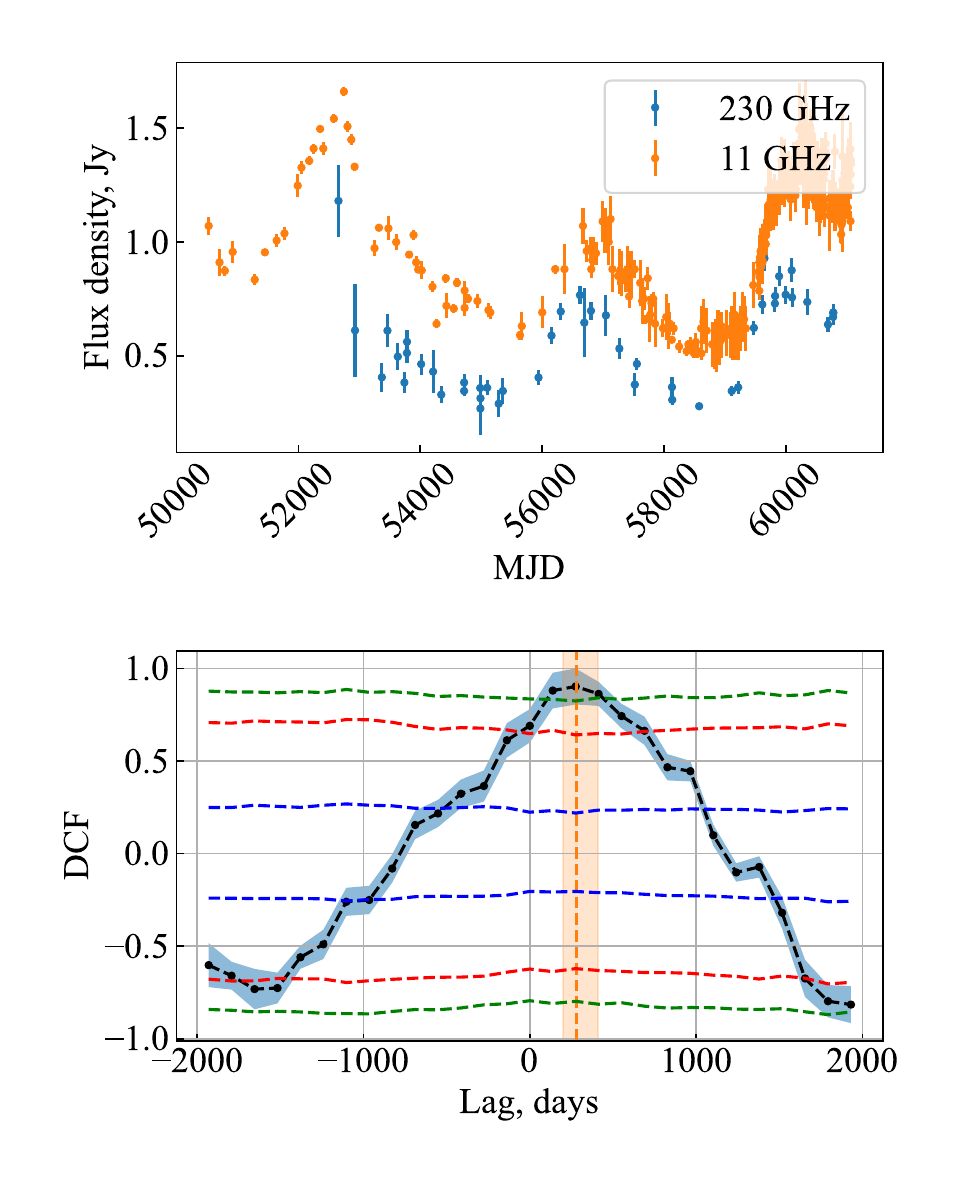}
\includegraphics[width=0.7\columnwidth]{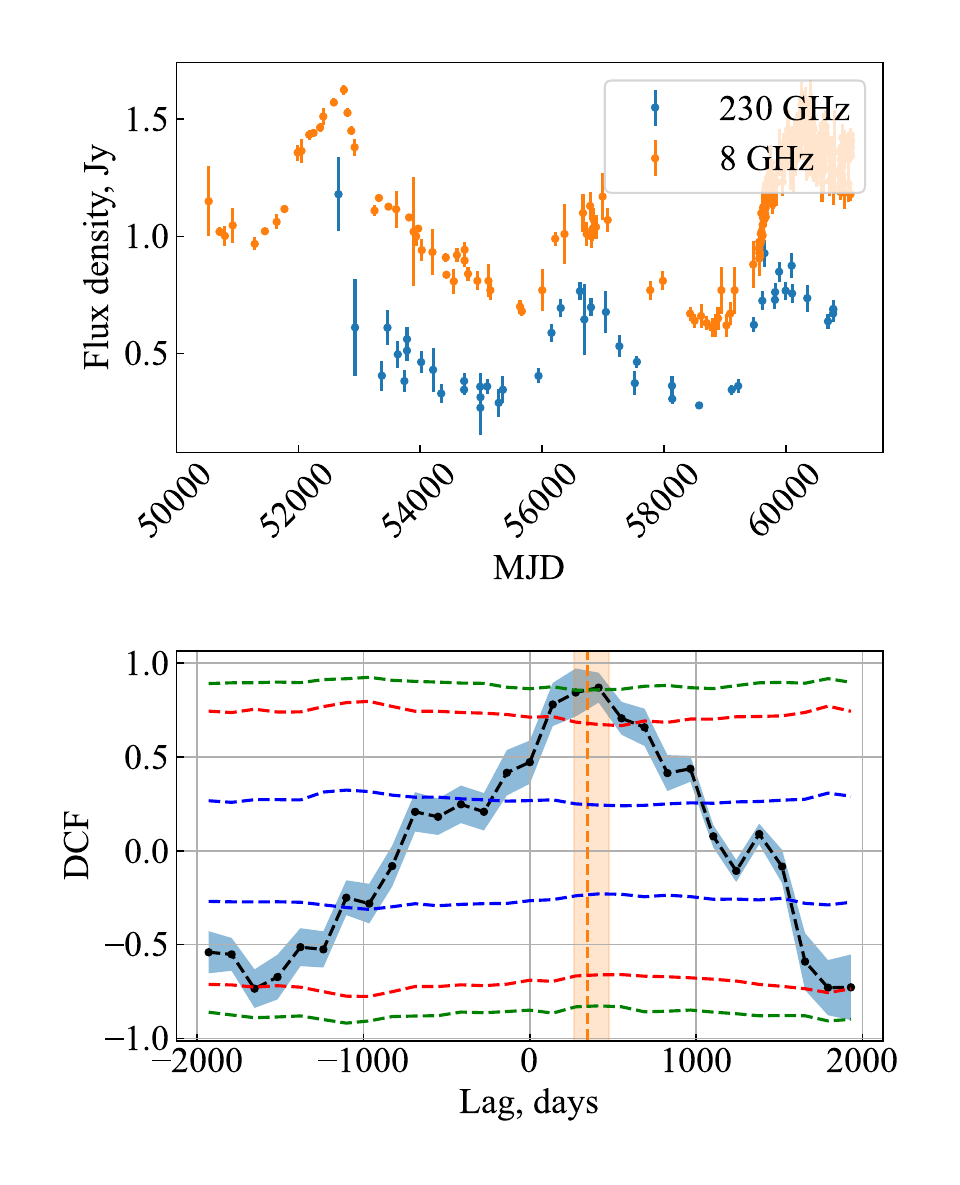}
}
\contcaption{} 
\end{figure*}

\begin{figure*}
\centerline{
\includegraphics[width=0.7\columnwidth]{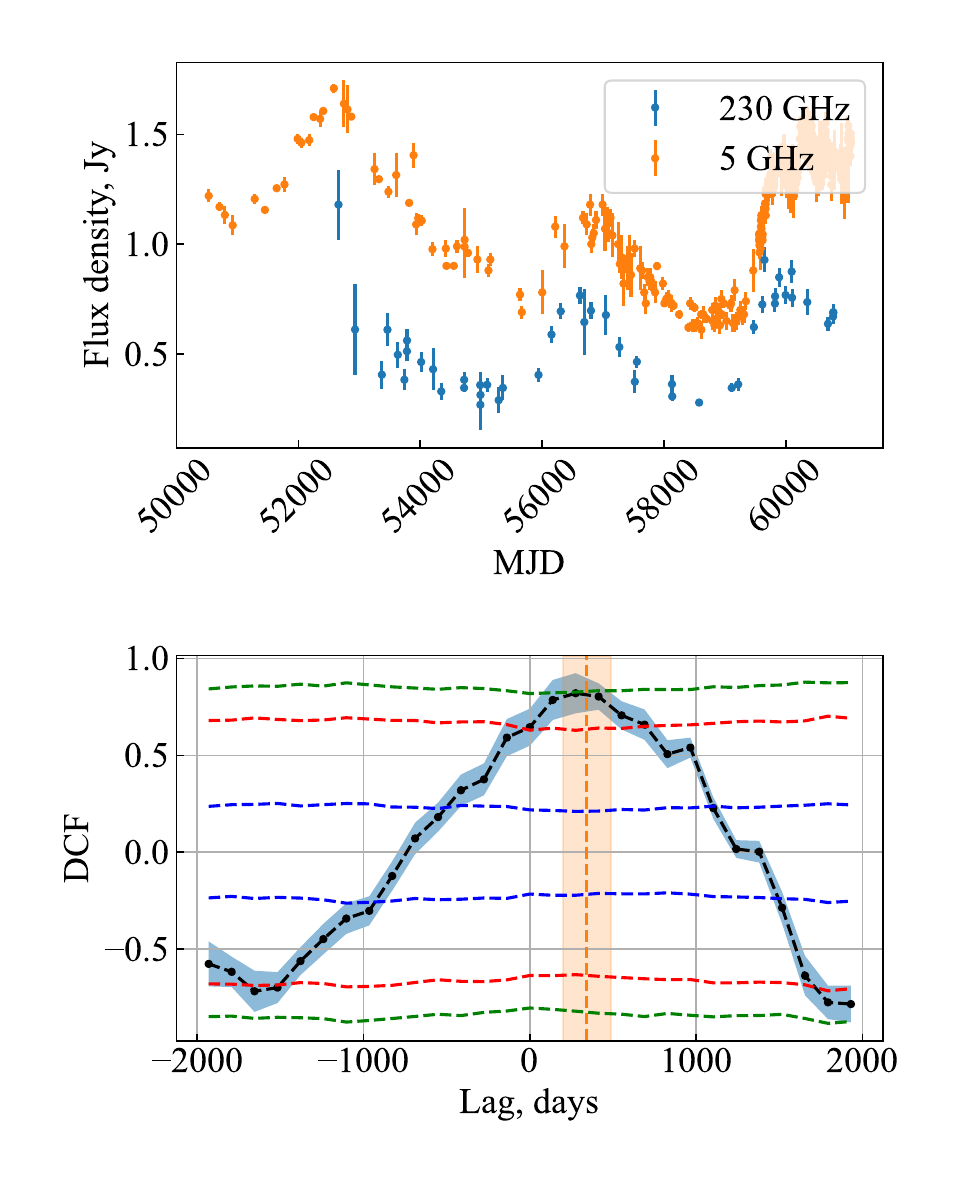}
\includegraphics[width=0.7\columnwidth]{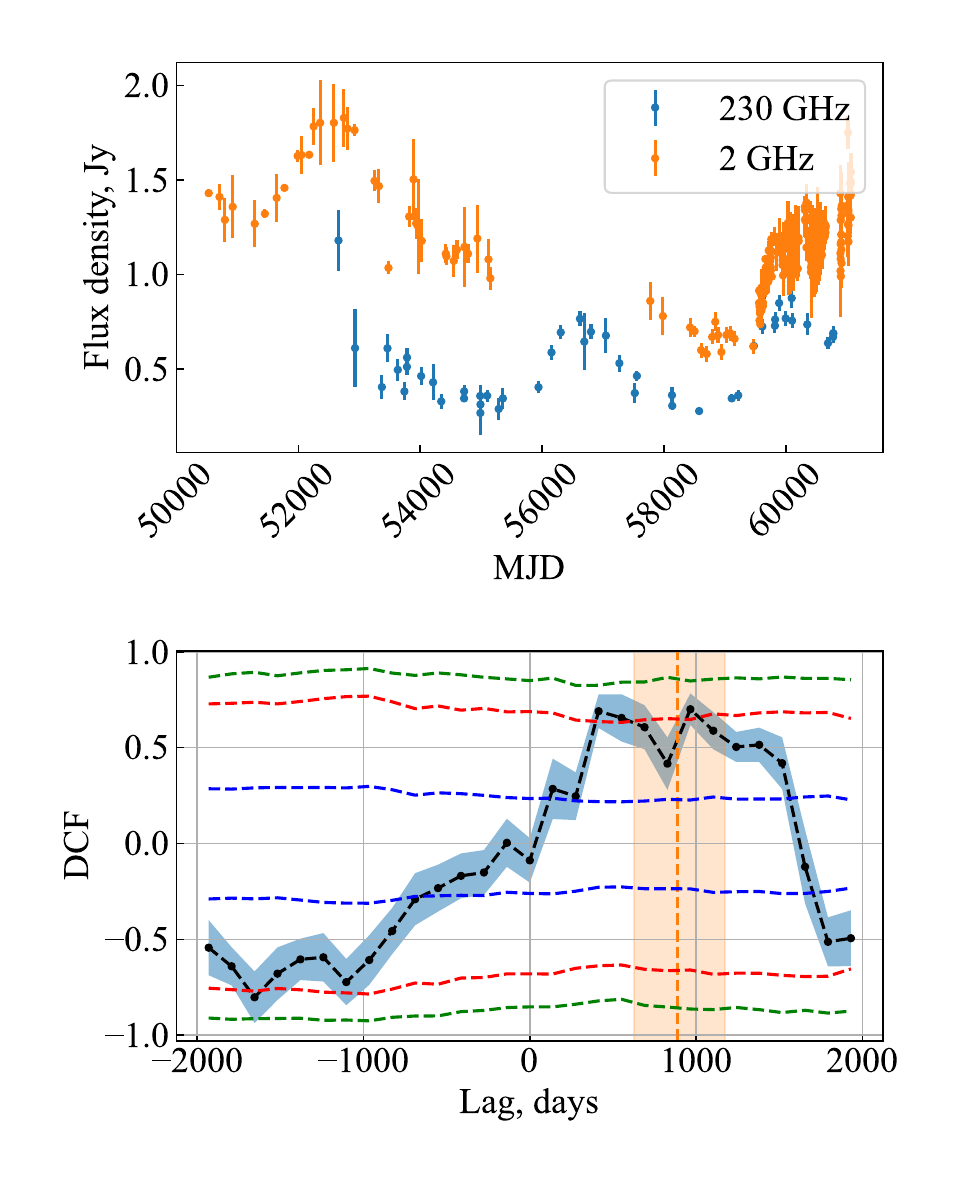}
\includegraphics[width=0.7\columnwidth]{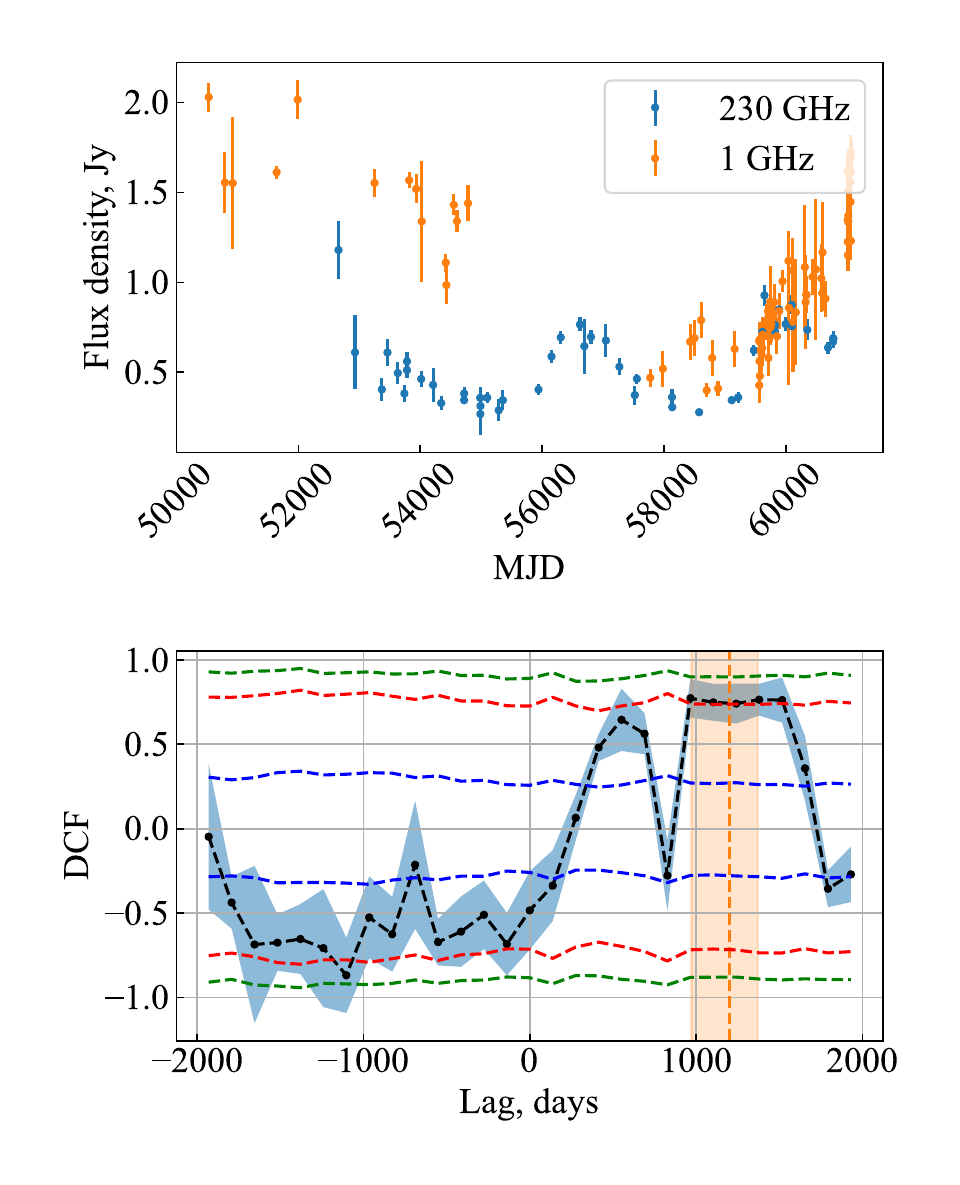}
}
\centerline{
\includegraphics[width=0.7\columnwidth]{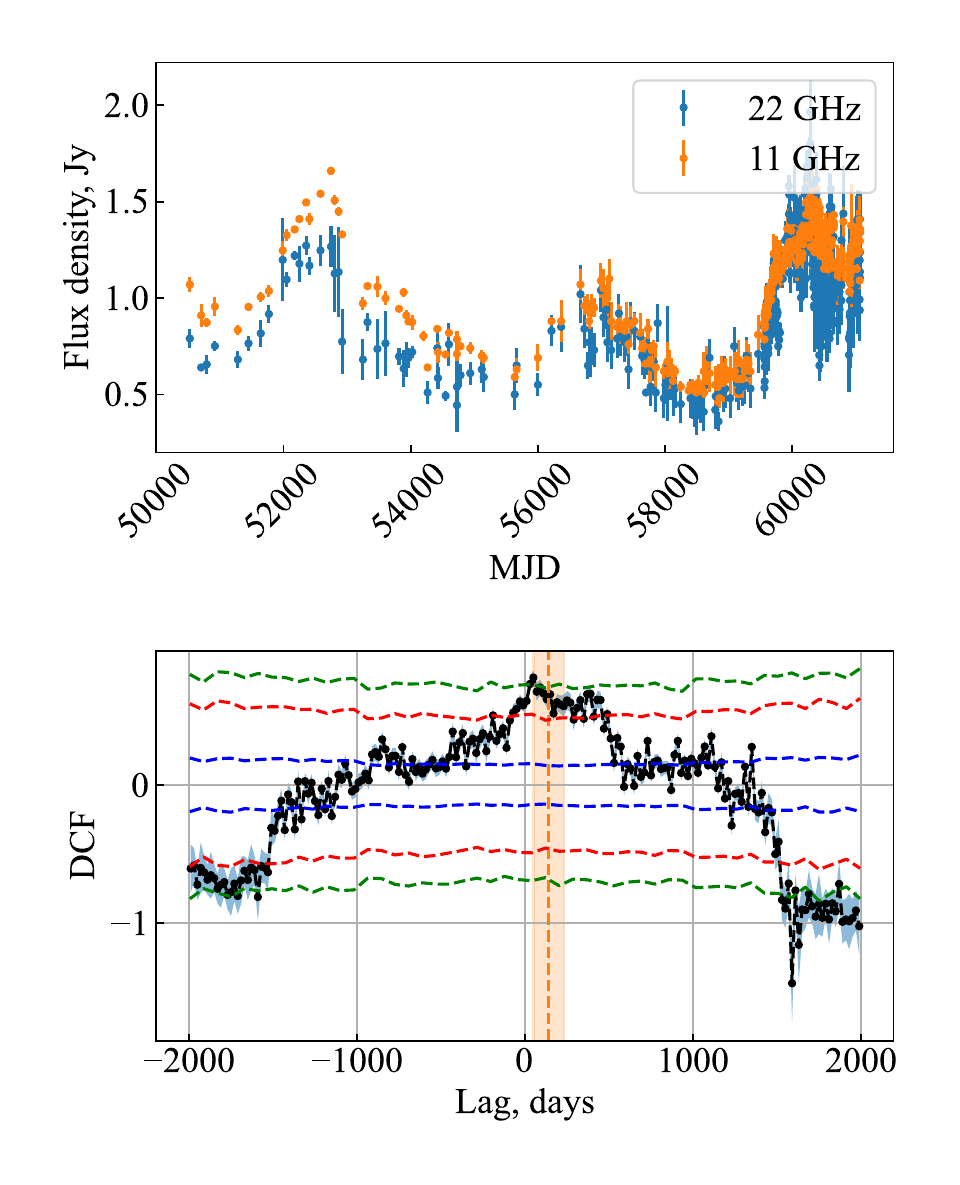}
\includegraphics[width=0.7\columnwidth]{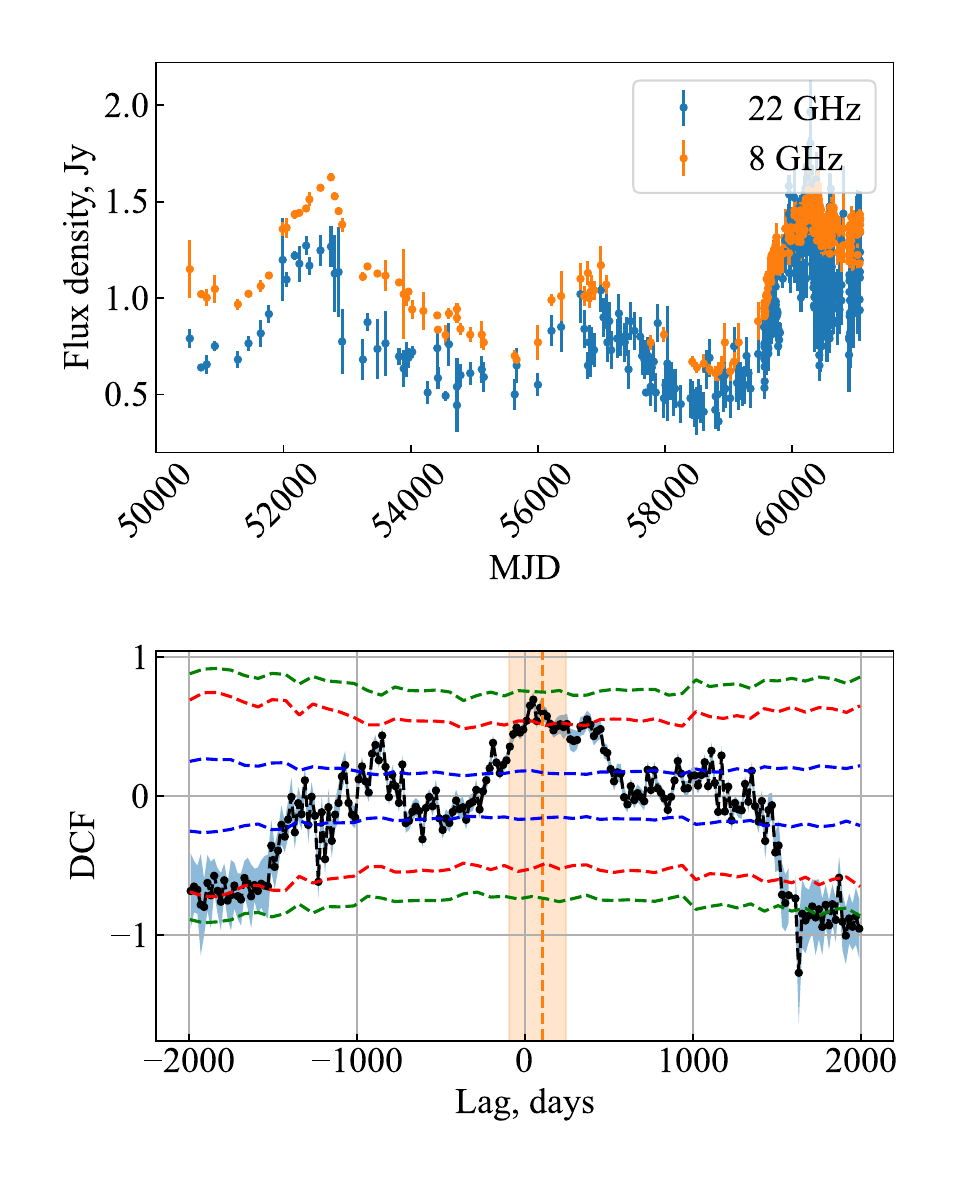}
\includegraphics[width=0.7\columnwidth]{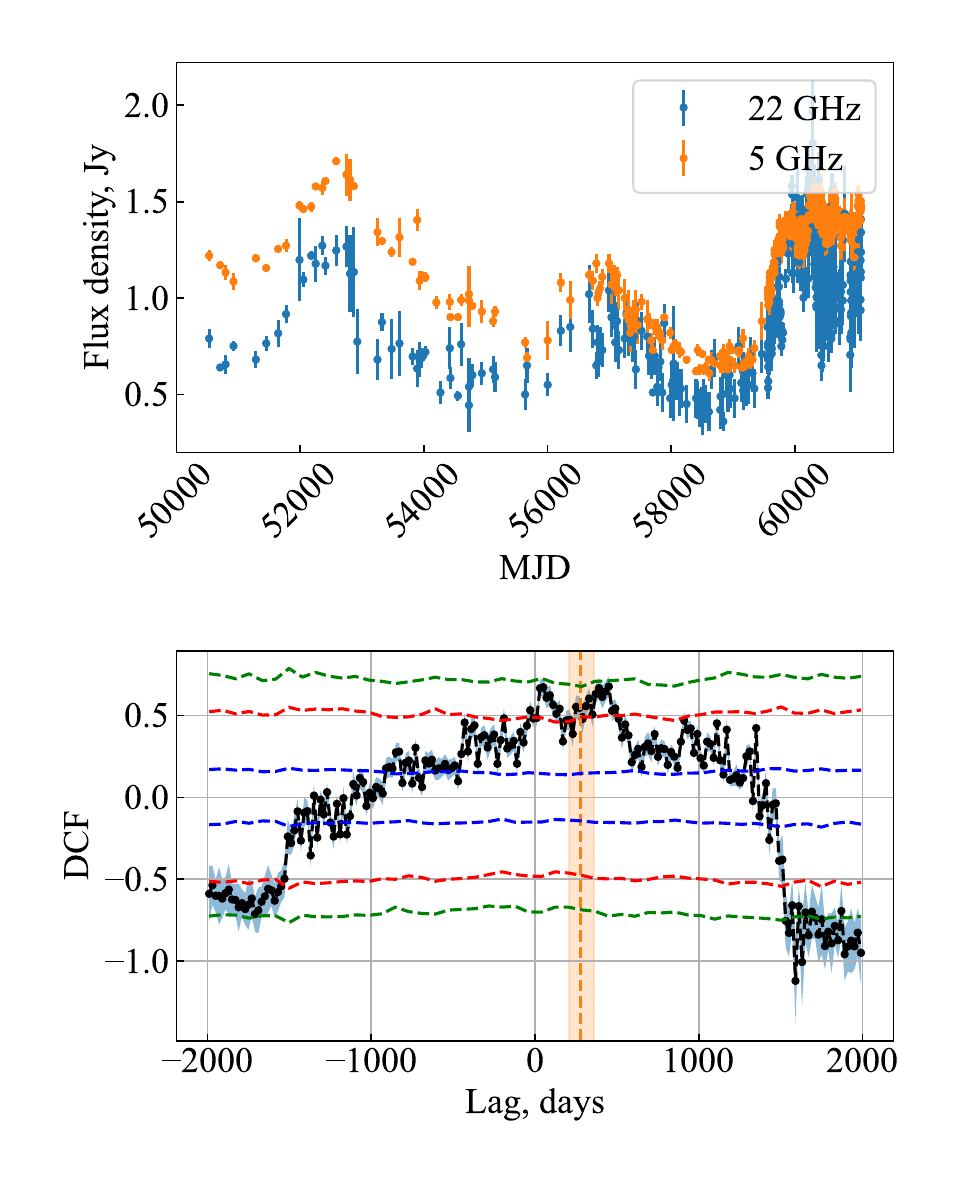}
}
\centerline{
\includegraphics[width=0.7\columnwidth]{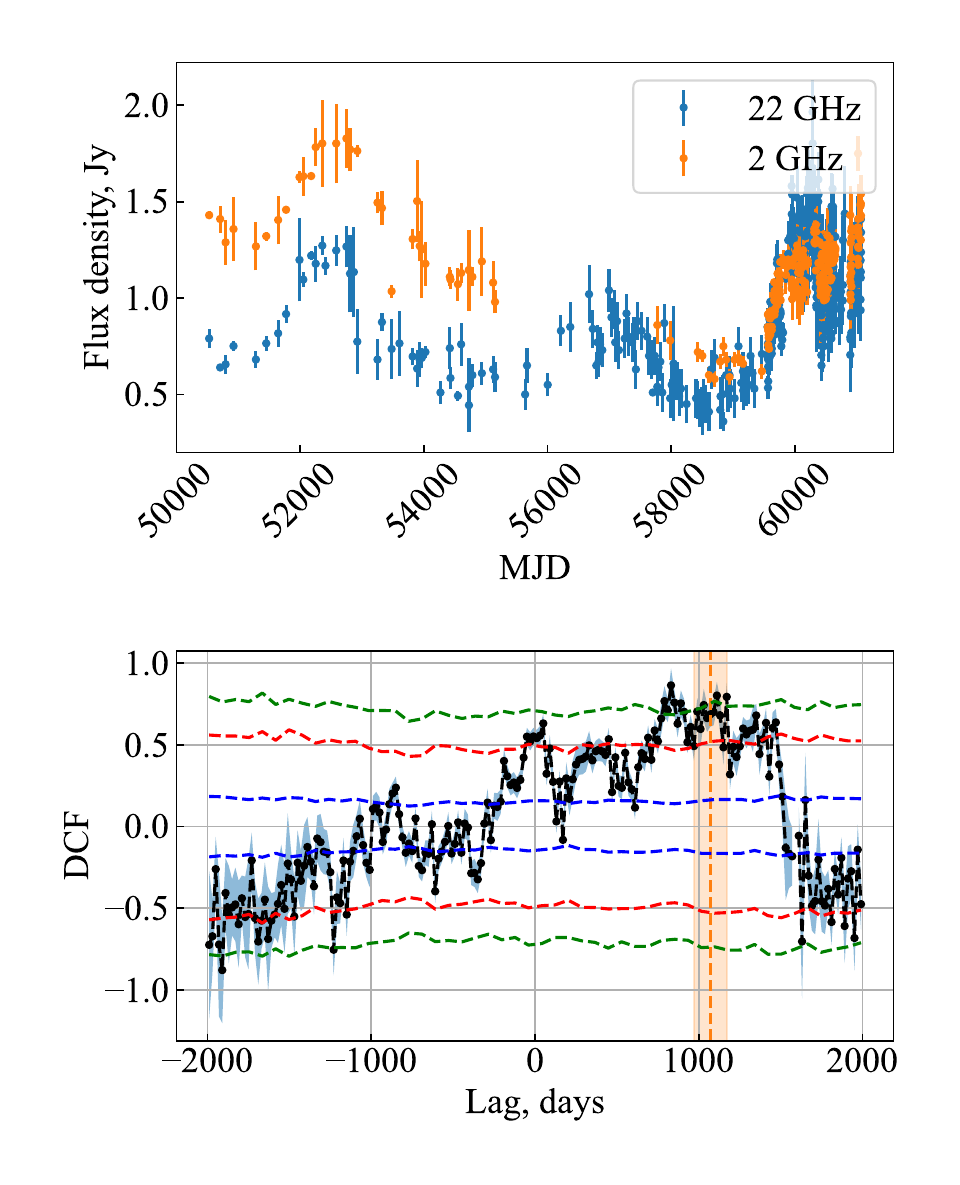}
\includegraphics[width=0.7\columnwidth]{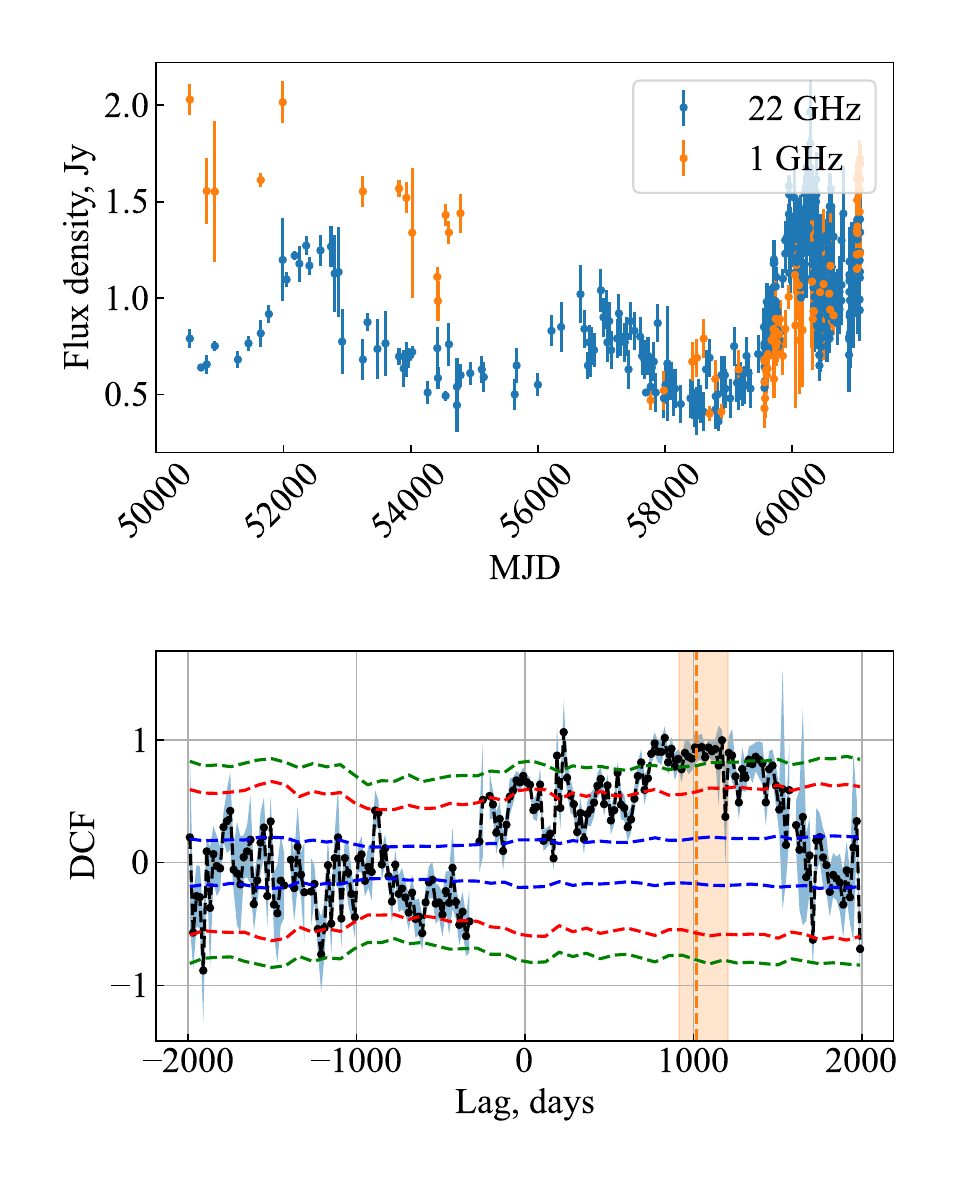}
\includegraphics[width=0.7\columnwidth]{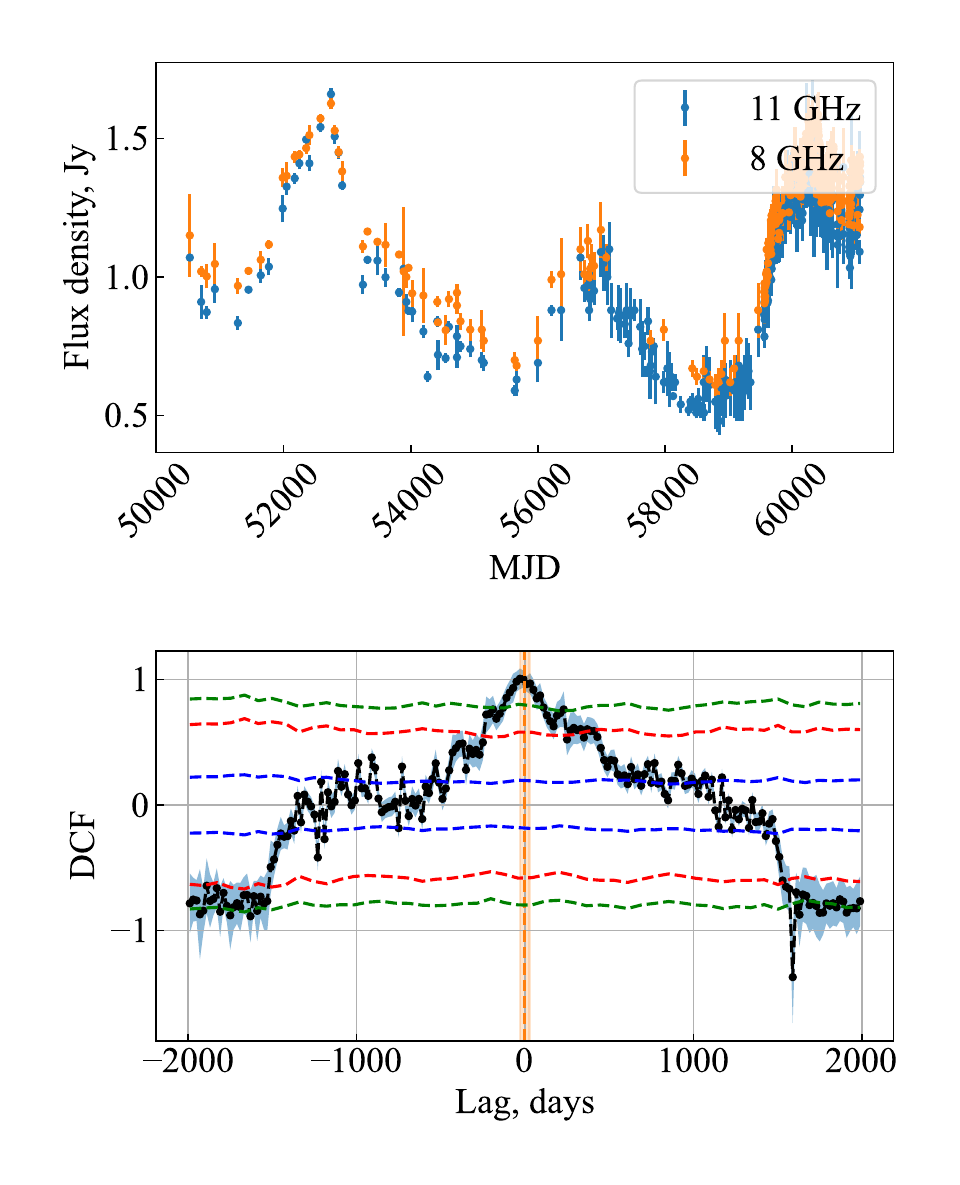}
}
\contcaption{} 
\end{figure*}

\begin{figure*}
\centerline{
\includegraphics[width=0.7\columnwidth]{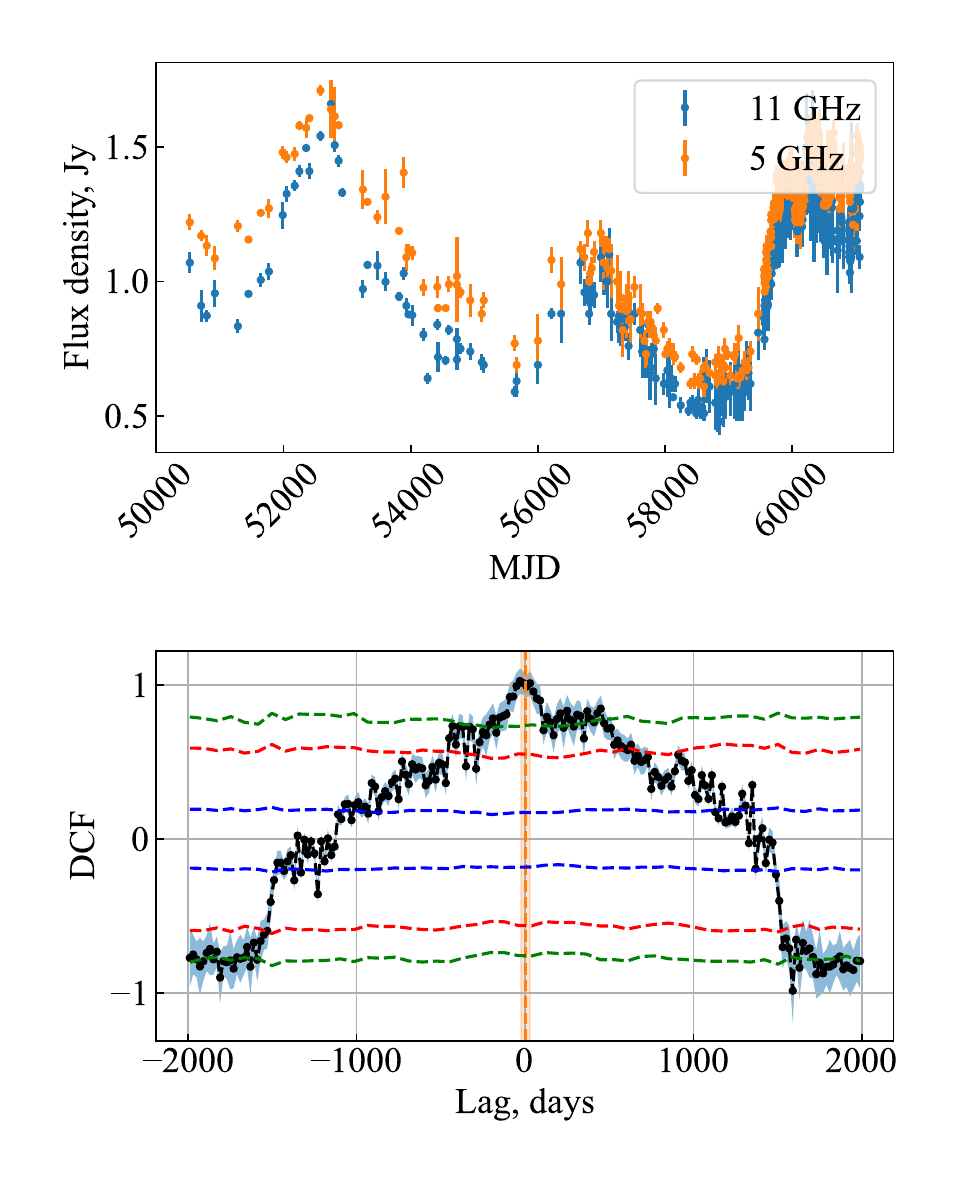}
\includegraphics[width=0.7\columnwidth]{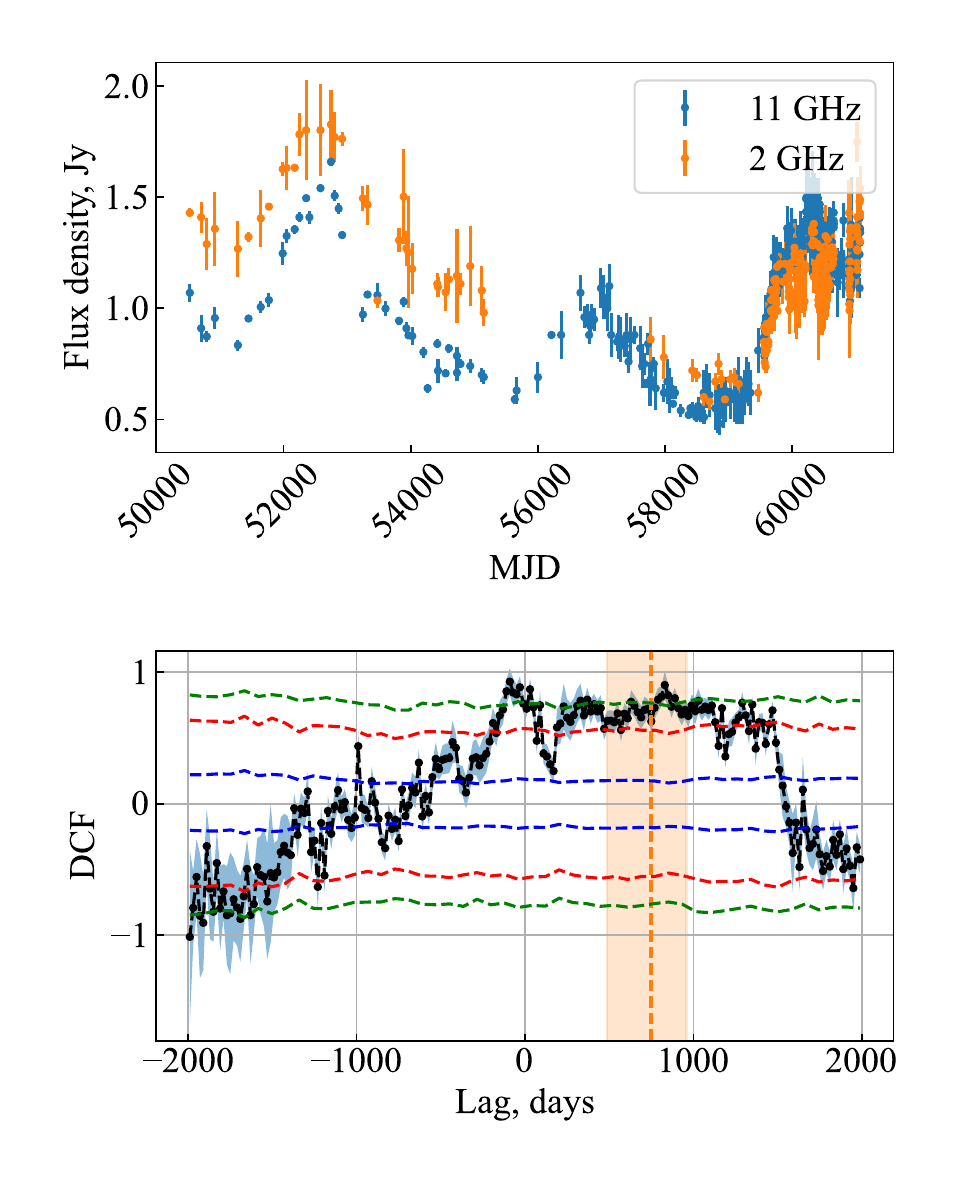}
\includegraphics[width=0.7\columnwidth]{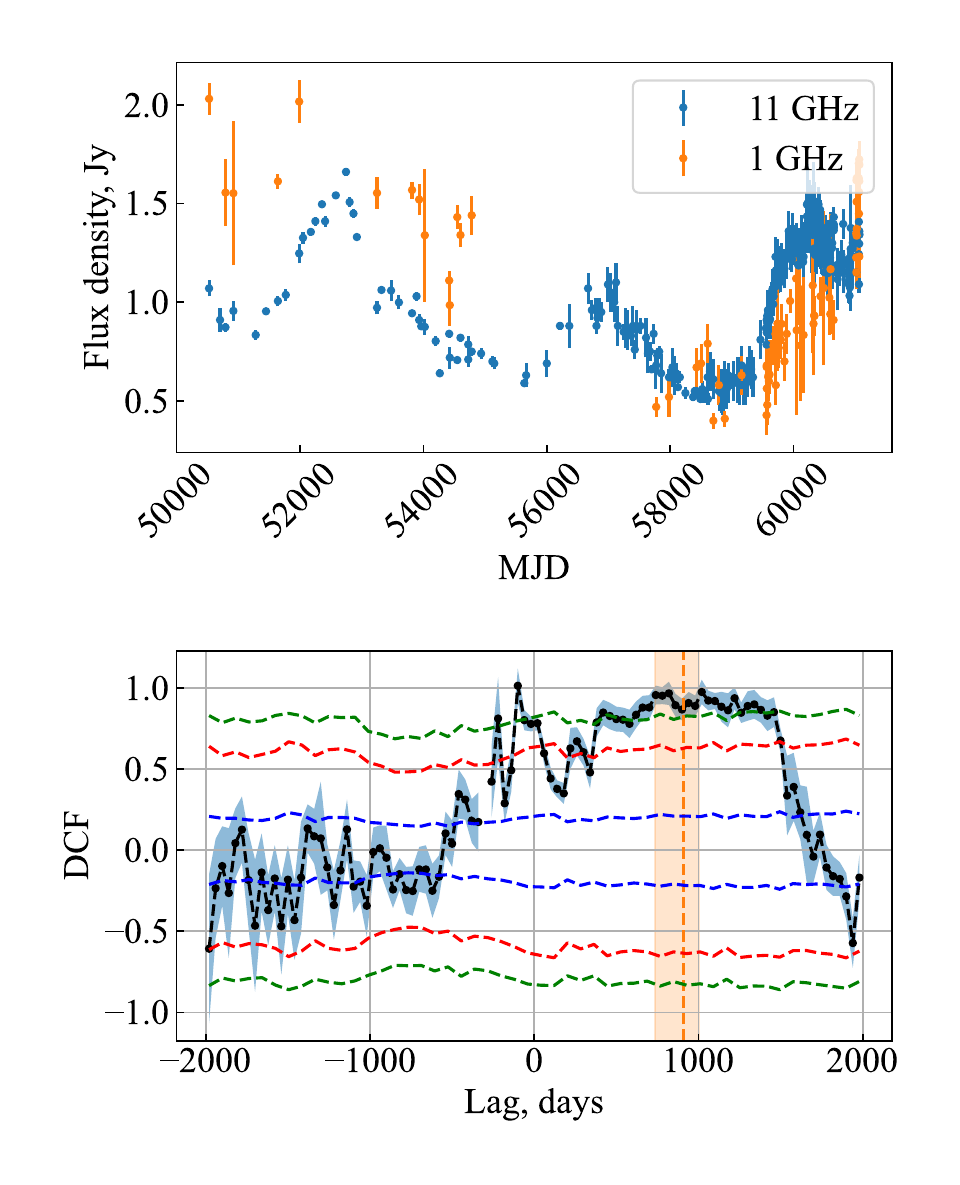}
}
\centerline{
\includegraphics[width=0.7\columnwidth]{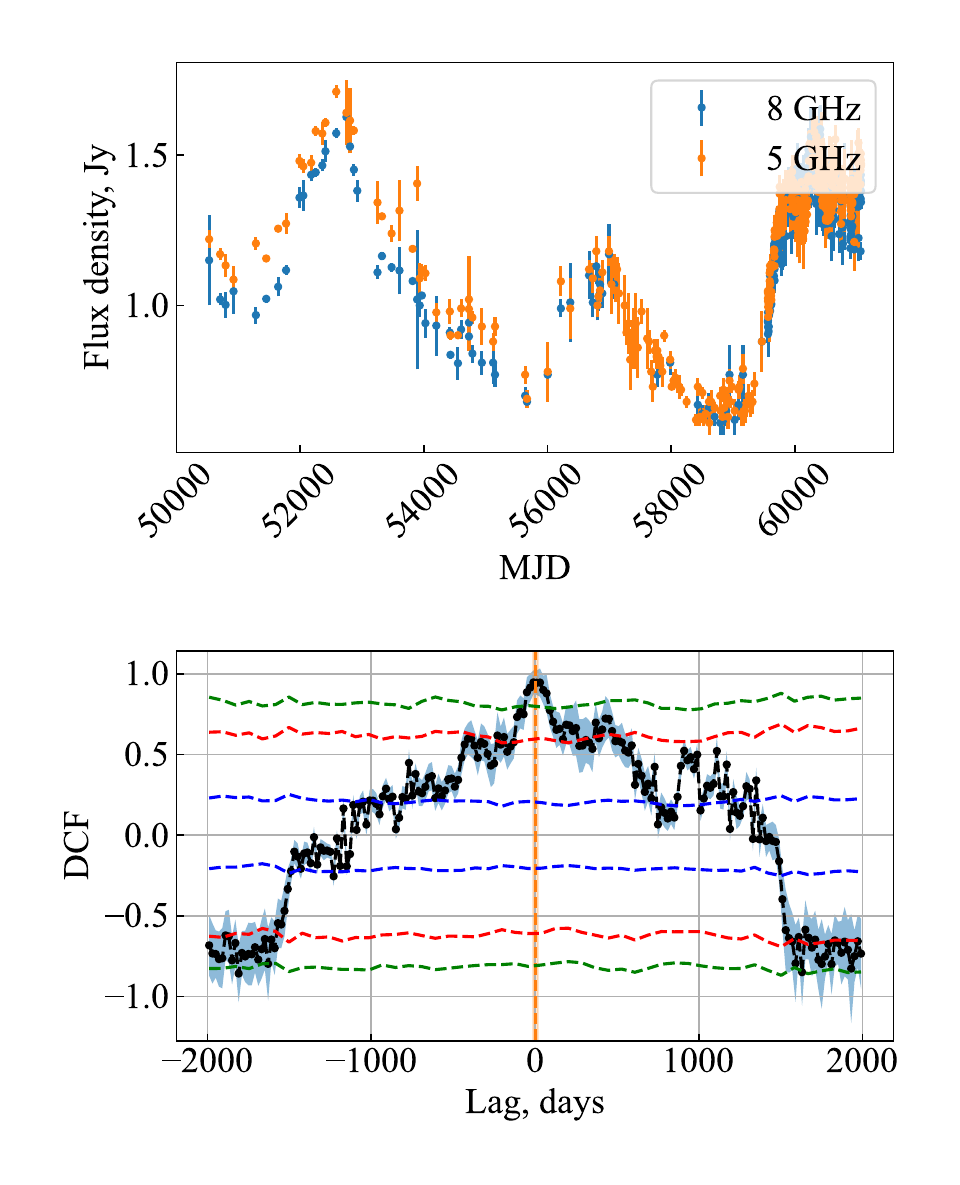}
\includegraphics[width=0.7\columnwidth]{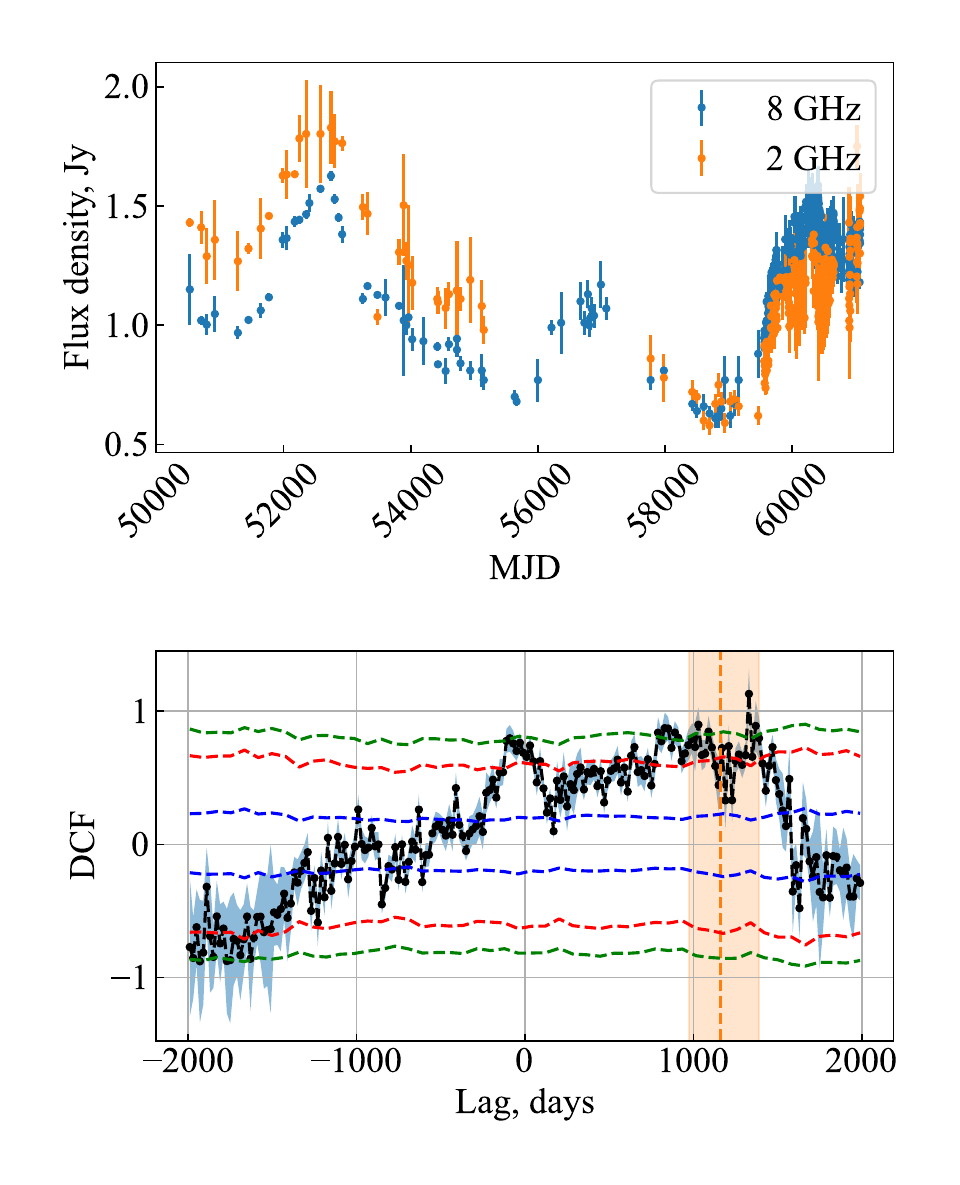}
\includegraphics[width=0.7\columnwidth]{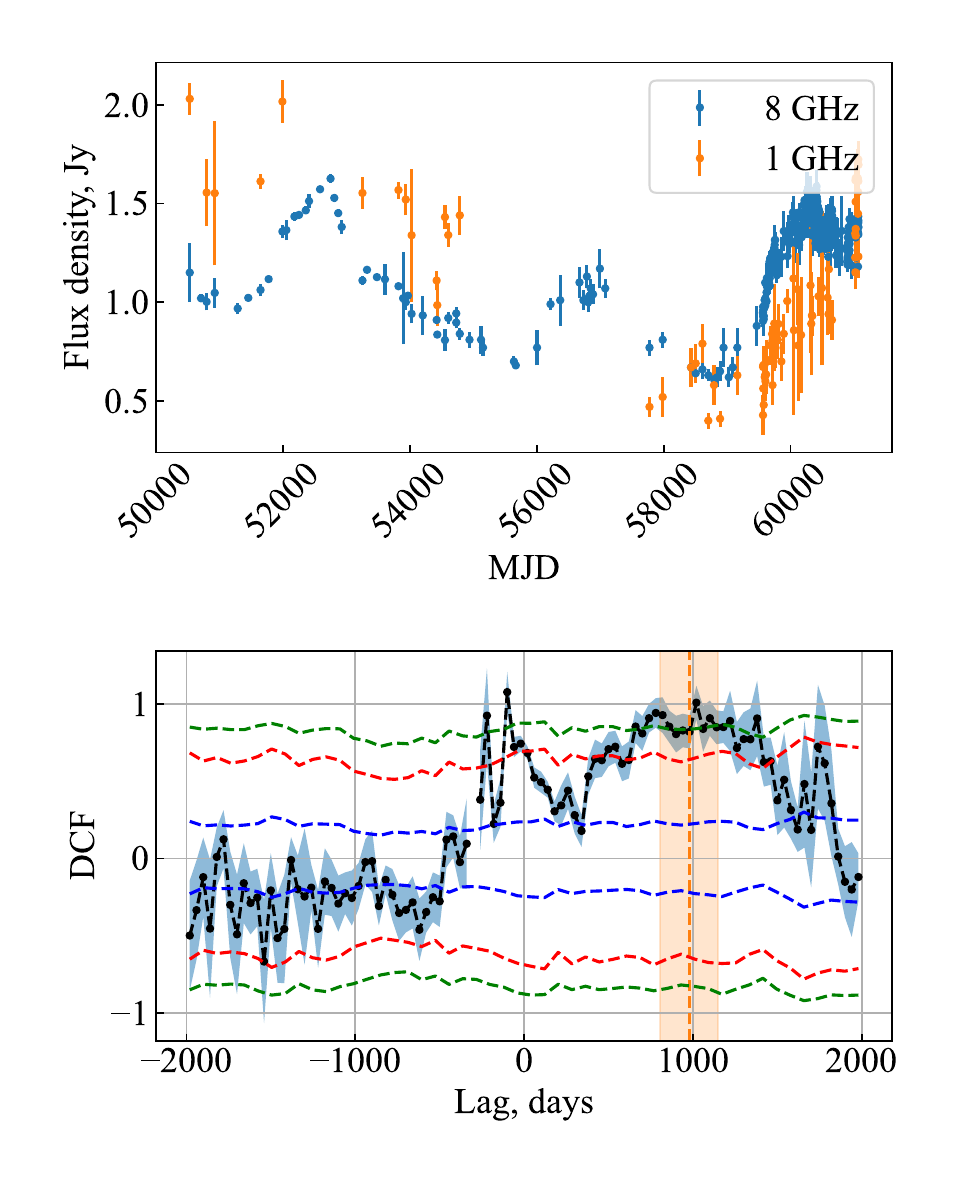}
}
\centerline{
\includegraphics[width=0.7\columnwidth]{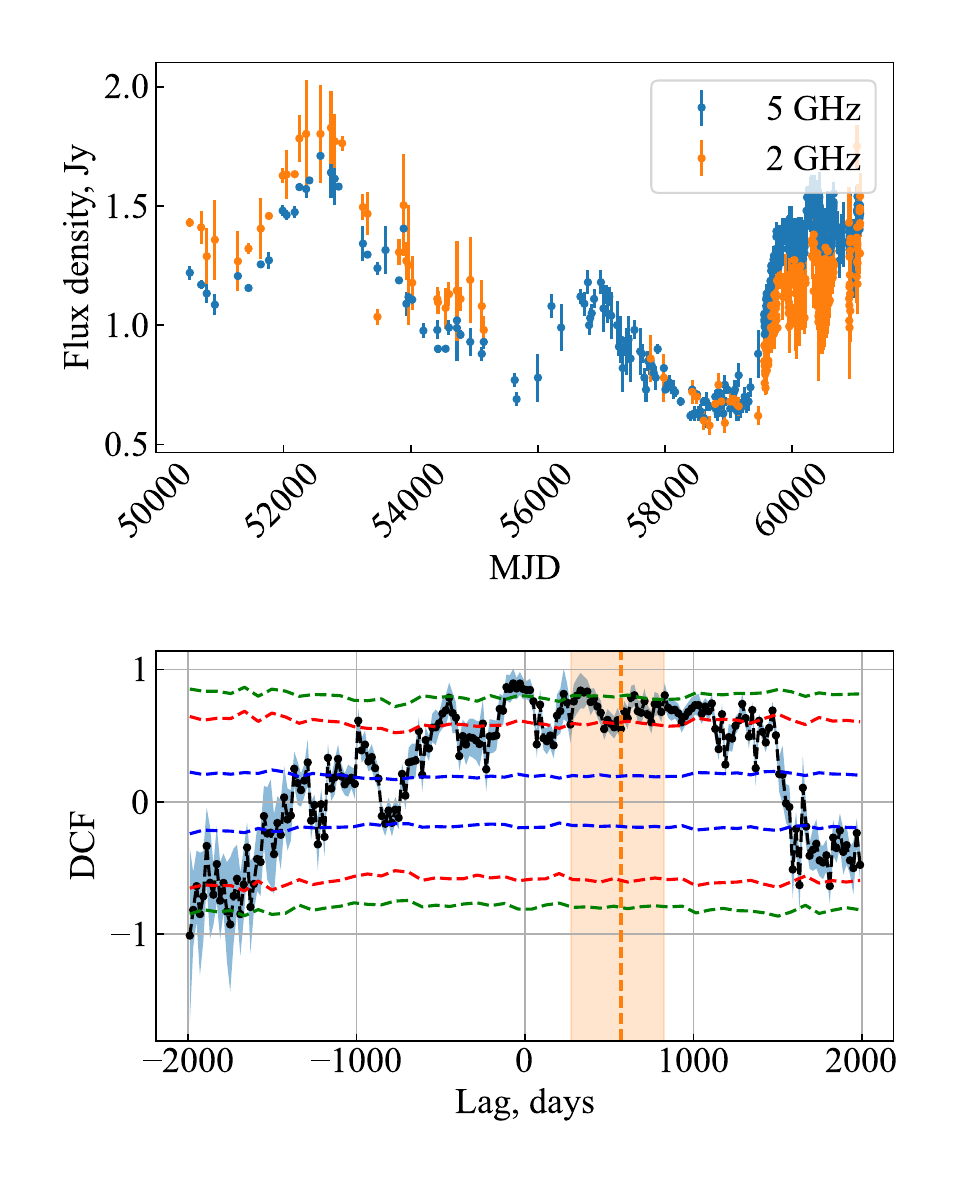}
\includegraphics[width=0.7\columnwidth]{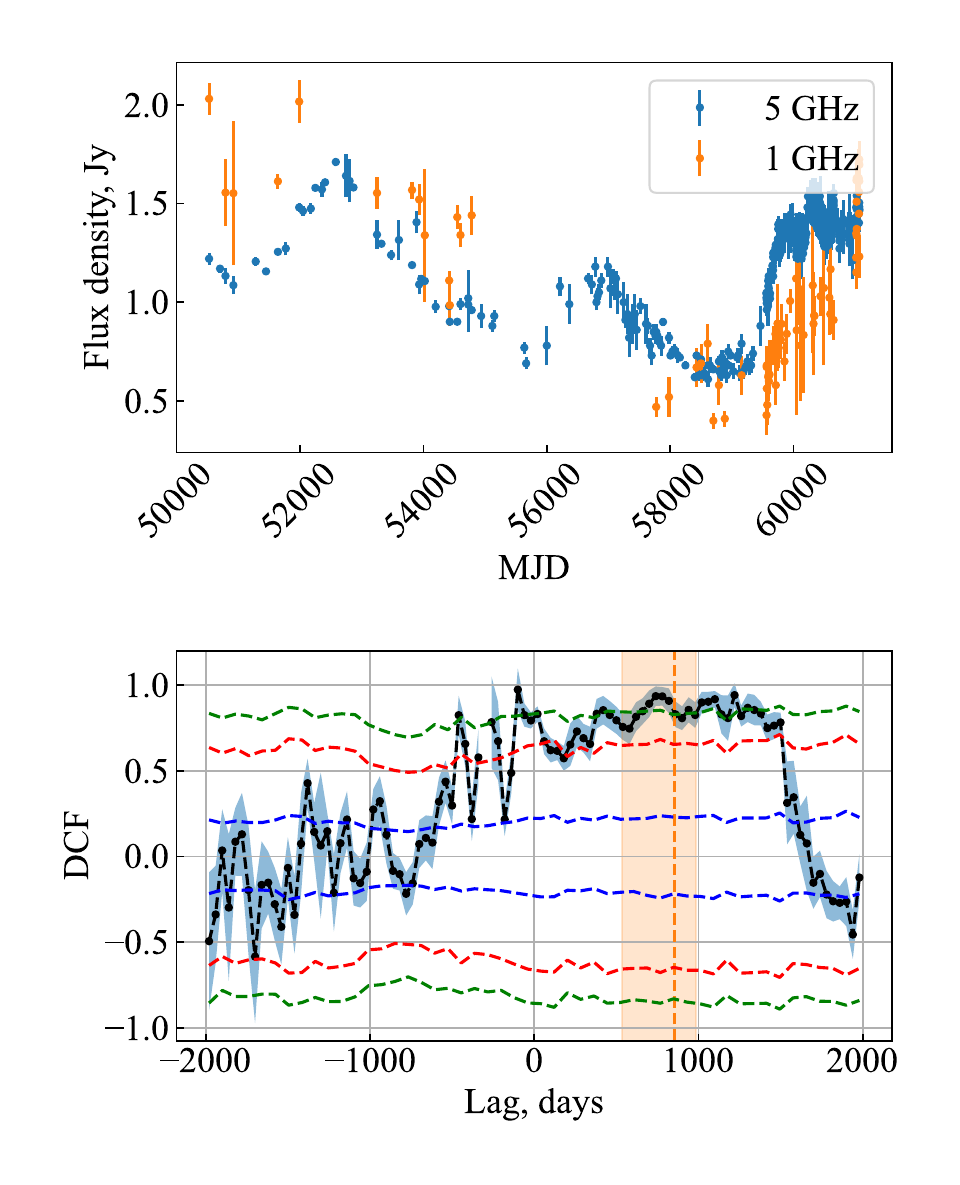}
\includegraphics[width=0.7\columnwidth]{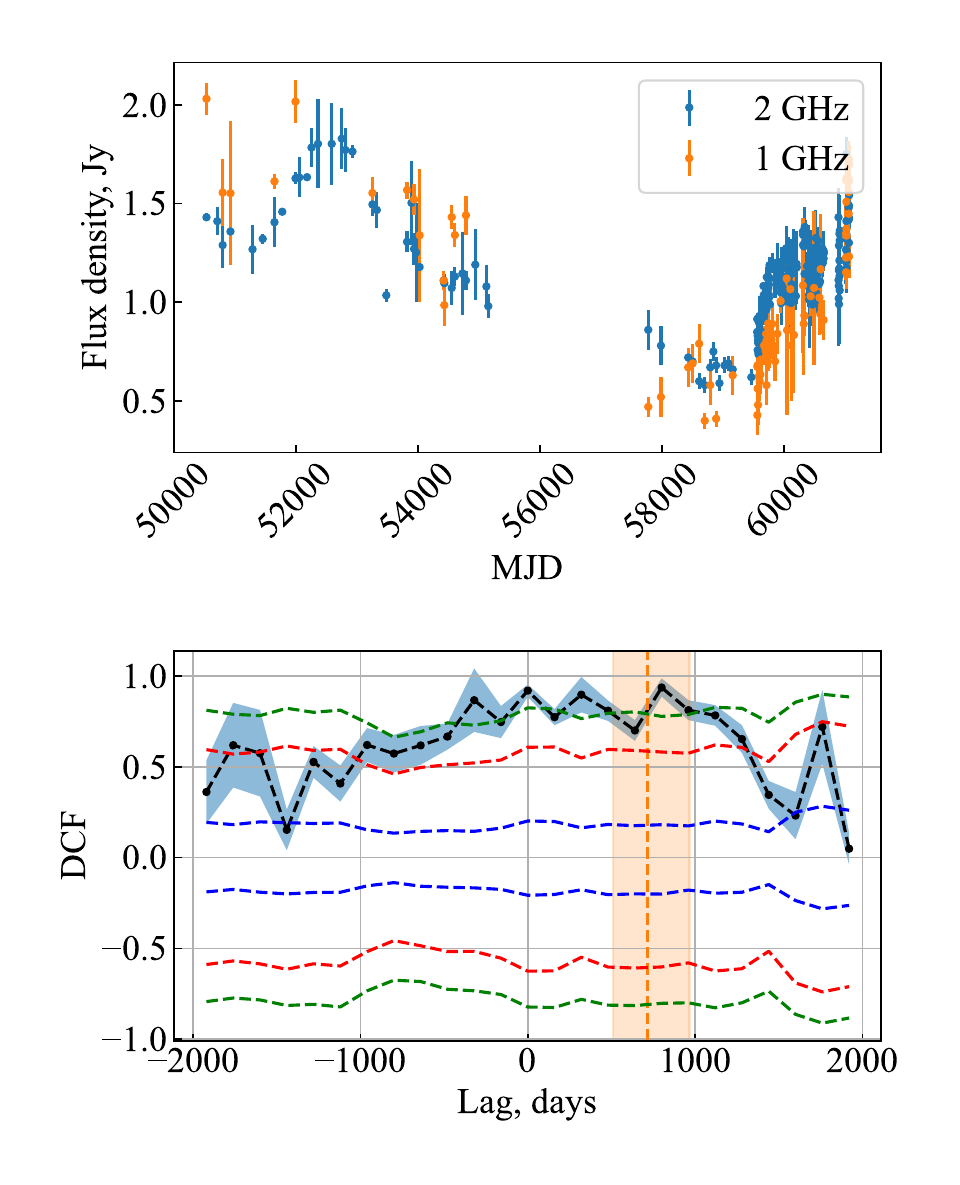}
}
\contcaption{} 
\end{figure*}

\section{Sinusoidal modelling results}
\label{app:sin}

This appendix presents the supplementary results of the sinusoidal modelling discussed in Section~\ref{sec:sin_model}. Fig.~\ref{fig:sin_model} shows the best-fitting sinusoidal curves for the individual light curves, while Fig.~\ref{fig:joint_period} presents the joint likelihood curve for the combined radio--millimetre fit. These figures are provided to illustrate the robustness of the characteristic $\sim 10$--11 yr modulation and to show how the fitted model compares with the data in each band.

\begin{figure*}
\centerline{
\includegraphics[width=\columnwidth]{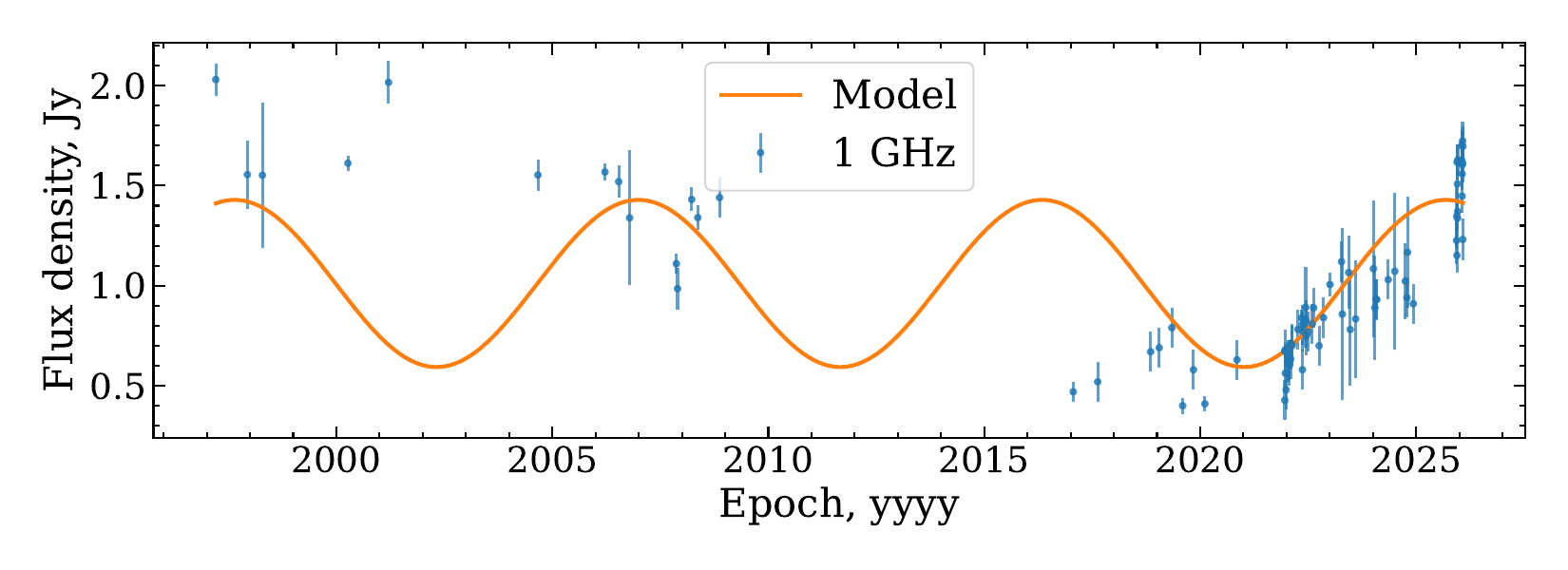}
\includegraphics[width=\columnwidth]{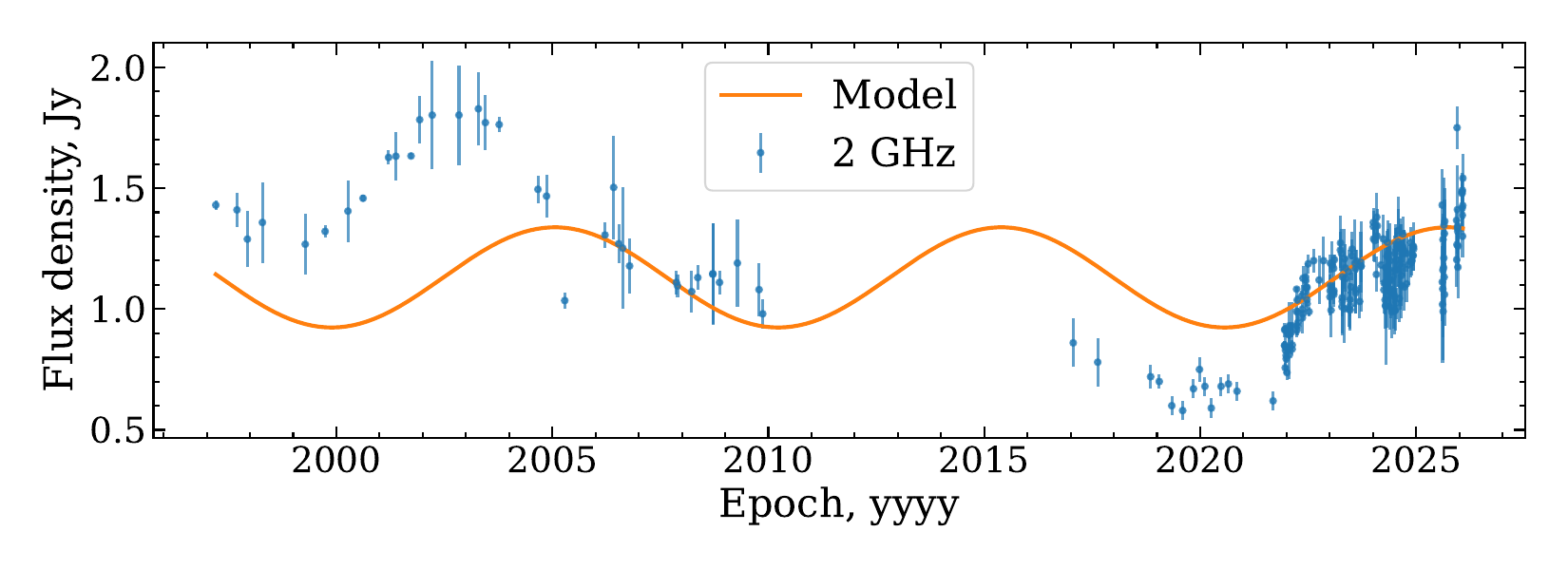}
}
\centerline{
\includegraphics[width=\columnwidth]{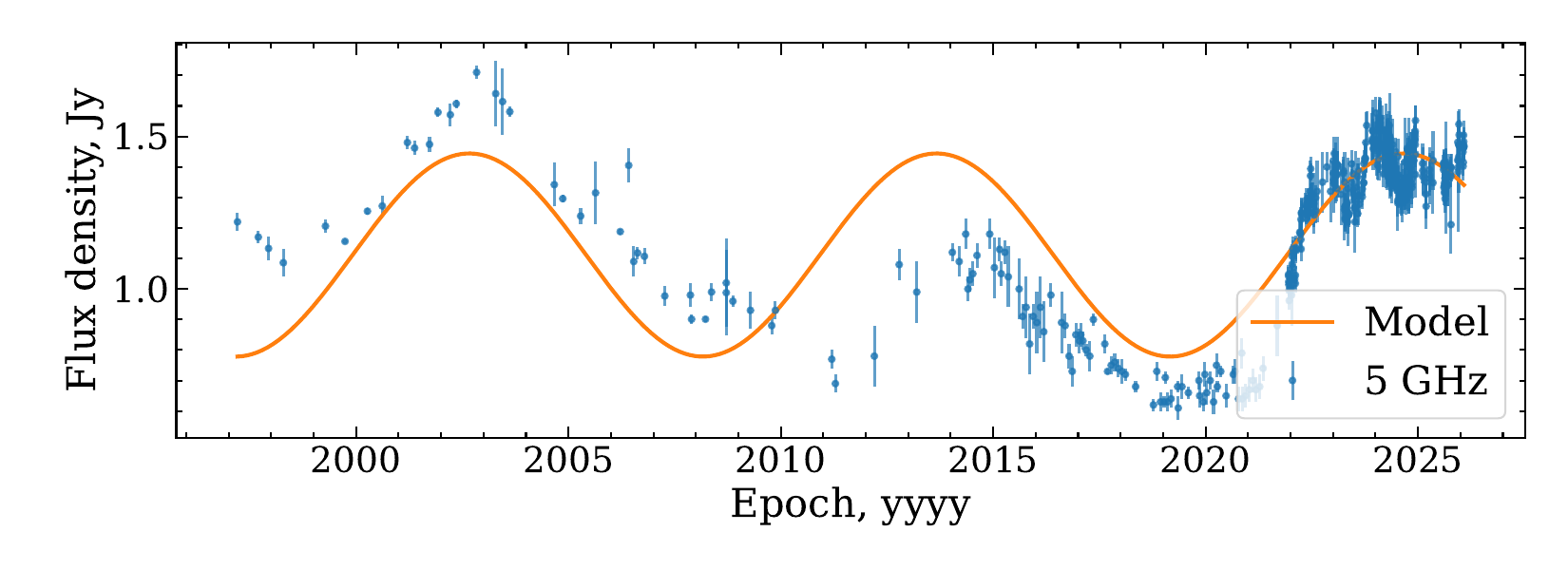}
\includegraphics[width=\columnwidth]{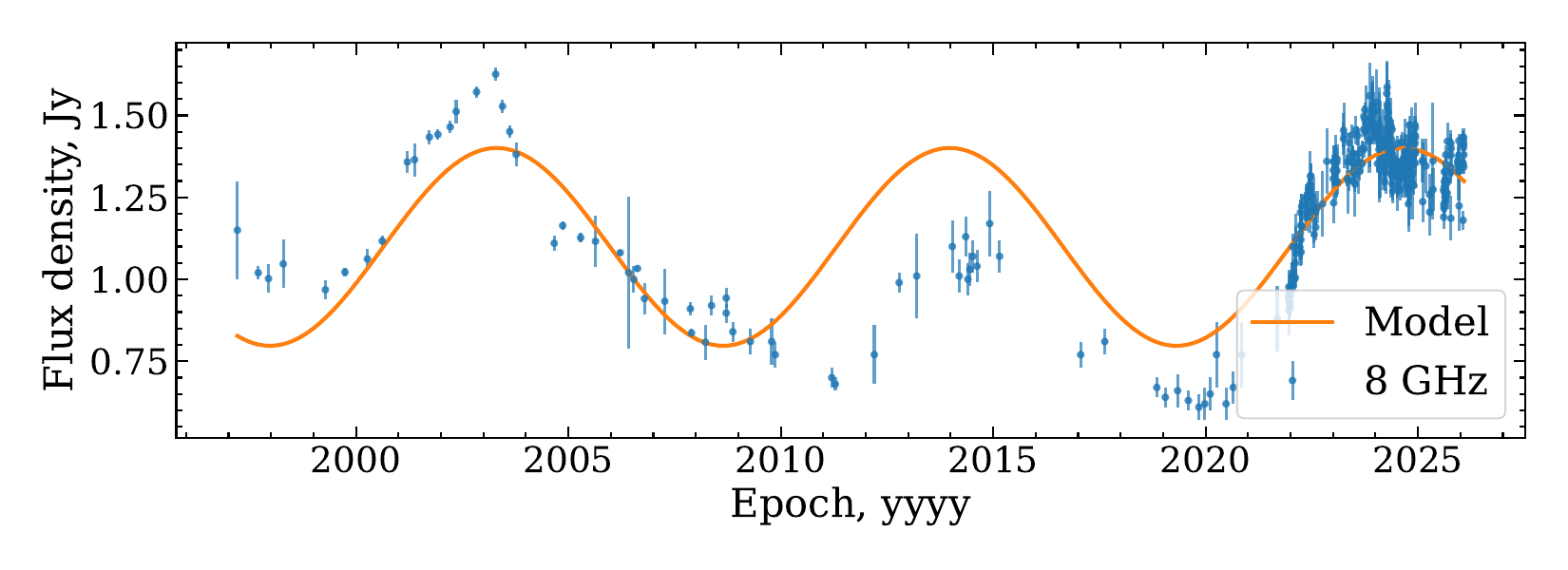}
}
\centerline{
\includegraphics[width=\columnwidth]{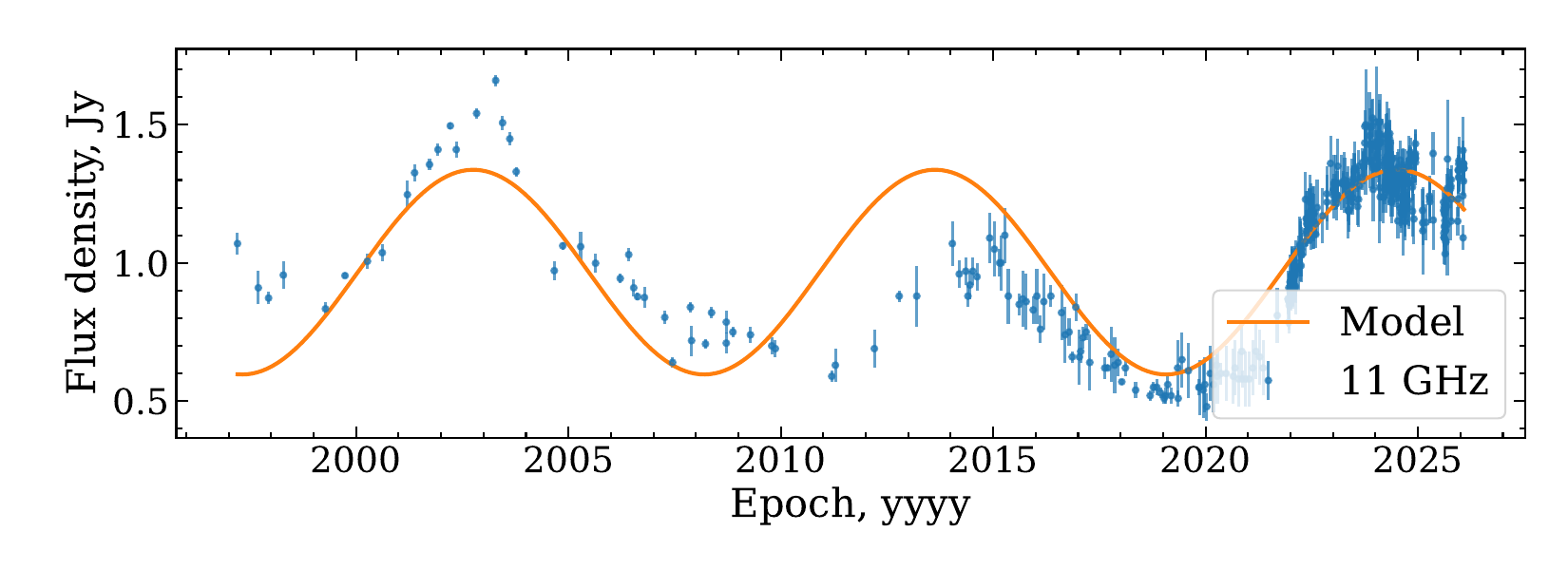}
\includegraphics[width=\columnwidth]{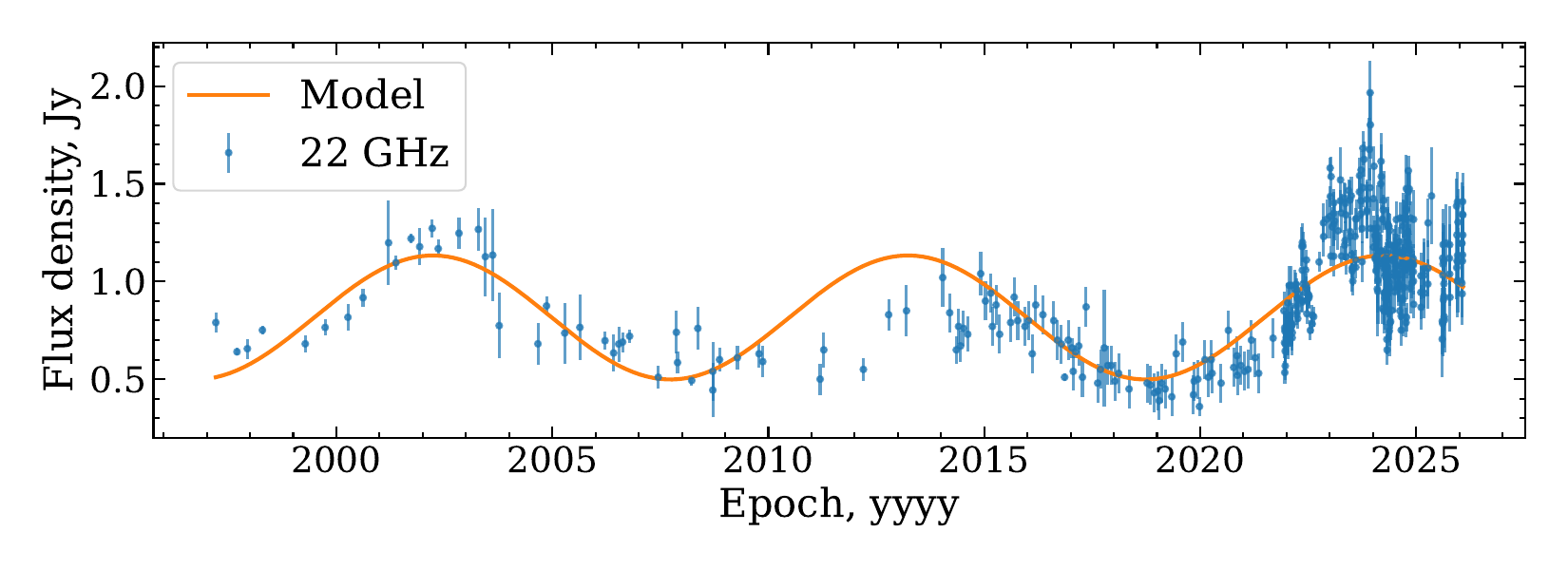}
}
\centerline{
\includegraphics[width=\columnwidth]{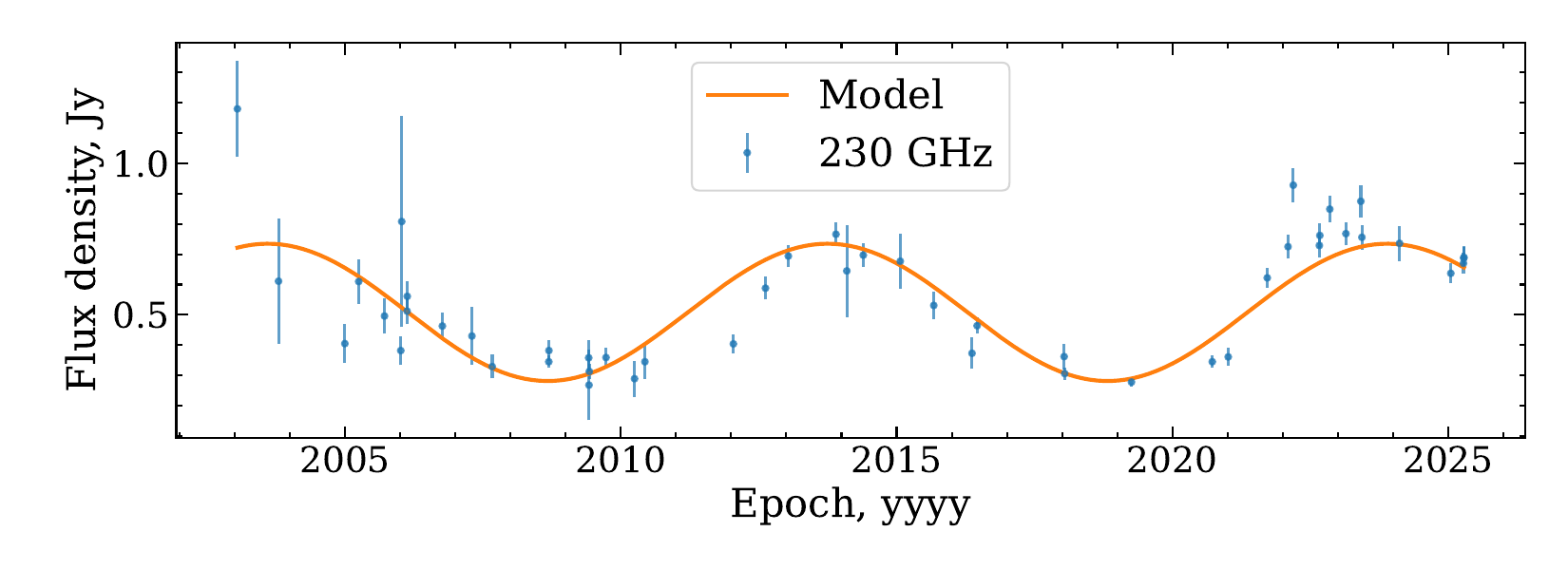}
\includegraphics[width=\columnwidth]{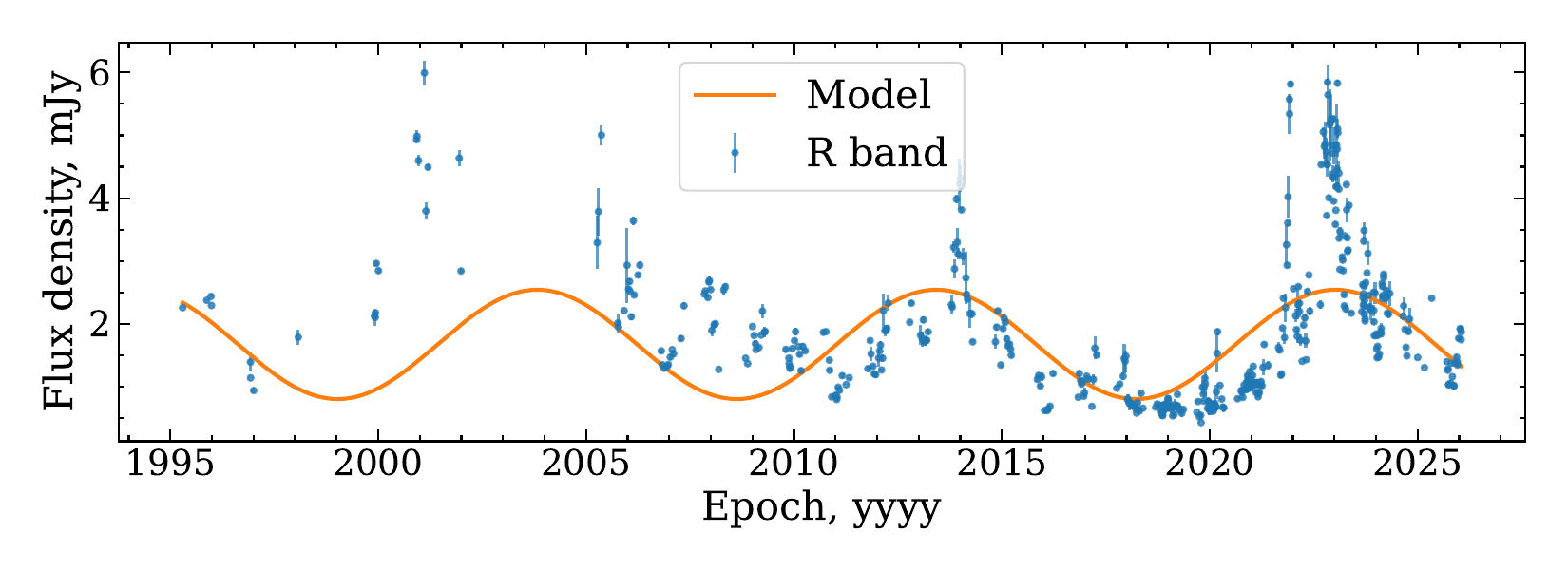}
}
\centerline{
\includegraphics[width=\columnwidth]{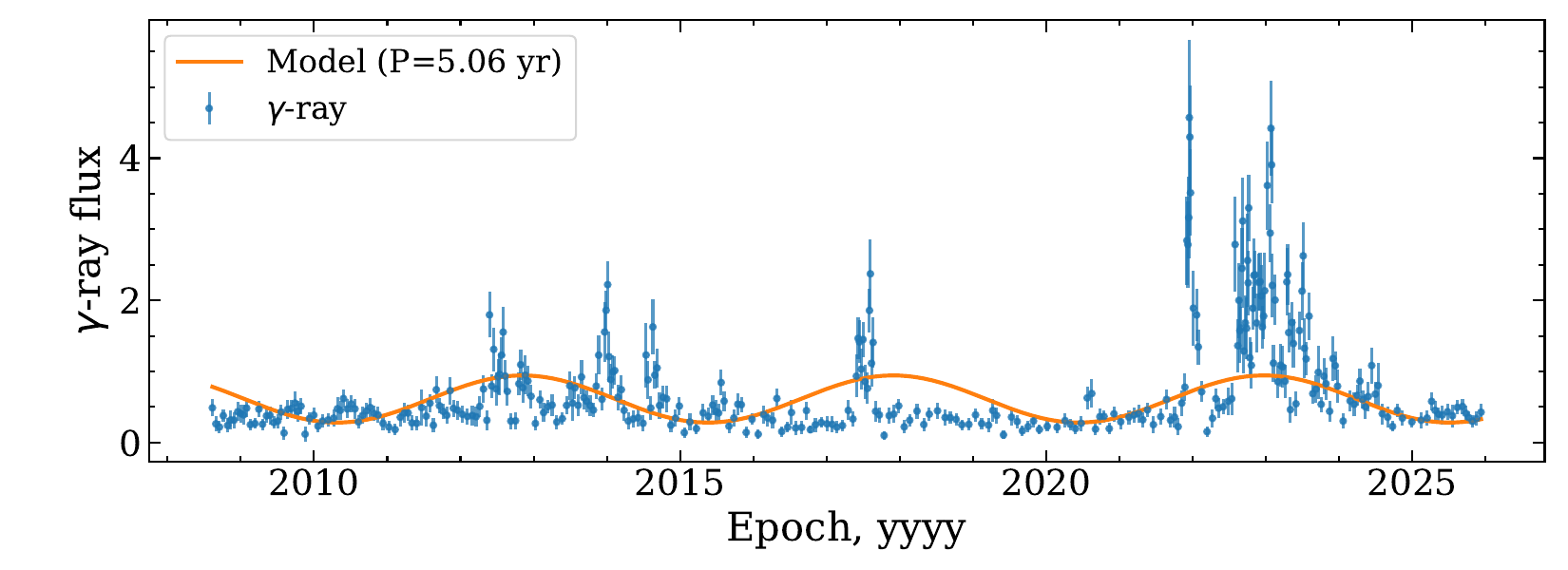}
\includegraphics[width=\columnwidth]{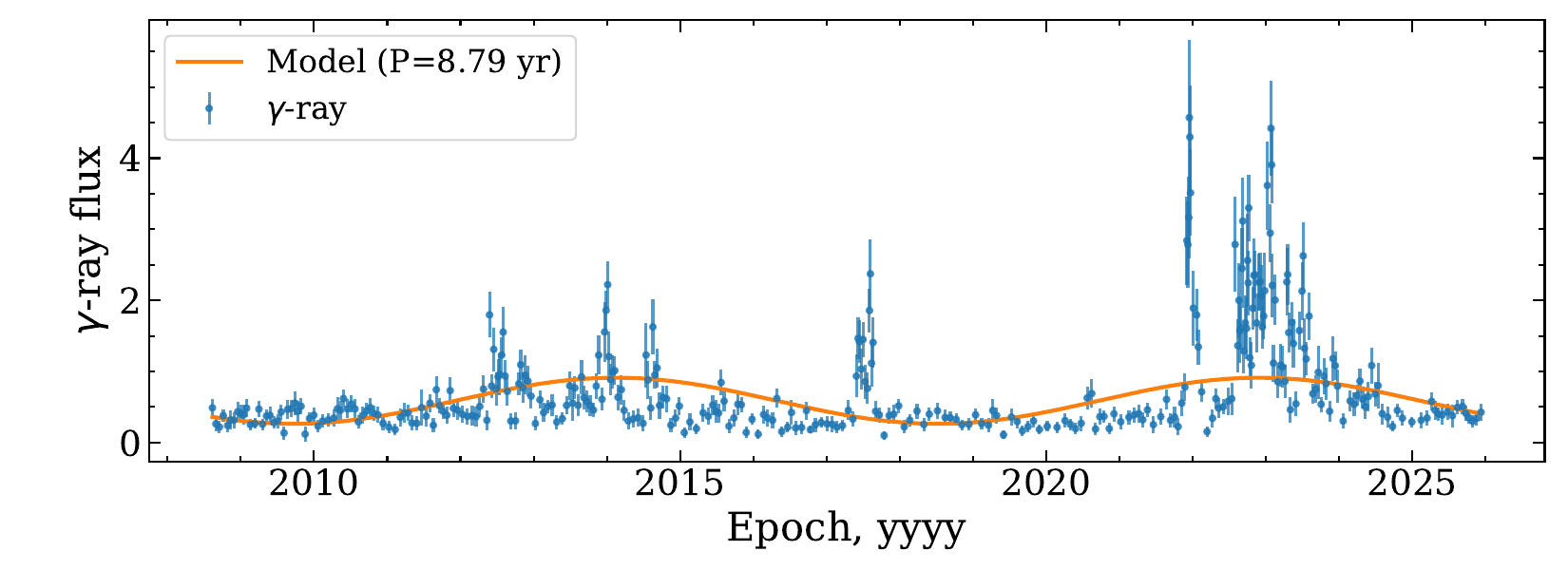}
}
\caption{Results of the sinusoidal modelling of the \pks multiband light curves.
The fitted curves are shown to illustrate the phenomenological modulation patterns inferred from the GLS and WWZ analyses. The radio bands exhibit consistent long-term modulation near $\sim$10--11 yr, while the optical and $\gamma$-ray bands show weaker and less stable shorter-timescale components.} 
\label{fig:sin_model}
\end{figure*}

\begin{figure}
\centering
\includegraphics[width=0.9\linewidth]{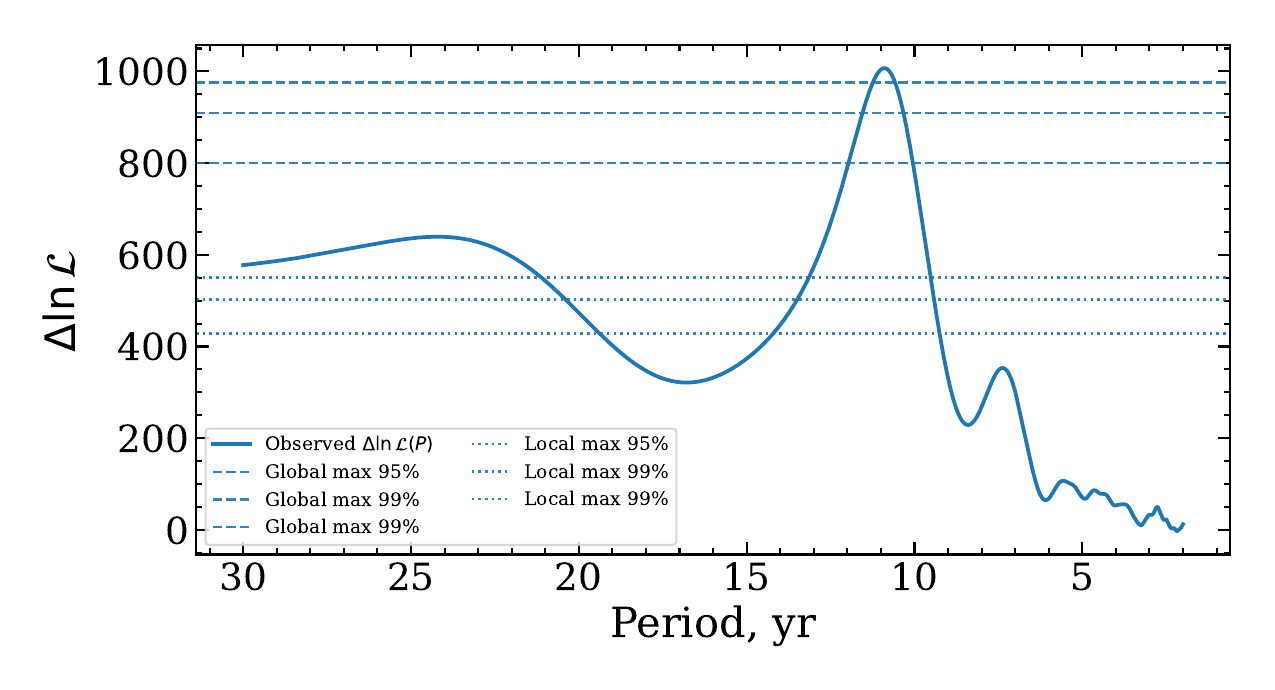}
\caption{Joint likelihood curve $\Delta\ln L(P)$ obtained from the combined sinusoidal fitting to the 5, 8, 11, 22, and 230 GHz light curves of \pks. The horizontal lines show the 95, 99, and 99.9 per cent significance thresholds derived from 5000 synthetic red-noise light curves generated with a power-law power spectral density ($P(f)\propto f^{-2}$). The likelihood peak occurs at $P = 10.9$ yr. The peak exceeds the red-noise thresholds, indicating that the observed quasi-periodic modulation is unlikely to be reproduced by stochastic red-noise variability.}
\label{fig:joint_period}
\end{figure}

\clearpage
\onecolumn 
\bsp
\label{lastpage}

\end{document}